\documentclass[12pt]{article}
\usepackage{amsmath}
\usepackage{euscript}
\usepackage[T1]{fontenc}
\usepackage[latin1]{inputenc}
\usepackage{graphics}
\usepackage{longtable}
\usepackage{color}
\usepackage{pslatex}
\usepackage{subfigure}
\usepackage{pstricks,epsfig,amssymb,axodraw}
\usepackage{cite}  
\begin{document}

\def\lb{\nextline}
 
\def\Order#1{{\cal O}($#1$)}
\def\Kuhn{K\"uhn}
\def\alphapi{\Bigl({\alpha\over\pi}\Bigr)}
\def\sovem{{s\over m^2_e}}
\def\Born{{\rm Born}}
\def\nubar{\bar{\nu}}
\def\nubarnu{\bar{\nu}\nu}
\def\nubart{\bar{\nu}_\tau}
\def\sstrut{$\strut\atop\strut$}

  \def\PL #1 #2 #3 {Phys. Lett. {\bf#1}           (#3)  #2}
  \def\NP #1 #2 #3 {Nucl. Phys. {\bf#1}           (#3)  #2}
  \def\PR #1 #2 #3 {Phys. Rev. {\bf#1}            (#3)  #2}
  \def\PP #1 #2 #3 {Phys. Rep. {\bf#1}            (#3)  #2}
  \def\PRL #1 #2 #3 {Phys. Rev. Lett. {\bf#1}     (#3)  #2}
  \def\CPC #1 #2 #3 {Comp. Phys. Commun. {\bf#1}  (#3)  #2}
  \def\ANN #1 #2 #3 {Annals of Phys. {\bf#1}      (#3)  #2}
  \def\APP #1 #2 #3 {Acta Phys. Pol. {\bf#1}      (#3)  #2}
  \def\ZP  #1 #2 #3 {Z. Phys. {\bf#1}             (#3)  #2}

\def\uncatcodespecials{\def\do##1{\catcode`##1=12 }\dospecials}
\def\setupverbatim{\tt
  \def\par{\leavevmode\endgraf} \catcode`\`=\active
  \obeylines \uncatcodespecials \obeyspaces \parindent=5mm \parskip=0pt}
{\obeyspaces\global\let =\ } 
{\catcode`\`=\active \gdef`{\relax\lq}}
\def\beginverbatim{\par\begingroup\setupverbatim\doverbatim}
{\catcode`\|=0 \catcode`\\=12 
  |obeylines|gdef|doverbatim^^M#1\endverbatim{#1|endgroup}}
\def\Was{\hbox{W\c as\;}}
\def\M{\hbox{\cal M}}
\def\lips{\hbox{Lips}}
\def\Im{\hbox{Im}}
\def\GeV{\hbox{GeV}}
\def\Maj{M_{R}}  \def\Gaj{\Gamma_{R}}
\def\beq{\begin{equation}} 
\def\eqiv{\sim}
\def\eeq{\end{equation}} 
\def\eps{\epsilon}
\begin{titlepage}
 
\begin{flushright} CERN-PH-TH/2006-056, \\
{ IFJPAN-IV-2006-3} 
\end{flushright}
 
\vspace{0.5cm}
\begin{center}
{\bf\Large
Next to Leading Logarithms  and the PHOTOS Monte Carlo 
}\end{center}
 
\begin{center}
   {\bf  P. Golonka}  and  {\bf  Z. W\c{a}s }\\
   {\em  
       CERN, 1211 Geneva 23, Switzerland\\
        and \\
        Institute of Nuclear Physics, PAN,
        Krak\'ow, ul. Radzikowskiego 152, Poland}\\

\end{center}
\vspace{.6 cm}
\begin{center}
{\bf   ABSTRACT   }
\end{center}
With the approaching start-up of the experiments at LHC, the urgency to quantify systematic
 uncertainties of the generators,  used in the interpretation of the data, 
is becoming pressing. The PHOTOS Monte Carlo program is often used for the simulation
of experimental, selection-sensitive, QED radiative corrections in decays of
$Z$ bosons and other heavy resonances and particles. Thanks  to its complete
phase-space coverage it is possible, with no approximations  
  for any decay channel, to implement  the  matrix-element.
 The present paper will be devoted to  those parts of 
the next-to-leading order corrections for $Z$ decays which are normally missing
in PHOTOS.  The analytical form of the
exact and truncated, {\it standard}, kernel used in PHOTOS will be explicitly given. The correction, being the ratio of the exact to the approximate kernel, 
can be activated as an optional contribution  to the internal weight of PHOTOS.

 To calculate the weight, the  information  on
the effective Born-level $Z/\gamma^*$ couplings and even directions of the 
incoming beams, is needed. 
A universal implementation  would have made the PHOTOS solution less modular 
and less convenient for the users.
That is why, for the time being, we will keep the correcting weight as an extra
option, available for special tests only.

We will quantify the numerical effect of the approximation with the 
help of a multitude of distributions. 
The numerical size of the effect is in general below 0.1\%; 
however, in some corners of the phase-space 
(well defined and contributing less than 0.5\% to the total rate),
it may reach up to about 20\%{} of their relative size.

\vspace{.5cm}
\begin{flushleft}
{  CERN-PH-TH/2006-056, \\  IFJPAN-IV-2006-3 \\
 April, 2006}
\end{flushleft}
 
\vspace*{1mm}
\bigskip
\footnoterule
\noindent
{\footnotesize \noindent
Supported in part  by the EU grant MTKD-CT-2004-510126, 
in partnership with the CERN Physics Department,
and the Polish State Committee for Scientific Research 
(KBN) grant 2 P03B 091 27 for years 2004--2006.
}
\end{titlepage}
\vspace{.5cm}
\begin{center}
{\bf 1. Introduction}
\end{center}
\vspace{.5cm}

Analysing the data from high-energy physics experiments, we try  to 
solve the {\it ``experiment = theory''} equation.  This non-trivial task
requires many different effects to be considered simultaneously.
From the experimental side, these are mainly detector acceptance and cuts,
which are dictated by the construction and physical properties of the detector: 
the shapes of distributions may be distorted by, say, misidentification 
and residual background contamination; these effects need to be discriminated 
in an appropriate and well-controlled way.
From the theoretical side, {\it all} effects of known physics have to be included in 
predictions as well. Only then can experimental data and theoretical predictions be 
confronted to determine numerical values of some coupling constants or 
effects of new physics (to be discovered). 

A well-defined class of theoretical effects contains the QED radiative 
corrections.
PHOTOS is a universal Monte Carlo algorithm  that simulates the effects of these 
 corrections in decays of particles and resonances.
It is a project with a rather long history: the first version was
released in 1991 \cite{Barberio:1990ms}, followed by version 2.0 
\cite{Barberio:1994qi} (double emission, threshold terms for fermions). 
The package is in wide use \cite{Dobbs:2004qw}:
it was applied as a precision simulation tool for 
$W$ mass measurement at the Tevatron \cite{Abazov:2003sv}
and LEP  \cite{Abbiendi:2003jh,Abdallah:2003xn},
and for CKM  matrix measurements 
in decays of $K$ and $B$ resonances 
(NA48 \cite{Lai:2004bt}, KTeV\cite{Alexopoulos:2004up}, Belle \cite{Limosani:2005pi},
BaBar \cite{Aubert:2004te} and at Fermilab \cite{Link:2004vk}). 

Throughout the years the core algorithm for the generation of $O(\alpha)$ corrections
did not change much; however, its precision, applicability to various 
processes, and numerical stability improved significantly. 
New features, such as multiple photon radiation
or interference effects for all possible decays, were also introduced.
 
Growing interest in the algorithm expressed by the experimental
collaborations (including the future LHC experiments) was a motivation to
perform a more detailed study of the potential and precision of the PHOTOS
algorithm. The present paper is the third in the series \cite{Golonka:2005dn,Golonka:2005pn}. It is devoted
to the $Z$ boson decay and to simplifications in the matrix element used 
in PHOTOS for  that channel.
We also explore the limitations originating from compromises
introduced into  PHOTOS bremsstrahlung kernels, which assured convenience
 of use; no process-dependent weight need be involved.

In that respect, the  study of the PHOTOS {\it matrix element} can be understood as 
a part of the  on-going effort to find 
 the practical solutions of the improved expansions. Some aspects of our solution
resemble those of classical exclusive exponentiation as described in \cite{kkcpc:1999,Jadach:2000ir};
in an other, the parton shower may be identified. The solution may be understood as a rearrangement 
of the QED perturbation expansion, yet this point will not be discussed here. Instead, let us point to some
similarities of the PHOTOS solution to the methods discussed elsewhere:  
interaction picture  of Quantum Mechanics, expansion of 
special functions around asymptotic solutions \cite{Nikiforov} or in field theory;
see eg. \cite{Tkachov:1997gz}. In PHOTOS the  expansion is performed 
in terms of multidimensional operators.

The paper is organized as follows. In Section 2 the main properties used in the PHOTOS design 
and, in particular, the analytical form of the (NLO) weight, necessary to introduce the complete
first-order matrix-element, are presented.
It is also explained there how the complete matrix elements break the requirement
 of separation of the calculation of the final-state bremsstrahlung 
from the properties of the Born-level matrix elements and the phenomena affecting
the $Z$ production. To support the discussion and visualize the results, 
 a multitude of numerical 
comparisons and tests will be presented.
Section 3 provides the definition of the method used in those comparisons. 
The method is particularly suitable to visualize the results in the non-collinear regions of the phase space.
 Section 4 presents numerical tests performed at fixed first order of the QED expansion. 
Since
PHOTOS uses the same building block for a part of the single-photon  generation algorithm
 and for the multiple bremsstrahlung, the results presented in this
section have implications for the multiple-photon option of PHOTOS.
Section 5 addresses these aspects of the program construction which are relevant 
to the use of the NLO weight
in the multiple-photon option. Section 6 collects the
results of the tests performed for the programs run with multiple-photon emission. 
Summary, Section 7, closes the paper.

\vspace{.5cm}
\begin{center}
{\bf 2. Phase space and matrix element}
\end{center}
\vspace{.5cm}

To discuss the implementation of the 
complete first-order QED radiative corrections in $Z$ decay, 
we must  start with the complete parametrization of the phase space.

Let us start with the explicit expression for the parametrisation of the $(n+1)$-body final state 
in the decay of  an  object of four-momentum $P$. 
To define iterative relations, let us denote the four-momenta 
of the first  $n$ decay products as $k_i$,  and  the last $(n+1)$-th as $q$.
For the case discussed here, the ($n+1$)-th particle will  always be the real and massless  
photon. However, the parametrization does not rely on this assumption and, in principle, can be 
applied to define other formulas for the phase space, such as the
emission of a (massive) pion, and could even be extended to the case of emission of pairs of heavy particles. 
In later steps of our construction the fact that photons are massless and the related properties 
of QED matrix elements will of course be used. 

In the following, the notations from refs. \cite{Was:1994kg,Jadach:1993hs} will be used. 
We will, however,  not rely on any particular results of those papers and only
point to the more detailed presentations of other, nonetheless 
quite similar, options for
the exact $n$-body phase-space parametrization to the one presented here.

Let us define the element of Lorentz-invariant phasespace ({\it Lips}) as follows:

\begin{eqnarray}
&&dLips_{n+1}(P)= \nonumber\\
&&
{d^3k_1 \over 2k_1^0 (2\pi)^3}\; . . .\;{d^3k_n \over 2k_n^0 (2\pi)^3}
{d^3q \over 2q^0 (2\pi)^3}
(2\pi)^4 \delta^4\Bigl(P -\sum_1^n k_i-q\Bigr)\nonumber\\
&=&
d^4p\delta^4(P -p-q){d^3q \over 2q^0 (2\pi)^3}
{d^3k_1 \over 2k_1^0 (2\pi)^3} \;. . .\;{d^3k_n \over 2k_n^0 (2\pi)^3}
(2\pi)^4 \delta^4\Bigl(p -\sum_1^n k_i\Bigr)\nonumber\\
&=&
d^4p\delta^4(P -p-q){d^3q \over 2q^0 (2\pi)^3} dLips_n(p\to k_1 ... k_n).
\label{Lips_n+1}
\end{eqnarray}
 Extra integration variables, the four-vector $p$, compensated with $\delta^4\bigl(p -\sum_1^n k_i\bigr) $, are first introduced.
In the next step, another integration variable $M_1$ and 
$\delta\bigl(p^2 -M_1^2\bigr) $ are introduced.
The element of the phase-space integration may thus be transformed into:

\begin{eqnarray}
&&dLips_{n+1}(P)=  \nonumber\\
&&
{dM_1^2 \over (2\pi)} dLips_2(P \to p\ q) \times dLips_n(p \to k_1 ... k_n)\nonumber\\
&=&
dM_1^2  \biggl[d\cos\theta d\phi {1 \over 8(2\pi)^3}
{\lambda^{1\over 2}(M^2, {M_1^2 },{m^2 })\over M^2}\biggr]
\times dLips_n(p \to k_1 ... k_n).
\label{Lips_n+1.3}
\end{eqnarray}

The part of the phase-space Jacobian corresponding to the integration over the direction 
and the energy of the last particle (or invariant mass of the remaining system) is explicitly given;
$\lambda(a, b,c)={a^2+b^2+c^2-2ab-2ac-2bc}$. 
The integration over the angles is defined in the  rest frame of $n+1$ particles;
the integration over the invariant mass $M_1$ is limited by the phase-space boundaries.
There is no need to choose the axes with respect to which the angles are oriented;  
we will not elaborate on that point here, as details can be found in Ref. \cite{Barberio:1990ms}.
 Formula (\ref{Lips_n+1.3}) may 
be iterated to  provide parametrization of the phase space with an arbitrary number of 
final-state 
particles. The question of the orientation of the reference frames used to define the angles
and the order of the choice for  limits in $M_i$ integrations, becomes particularly complex then; our choice is described in ref. \cite{Barberio:1994qi}. 
Since nothing new was introduced for the purpose of the present study
we will not discuss this interesting point further.  Except for the details mentioned above,
 the choice we made for the phase-space organization is  the same as in  
FOWL \cite{FOWL}, TAUOLA \cite{Jadach:1993hs}, and probably many other generators.

To simplify the formula for the phase space, let us finally take advantage of the zero mass of the photon. 
The invariant mass of the system of all particles but the first one may be replaced by the energy
of the first one (defined in the $P$ rest frame). The phase-space formula can then be written as:
 
\begin{eqnarray}
 && dLips_{n+1}(P)=  \nonumber\\
&&
\biggl[ 4 dk_\gamma {k_\gamma} d\cos\theta d\phi {1 \over 8(2\pi)^3}
\biggr]
\times dLips_n(p \to k_1 ... k_n) \nonumber\\
&=&
\biggl[ k_\gamma dk_\gamma  d\cos\theta d\phi {1 \over 2(2\pi)^3}
\biggr]
\times dLips_n(p \to k_1 ... k_n).
\label{Lips_n+1.5}
\end{eqnarray}
If we had $l$  photons accompanying $n$ other particles,
the factor in square brackets would be iterated.
A statistical factor ${1 \over l!}$ would complete the formula for the phase-space 
parametrization, which is quite similar to the formal expansion
of the exponent\footnote{The exact form of the functional 
exponent is achieved if the four-vector 
$p$  is  replaced by $P$ in formula (\ref{Lips_n+1.5}). In this way the tangent
space for the ($n+1$)-body phase space can be constructed. We use that space, together with 
an eikonal-like  form   of the matrix element 
(emissions from individual final-state charged products  are treated as independent),
for the  construction of the crude distribution of photon emission probability. 
Note that, in this space, the photons' four-momenta are unconstrained by energy--momentum conservation. 
The limits on the energies of the photons
are arbitrary. We checked that, at the $10^{-4}$ precision level, the results obtained from our
 simulations do not depend on the particular choice.   
We leave the underlying  formal aspect
of the algorithm to future papers.}.
The last formula, supplemented with the definition of the orientation of the angles, is used
to define the full kinematic configuration of the event. 
The four-momenta of all final-state particles may now be constructed
from the angles and energies ($k_{\gamma_i})$ 
of the photons, and the angles and masses of the other decay products.

Similarly, an inverse operation may be performed; the energies and angles for 
the parametrization could be reconstructed from the four-vectors (even though the  parametrization was not
 necessarily   used in the previous generation steps).
The  phase-space Jacobians may be easily  calculated as 
well. 
By replacing $dLips_n(p \to k_1 ... k_n)$ in formula (\ref{Lips_n+1.5}) by 
 $dLips_n(P \to k_1 ... k_n)$ we obtain a parametrization where the  photons do not affect the construction 
of other particles' momenta.  
This operation could be considered as treating the photon in 
an approximation valid only in the soft photon limit. 
This, however, {\it does not need} to be the case. In the first step, the photon may be 
constructed with an arbitrarily large momentum, as nothing else depends on it.  
The kinematical variables of the photon are generated with the help of the distribution defined 
by the factor 
$\bigl[ k_\gamma dk_\gamma  d\cos\theta d\phi {1 \over 2(2\pi)^3} \bigr] $;
which provides the photon variables of  the {\it tangent
space}.
 Fully constructed with four-momenta, an  event of the $n$-body decay can be turned back into a representation
of angles and invariant masses.   In the final step these angular variables, together with those
of the photon, can be used to define a new event in the ($n+1$)-body phase space.
In the case when the new kinematical variables do not fit the limits of available 
($n+1$)-body
phase space  the new event should be rejected and the original configuration 
(in the $n$-body phase space) kept.  An important property of the algorithm presented here
 is the full coverage of the  ($n+1$)-body phase-space being assured. 
 In this procedure, the difference between $n$-body and $n+1$ body phase-space Jacobians 
can be calculated in an unambiguous way and introduced in the same rejection step as for the phase-space
limits\footnote{The effects of 
 matrix elements, including those of virtual corrections, have to be introduced at this stage as well.
They are indispensable, for example, to calculate relative probabilities of configurations with distinct number 
of  final-state particles. }. 

The features and transformations of the phase-space parametrization presented here
 are at the heart of the construction 
of the PHOTOS kinematics and have been used since its beginning.
To complete the generation of photons, the exact phase-space parametrization must be completed with a
matrix  element, with both  virtual and  real QED corrections included. Careful 
regularization  of soft singularities must be performed\footnote{Volumes of the partial width attributed to the configurations
with $n$, $n+1$ particles, etc.,  have to be normalized to the total width, both at the 
level of the tangent and the correct (final) phase space.}.   

In the standard version of PHOTOS, as published in \cite{Barberio:1990ms,Barberio:1994qi},
the  expression
\begin{eqnarray}
X_{f}^{\mathrm{PHOTOS}}=&\frac{Q'^{2}\alpha(1-\Delta)}{4\pi^{2}s}s^{2} \hskip 3 mm \Bigg\{ \hskip 8 cm \nonumber \\
\frac{1}{k'_{+}+k'_{-}}\frac{1}{k'_{-}}&\bigg[(1+(1-x_{k})^{2})
\frac{{d}\sigma_{B}}{d\Omega}\Big(s,\frac{s(1-\cos\Theta_{+})}{2},
\frac{s(1+\cos\Theta_{+})
}{2}\Big)\bigg]\frac{(1+\beta\cos\Theta_{\gamma})}{2}\;\;\; \nonumber\\
+ 
\frac{1}{k'_{+}+k'_{-}}\frac{1}{k'_{+}}&\bigg[(1+(1-x_{k})^{2})
\frac{{d}\sigma_{B}}{d\Omega}\Big(s,\frac{s(1-\cos\Theta_{-})}{2},
\frac{s(1+\cos\Theta_{-})
}{2}\Big)\bigg]\frac{(1-\beta\cos\Theta_{\gamma})}{2}\Bigg\} \nonumber \\
\mathrm{where:} & \Theta_{+}=\angle(p_{+},q_{+}),\; \Theta_{-}=\angle(p_{-},q_{-}), 
\;\hskip 4 cm \nonumber\\
 & \Theta_{\gamma}=\angle(\gamma,\mu^{-})\;  \textrm{is\, defined\,
in}\;(\mu^{+},\mu^{-})\textrm{-pair\, rest\, frame,} \hskip 1.2 cm
\label{X-fotos}
\end{eqnarray}
is used for the real-photon matrix element. The virtual corrections 
are requested to be such that the total decay rate remains unchanged after
complete QED corrections are included. The expression, without approximation, reads:
\begin{eqnarray}
X_{f}=\frac{Q'^{2}\alpha(1-\Delta)}{4\pi^{2}s}s^{2} &
\Bigg\{\frac{1}{(k'_{+}+k'_{-})}\frac{1}{k'_{-}}\bigg[\frac{{d}\sigma_{B}
}{{d}\Omega}(s,t,u')+\frac{{d}\sigma_{B}}{{d}\Omega}(s,t',u
)\bigg]\nonumber \\
 &
+\frac{1}{(k'_{+}+k'_{-})}\frac{1}{k'_{+}}\bigg[\frac{{d}\sigma_{B}}{{d}
\Omega}(s,t,u')+\frac{{d}\sigma_{B}}{{d}\Omega}(s,t',u)\bigg
]\Bigg\}.
\label{X-mustraal}
\end{eqnarray}
The combined effect of the virtual and real corrections on the total rate is its 
increase by a factor of
$1+ \frac{3}{4} \frac{\alpha}{\pi}$.

The notation from ref.~\cite{Berends:1982ie} are used:
\begin{eqnarray}
s=2p_+\cdot p_-,  & s'=2q_+\cdot q_-, \nonumber \\
t=2p_+\cdot q_+,  & t'=2p_+\cdot q_-, \nonumber\\
u=2p_+\cdot q_-,  & u'=2_-\cdot q_+, \nonumber \\
k'_{\pm}=q_\pm\cdot k, &  x_k= 2E_\gamma/\sqrt{s}.
\end{eqnarray}
This paper
collects complete first-order radiative corrections for the process 
$e^+e^-\to \mu^+\mu^-(\gamma)$. Final-state bremsstrahlung 
constitutes part of these results, where nonetheless reference is made to the incoming 
electron beam momenta.

 The $\Delta$ term encapsulates final-state mass-dependent terms,
$p_+$, $p_-$, $q_+$, $q_-$, $k$ denote four-momenta of:
 incoming $e^+$, $e^-$, 
outcoming $\mu^+$, $\mu^-$ and the bremsstrahlung photon respectively. 
Expression (\ref{X-mustraal})
 is explicitly taken from ref.  \cite{Berends:1982ie}, on Monte Carlo MUSTRAAL, 
which is where  the interested reader will find the   details of the definitions
of variables and expressions such as $\Delta$, used in the 
formulae (\ref{X-fotos}) and (\ref{X-mustraal}). 

The ratio of (\ref{X-mustraal}) to (\ref{X-fotos}) constitutes the basic element of upgrading 
PHOTOS functionality to the complete first order\footnote{This is only true for PHOTOS 
being run at first order. For the multiple-photon radiation option,
the iteration of the single-photon emission kernel (and thus also its weight) is performed; see section 5.}.
Nothing needs to be changed in the phase-space parametrization.
The effects of the virtual corrections have to be included as well and have to be properly introduced
in the normalization.
The expression for the correcting weight could be chosen\footnote{ Alternatively, a factor $(1+ \frac{3}{4} \frac{\alpha}{\pi})$ can be included in the definition 
of the crude distribution.}  simply as
\begin{equation}
wt=\frac{X_{f}}{X_{f}^{\mathrm{PHOTOS}}}\frac{1}{(1+ \frac{3}{4} \frac{\alpha}{\pi})}.
\label{wgt}
\end{equation}
For the purpose of constructing a Monte Carlo algorithm, however, it is more convenient
 to separate it into a sum of two generation branches (with slightly different angular variable mapping).
Then, the expression for the distribution and those for the weight take the form
\begin{eqnarray}
X_{f}=& X_{f}^1+X_{f}^2 \hskip 13 cm \nonumber \\
X_{f}^1=& WT_1 \;\; \frac{Q'^{2}\alpha(1-\Delta)}{4\pi^{2}s}s^{2} \frac{1}{k'_{+}+k'_{-}}\frac{1}{k'_{-}}\bigg[(1+(1-x_{k})^{2})
\frac{{d}\sigma_{B}}{d\Omega}\Big(s,\frac{s(1-\cos\Theta_{+})}{2},
\frac{s(1+\cos\Theta_{+})
}{2}\Big)\bigg]\frac{(1+\beta\cos\Theta_{\gamma})}{2},\;\;\nonumber \\
X_{f}^2=& WT_2 \;\; \frac{Q'^{2}\alpha(1-\Delta)}{4\pi^{2}s}s^{2} \frac{1}{k'_{+}+k'_{-}}\frac{1}{k'_{+}}\bigg[(1+(1-x_{k})^{2})
\frac{{d}\sigma_{B}}{d\Omega}\Big(s,\frac{s(1-\cos\Theta_{-})}{2},
\frac{s(1+\cos\Theta_{-})
}{2}\Big)\bigg]\frac{(1-\beta\cos\Theta_{\gamma})}{2}, \nonumber \\
 WT_1=&  \frac{\frac{{d}\sigma_{B}
}{{d}\Omega}(s,t,u')+\frac{{d}\sigma_{B}}{{d}\Omega}(s,t',u
)}{\bigg[(1+(1-x_{k})^{2})
\frac{{d}\sigma_{B}}{d\Omega}\Big(s,\frac{s(1-\cos\Theta_{+})}{2},
\frac{s(1+\cos\Theta_{+})
}{2}\Big)\bigg]\frac{(1+\beta\cos\Theta_{\gamma})}{2}\; \big(1+ \frac{3}{4} \frac{\alpha}{\pi}\big)},    \hskip 5 cm    \nonumber \\
 WT_2=&   \frac{\frac{{d}\sigma_{B}}{{d}
\Omega}(s,t,u')+\frac{{d}\sigma_{B}}{{d}\Omega}(s,t',u)}{\bigg[(1+(1-x_{k})^{2})
\frac{{d}\sigma_{B}}{d\Omega}\Big(s,\frac{s(1-\cos\Theta_{-})}{2},
\frac{s(1+\cos\Theta_{-})
}{2}\Big)\bigg]\frac{(1-\beta\cos\Theta_{\gamma})}{2}\; \big(1+ \frac{3}{4} \frac{\alpha}{\pi}\big)}. \hskip 5 cm
\label{wgt1}
\end{eqnarray}
At this point, let us make the following remark.
Event though the introduction of the NLO weight into PHOTOS is  trivial,  the developed 
approximation \cite{Barberio:1990ms} at the heart of PHOTOS design is not. It enabled 
universality of the program\footnote{Indeed, after  inspection, the differences between
formulae (\ref{X-fotos})  and (\ref{X-mustraal}) are quite significant. 
The exact expression does 
not allow a transfer of the complete Born-level angular dependence to the host generator. In the
correction weight, the two contributions, one depending on the angle $\Theta_{+}$ and another on  $\Theta_{-}$,
have to be simultaneously included. The dependence on the Born-level (effective) couplings 
thus need 
to be  known at the level of the calculation of the final-state bremsstrahlung weight. This would make
the modular structure of PHOTOS design more difficult to keep. Also, the direction
of the (effective) beam need to be provided for the calculation of  $\Theta_{+}$ and $\Theta_{-}$ angles. 
This exhibits another difficulty
in the separation of the final-state bremsstrahlung and the dynamics  of the initial state for $Z/\gamma^*$ 
production.}. Simplification was not necessary to  attribute the generation 
of bremsstrahlung photons to individual charged particles%
\footnote{ For  other decays, it will probably
not be necessary  to find an explicit form of such NLO 
separation. 
Starting from  the NNLO, such separation was shown to be impossible 
\cite{Kleiss:1990jv} anyway.  }. The separation  holds for the complete  NLO as well.
The simplified emission
kernel, which we used for other decays as well, reads:
\begin{eqnarray}
&  \frac{Q'^{2}\alpha(1-\Delta)}{4\pi^{2}s}s^{2} \frac{1}{k'_{+}+k'_{-}}\frac{1}{k'_{-}}(1+(1-x_{k})^{2})                  
\frac{(1+\beta\cos\Theta_{\gamma})}{2}\;\;\nonumber \\
&  \frac{Q'^{2}\alpha(1-\Delta)}{4\pi^{2}s}s^{2} \frac{1}{k'_{+}+k'_{-}}\frac{1}{k'_{+}}(1+(1-x_{k})^{2})
\frac{(1-\beta\cos\Theta_{\gamma})}{2}.
\label{dist}
\end{eqnarray}
It depends  
on the spin and charge of the ``emitting particle'' only\footnote{ In the original PHOTOS 
documentation we called this well-controlled truncation of the
kernel a ``property (such) that  leading-log (collinear) and infrared limits are properly reproduced''. 
This explanation turned out to be misleading for many readers. 
One can get biased, and  expect the  collinear approximation not only
for the kernel, but for the whole design of the algorithm. 
This would be a serious limitation of our program design
if indeed, as  suggested in ref. \cite{Hamilton:2006xz}, 
``PHOTOS was based on collinear approximation''. Fortunately it is not the case.  
Such confusion was not a concern  for the users, until now: precision 
requirements  were not as high.
}. It does not depend on the properties of the other decay products, which only define the phase-space 
limits. To obtain the universal
form of the photon emission kernel, the  interference was eliminated with the help of 
the factor, either  $\frac{2}{(1+\beta\cos\Theta_{\gamma})}$ or $\frac{2}{(1-\beta\cos\Theta_{\gamma})}$. 
The interference is  recovered later,  using the weight given in formula (17)   of ref. \cite{Barberio:1994qi}, that is with approximation. 
On the other hand, having paid the price of the approximated solution, both the kernel 
and the interference weight  can then  be used  
for   decay of any particle or resonance.

In our present study the
analytical expression for the matrix element for the 
$e^{+} e^{-} \to Z^{0}/\gamma^{*} \to \mu^{+} \mu^{-}$ is used 
(and compared to the approximated, yet process-independent, solution of the standard 
PHOTOS). 
If necessary, a matrix element for other decay processes (if available) could be used.

\vspace{.5cm}
\begin{center}
{\bf 3. Method used in numerical tests }
\end{center}
\vspace{.5cm}

In the comparison of the multitude of final states generated, 
at  different levels of physics sophistication by two distinct Monte Carlo programs,
 the choice of a method  is of great importance. 
To compare the Monte Carlo programs it is quite common to present the distributions generated by the programs superimposed on a single plot,
often in logarithmic scale. 
Such a method was used, for example, in ref. \cite{Hamilton:2006xz}. The method is
unquestionably sufficient, if one's interest is limited to, say, the collinear content
of the results or other distributions of the intrinsically logarithmic type.  

For instance, applying this method for the comparison
of the total energy carried out by all bremsstrahlung photons,
we would obtain a distribution such as those presented in Fig. \ref{fig:classical}.
We could conclude that there is excellent agreement, and the non-leading
effects, which are essential for estimating systematic errors for generators 
like PHOTOS,  would be marginalized in the presentation.  
\begin{figure*}
\caption{\label{fig:classical} \it \small A typical plot for comparisons, as 
described in ref. 
\cite{Hamilton:2006xz}. We use it to illustrate our method.
The histograms presented in the left and right plots have identical content,
and  show a comparison of KKMC \cite{kkcpc:1999} and PHOTOS
used in $Z$ decay.  
The total energy carried out by all  final-state photons is presented. 
Red (darker grey) colour represents the results of KKMC, green (lighter grey) of
 PHOTOS with the NLO weight activated. Samples of $10^7$ events were used in this 
comparison. The results are overwhelmed  
by the collinear/soft  content of the predictions.
\newline
If the $W$ instead of  $Z$ decay was chosen,
it would not be 
the case. There, NLO effects would be dominant 
for the part of the spectrum above 
$M_W/2$.
However, in that case, we would not profit from the second-order matrix element 
Monte Carlo, available  for tests. }

\begin{center}
\subfigure[A logarithmic scale is used. 
Excellent agreement between the two programs is
visible all over the energy range from 0 to $M_Z$. 
The presence of the two lines can be spotted  at the high end of the spectrum,
 mainly thanks to the statistical errors.
 The kink at the limit of the phase-space, where  single 
hard-photon configuration ceases to contribute,  dominates the content of the whole picture. 
 ]{\includegraphics[%
  width=0.49\columnwidth,
  keepaspectratio]{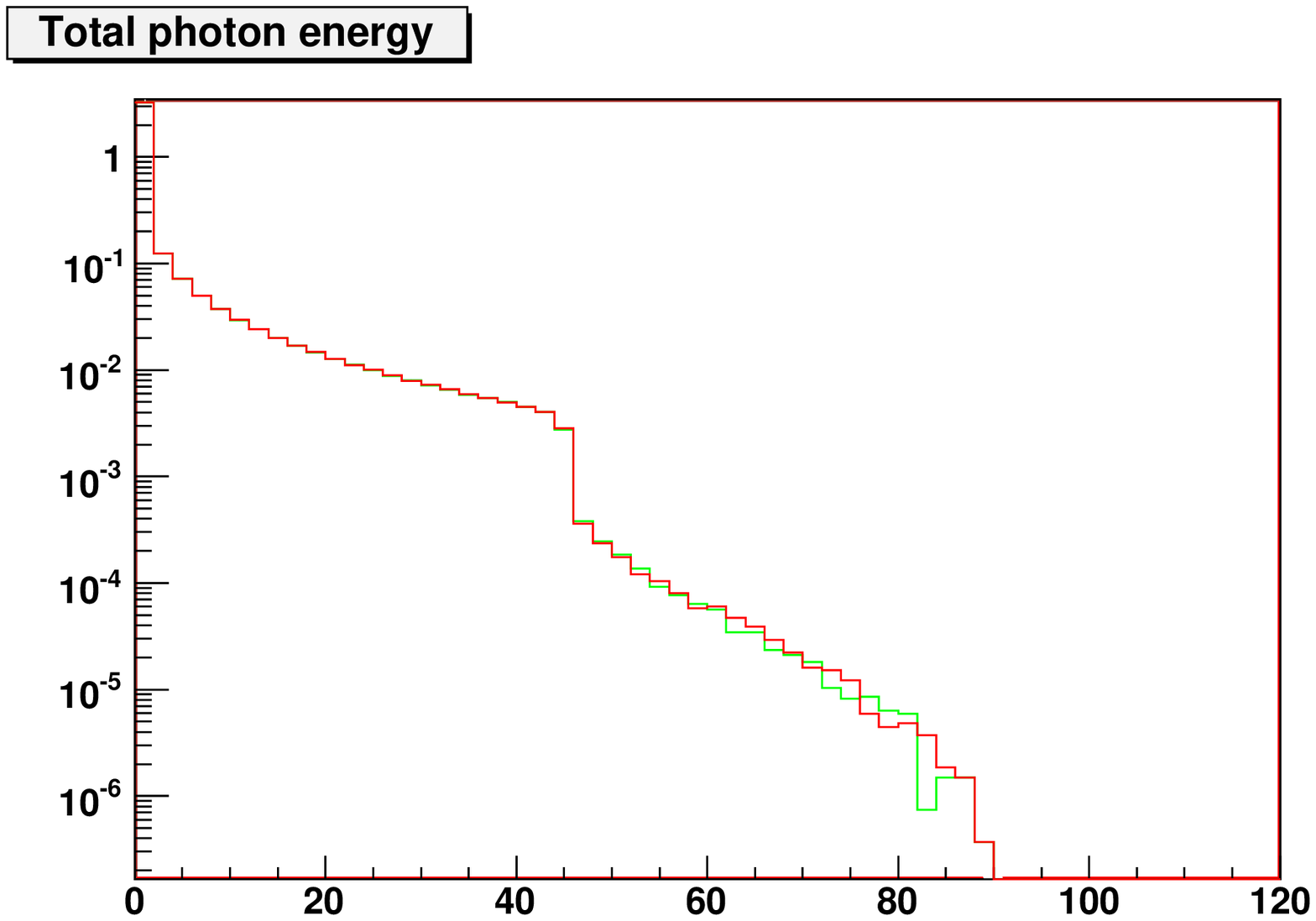}}
\subfigure[ {\it The method of MC-TESTER is applied. 
Both individual distributions, from PHOTOS and KKMC, are 
presented, but  overlap. The samples populate the first few bins of the histograms. 
The differences would normally be  
visible on the black histogram, which  presents the ratio of the results from
 PHOTOS and KKMC. The  agreement is perfect all over the spectrum.
No structure can be spotted in the vicinity of the kink (total photon energy, 
close to half of the $Z$ mass). If present,
the structure of possible differences, would be well separated from this of the
 shapes themselves.    } ]{\includegraphics[%
 width=0.49\columnwidth,
  keepaspectratio]{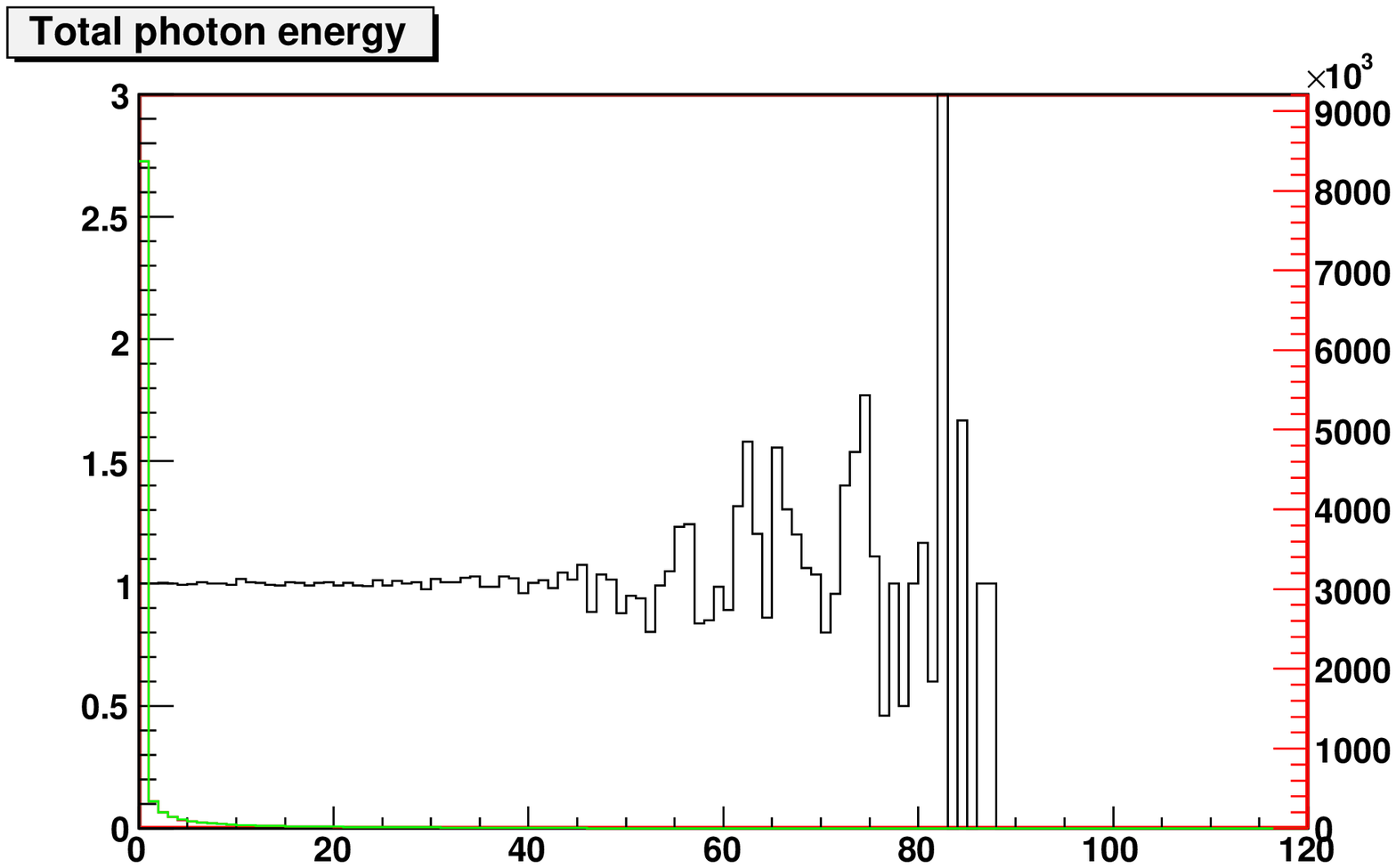}}
\end{center}
\end{figure*}
This is also  the case if for the same distribution (see right-hand side 
 of Fig. \ref{fig:classical})
 the method \cite{Golonka:2005pn} based on MC-TESTER is used.
The distributions are indeed dominated by the collinear content 
of the programs! For other distributions, sensitive to second-order matrix-element parts, missing 
in PHOTOS, the differences would become
visible on the plots obtained normally from MC-TESTER. 
That is why, in the present paper, we will 
keep to that  class of comparison
plots. 
The comparisons  are automated and standardized. This not only 
reduces the time needed for debugging  the tests, but also allows for
easy cross-comparisons of the results presented in our consecutive papers. 

For a selected decay process, such as the $Z/\gamma^*$ decay,
the four-momenta of the decay products and their flavours 
are extracted from the event record in an automated way (thereby
limiting the effort of setting up the appropriate analysis code and
also the risk of accidental errors). The decay 
events obtained that way are classified in distinct  decay channels,
 according to the  particles present in the final state.
 The histograms
of all possible invariant masses, which can be formed from the decay
products, are defined and filled for each identified
decay channel. At the end of the run they are
stored in output files.
Two output files (from distinct runs of event generators instrumented
with MC-TESTER) are then analysed, and the results are presented in
a form visualized  as a ``booklet'' made of plots and summary tables.
The user is given some general information concerning the comparison of the two
runs with different
Monte Carlo generators, a list of the decay channels with their branching
fractions, and the maximum values (for each decay channel)
 of the shape difference parameter (SDP)%
\footnote{ The shape difference parameter, defined in \cite{Golonka:2002rz},
quantifies the difference in shape of the histograms coming from the
two runs being compared. The SDP value is calculated separately for each histogrammed
mass: it quantifies the exclusive surface between the 
(normalized to unity) corresponding histograms obtained from the two runs.
The effects of statistical fluctuations are appropriately subtracted.
The maximum  SDP over all distributions for a given decay channel is taken
and printed in the table.}. 

For each decay channel the plots of histogrammed values are then included;
each plot presents two distributions from the two distinctive runs and a
curve, which is the  ratio of the two normalized distributions. The value of the SDP
is also printed for each plot.  
In practice, as in paper \cite{Golonka:2005pn},
 the histograms obtained from the compared programs will often overlap.
The differences will then be visible only in the plot of the ratio
of histograms.

The testing approach implemented in MC-TESTER could be used directly in the 
case of validation of the TAUOLA package. Nevertheless, for the purpose
of studies presented here, it needed an extension.
It is necessary, according to the particular method of handling soft photon cancellations,
to consistently treat the soft final-state QED bremsstrahlung photons, which may or may not be present in the 
event.
If results of different programs 
were compared blindly,
ambiguities due to differences in the treatment
of  the soft emission region and of the different boundaries for the photon 
phase-space (integrated analytically) would arise. 
To prevent these ambiguities, the most convenient solution was to introduce a technical
regulator {\it in the test itself}.

For our  comparisons to make physical sense and
remain automatic, we had to remove the softest photons from the final states.
We defined zero-, one-, and two-photon topologies in
the following way: we called the event ``zero photon''
if there was no photon of energy (in a decaying particle's rest frame)
larger than  $E_\mathrm{test}$.
The ``one-photon'' event had to have one (and only one) photon of 
energy larger than $E_\mathrm{test}$. 
If there were more than one such photons, we called it a ``two-photon'' event.
In the case where there were more than two photons of energy larger than $E_\mathrm{test}$, 
we considered only the two most energetic ones, and treated
the remaining (softer) ones as if they had not passed the $E_\mathrm{test}$ threshold.
For all the photons that did not pass the $E_\mathrm{test}$ threshold we
summed their four-momenta with the momentum of the outgoing fermion of smaller
angular separation. With the help of our test we divide the 
phase space for two fermions and an arbitrary number of photons into {\it slots}
of 0, 1 and 2 distinguished final-state photons. 

In the paper we will use two variants of this test definition:
\emph{test1} and \emph{test2}. The \emph{test2} is
exactly as explained above. In \emph{test1}, only one photon (the
most energetic one) will be accepted. The free parameter,
$E_\mathrm{test}$ is chosen to be $1$ GeV  for all results presented
in this paper.

Systematic histogramming of all possible invariant masses that can be
constructed from a combination of final-state four-vectors,
 and storing them as  
one-dimensional histograms, does not define a test of ultimate sensitivity.
The method is blind to the $P$-parity-sensitive effects, important for $\tau$ lepton
physics, for instance. Also important effects, such as  coherence between the photons 
are to a large degree washed out. Nonetheless we believe that the advantages
of the method are prevailing, and we decided to use it in this study.

\vspace{.5cm}
\begin{center}
{\bf 4. Results of the  tests performed at  first order}
\end{center}
\vspace{.5cm}

Let us start with a comparison of PHOTOS and
 KORALZ \cite{Jadach:1991ws}, both run at  first order and 
without exponentiation. For KORALZ, the complete first-order matrix element, as
 in Section 2, is used.
 The results from KORALZ are given by a 
red (darker grey) line and from PHOTOS by a green (lighter grey) line.
(In the presentation of the results we use the colour coding consistently in the plots and the summary tables, following the methodology of MC-TESTER).
The lines overlap almost completely on all plots and 
only the ratios of the distributions shown as  black histograms indicate that there is some  difference.
The actual plots for  MC-TESTER comparison, are prepended with a summary table giving  the fractions 
of event with and without photons (of energy above 1 GeV). 
In all comparisons samples of $10^8$ events were used.

\vspace{0.3cm} 
{\centering \begin{longtable}{|c|c|c|c|} 
\hline 
Decay channel &\multicolumn{2}{|c|}{ Branching ratio $\pm$ rough errors} & Max.  SDP\\ 
      & \textcolor{red}{KORALZ} & \textcolor{green}{PHOTOS} & \\ 
\hline 
\hline \( Z^{0} \rightarrow \mu^{-} \mu^{+}  \) & \textcolor{red}{82.5137 $\pm$  0.0091\%} &\textcolor{green}{  82.3622 $\pm$  0.0091\%} & 0.00000 \\ 
\hline 
\hline \( Z^{0} \rightarrow \mu^{-} \mu^{+} \gamma  \) & \textcolor{red}{17.4863 $\pm$  0.0042\%} &\textcolor{green}{   17.6378 $\pm$  0.0042\%} & 0.00534 \\ 
\hline 
\end{longtable}\par} 
\vspace{0.3cm} 

As can be seen, the difference in the fraction of events with 
photon (of energy above 1 GeV)  is about 0.15\%. 
Although noticeable (thanks to our method), this is not a large discrepancy.
Let us now turn to the distributions. 
\section*{Decay Channel: $Z^{0} \rightarrow \mu^{-} \mu^{+} \gamma $}
{ \resizebox*{0.49\textwidth}{!}{\includegraphics{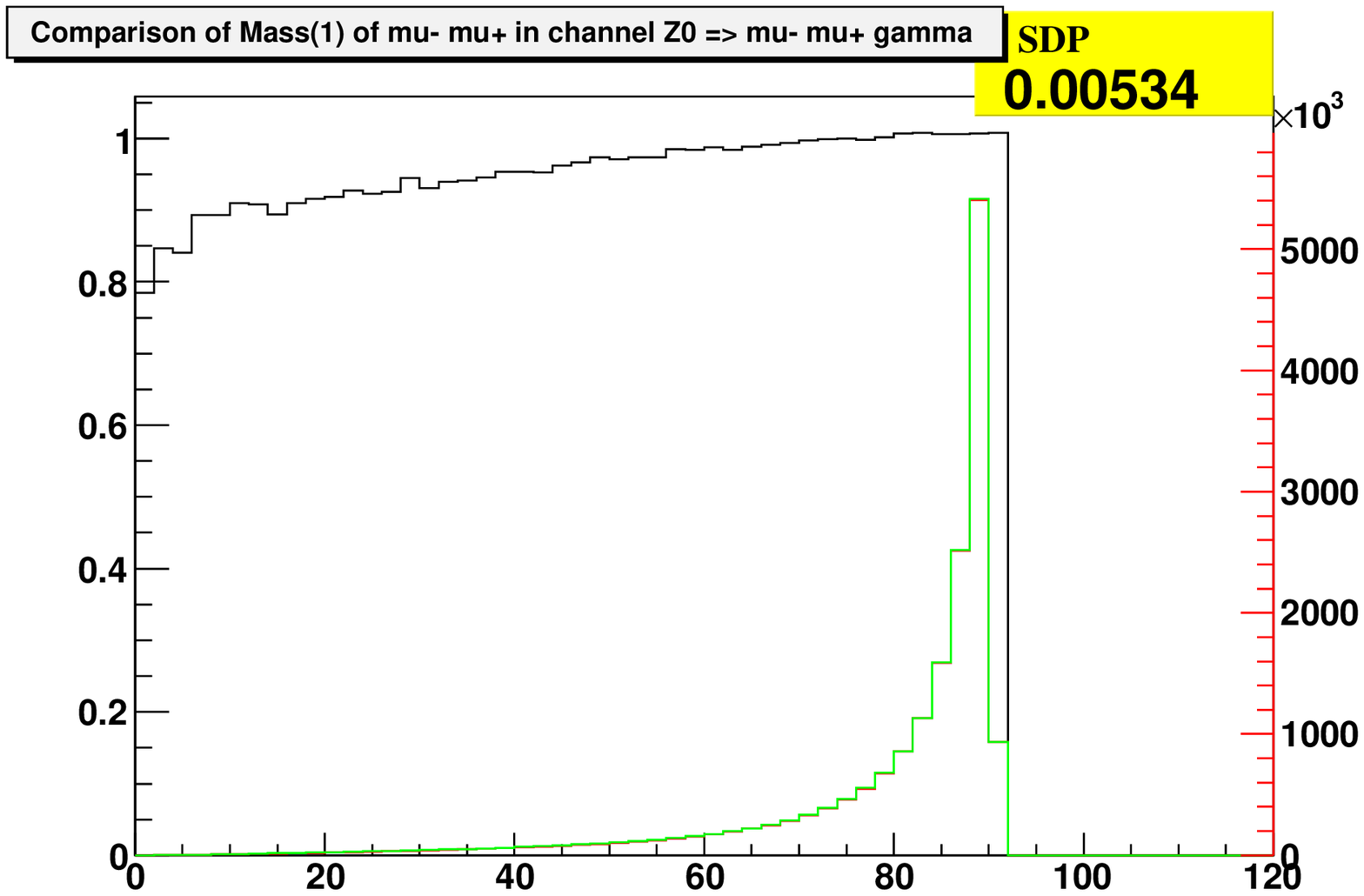}} }
{ \resizebox*{0.49\textwidth}{!}{\includegraphics{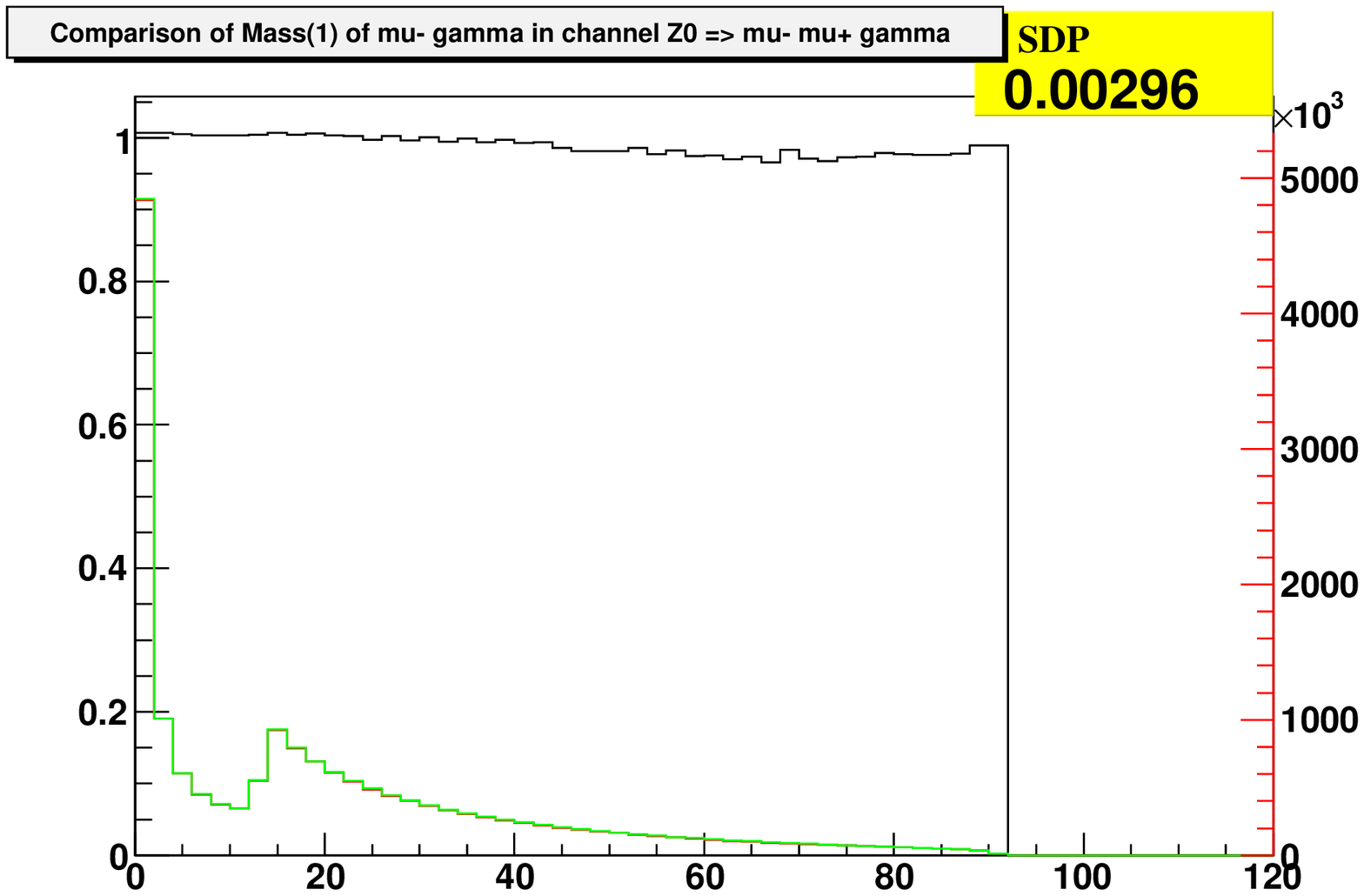}} }
{ \resizebox*{0.49\textwidth}{!}{\includegraphics{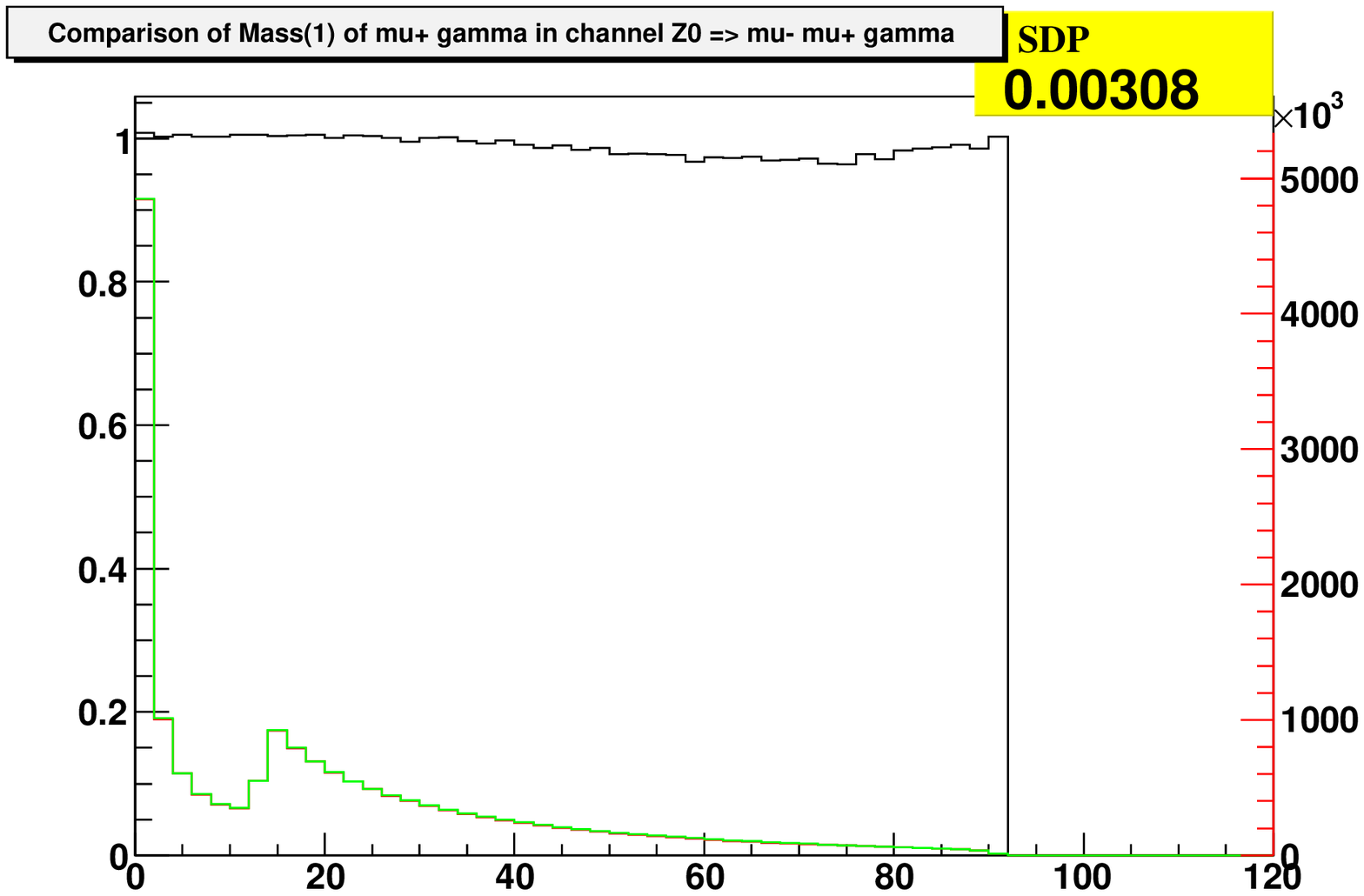}} }
{ \resizebox*{0.49\textwidth}{!}{\includegraphics{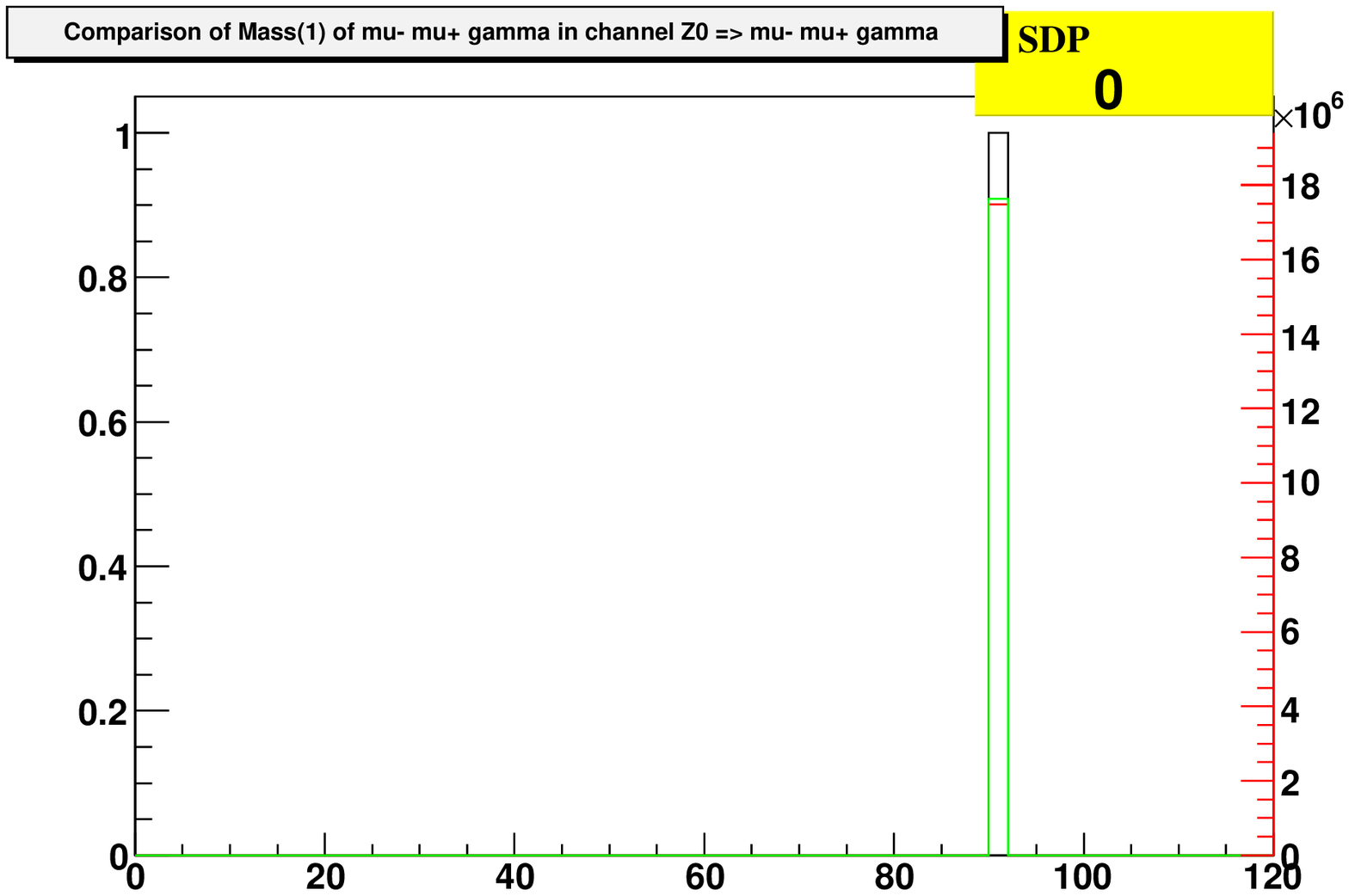}} }

Analysing the values of the SDP, 
printed in the upper right corners of the plots,
we conclude that the  surfaces between green and red 
histograms (both normalized to unity before calculation of the ratio) 
differ by at most 0.005; this low value quantifies the fact that
the histograms for the two programs overlap almost completely. 
The ratio of the two lines (black histogram) nonetheless reveals 
the difference,
which  is located at the far end of the spectra, sparsely
populated by configurations with  photons of extremely large energies and 
 away from the direction of the muons.  

Even though the agreement is  amazing, it is yet improved once the NLO term is included into the correction weight of PHOTOS:
 the differences disappear  below the statistical 
error of $10^8$ event samples and they are not noticeable in the histograms'
ratio curve either!
In the following table and figures, we collect the results as 
previously discussed,  but for the runs with the  NLO correcting weight 
 activated in PHOTOS.

\vspace{0.3cm} 
{\centering \begin{longtable}{|c|c|c|c|} 
\hline 
Decay channel &\multicolumn{2}{|c|}{ Branching ratio $\pm$ rough errors} & Max. SDP\\ 
      & \textcolor{red}{KORALZ} & \textcolor{green}{PHOTOS} & \\ 
\hline 
\hline \( Z^{0} \rightarrow \mu^{+} \mu^{-}  \) & \textcolor{red}{82.5110 $\pm$  0.0091\%} &\textcolor{green}{  82.5074 $\pm$  0.0091\%} & 0.00000 \\ 
\hline 
\hline \( Z^{0} \rightarrow \mu^{+} \mu^{-}  \gamma \) & \textcolor{red}{17.4890 $\pm$  0.0042\%} &\textcolor{green}{  17.4926 $\pm$  0.0042\%} & 0.00000 \\ 
\hline 
\end{longtable}\par} 
\vspace{0.3cm}


\section*{Decay Channel: $Z^{0} \rightarrow \mu^{+} \mu^{-}  \gamma$}
{ \resizebox*{0.49\textwidth}{!}{\includegraphics{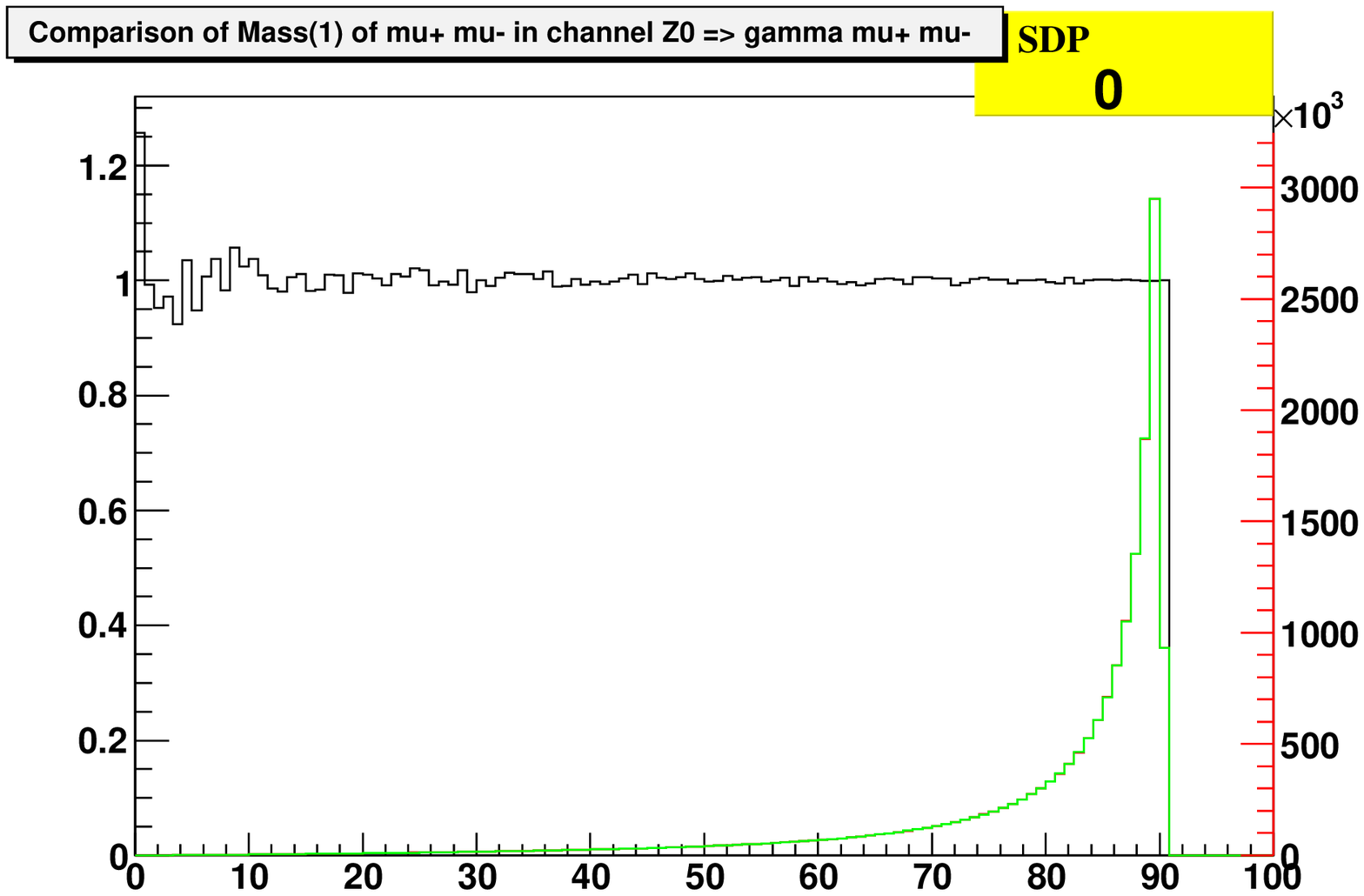}} }
{ \resizebox*{0.49\textwidth}{!}{\includegraphics{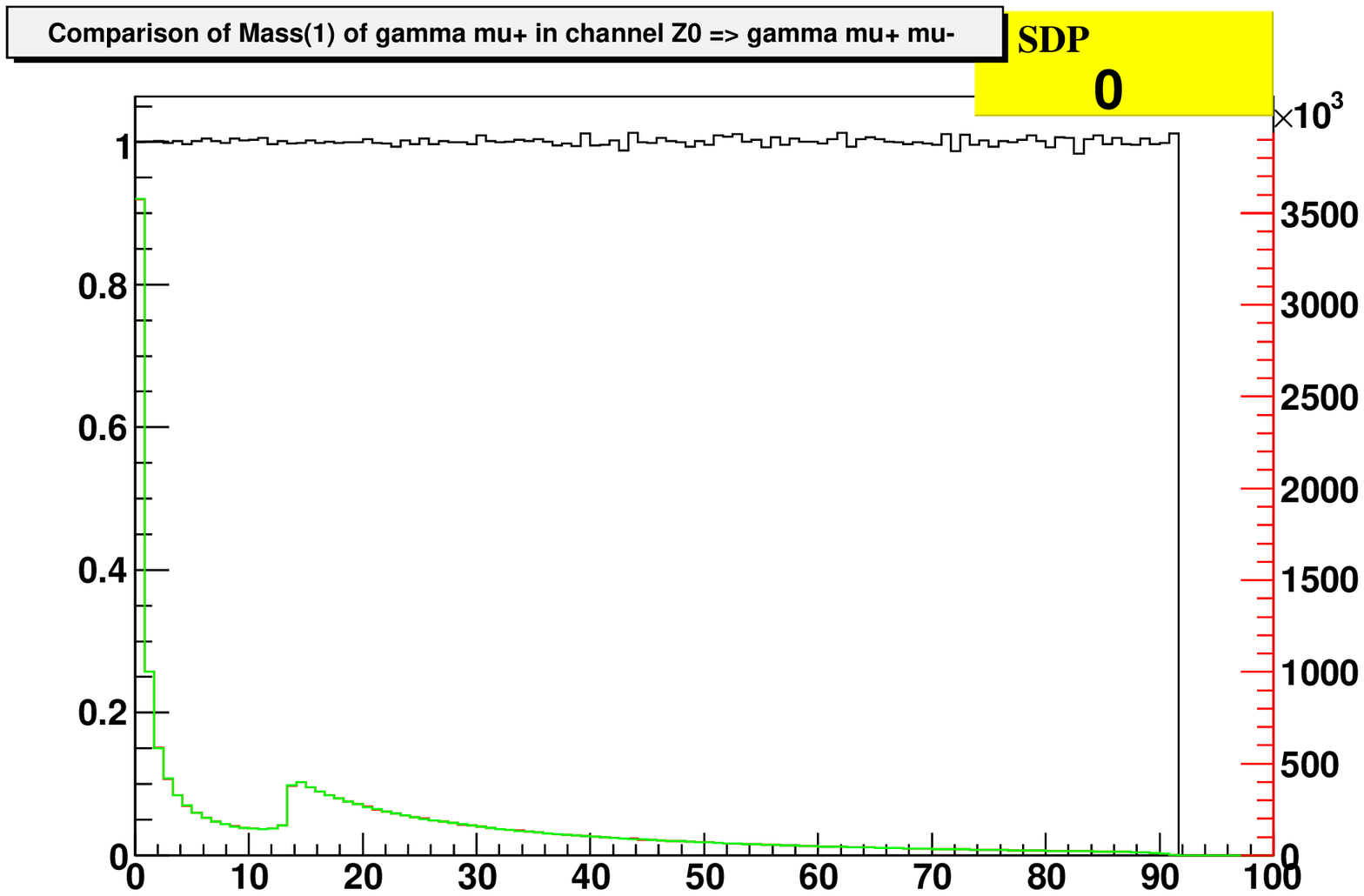}} }
{ \resizebox*{0.49\textwidth}{!}{\includegraphics{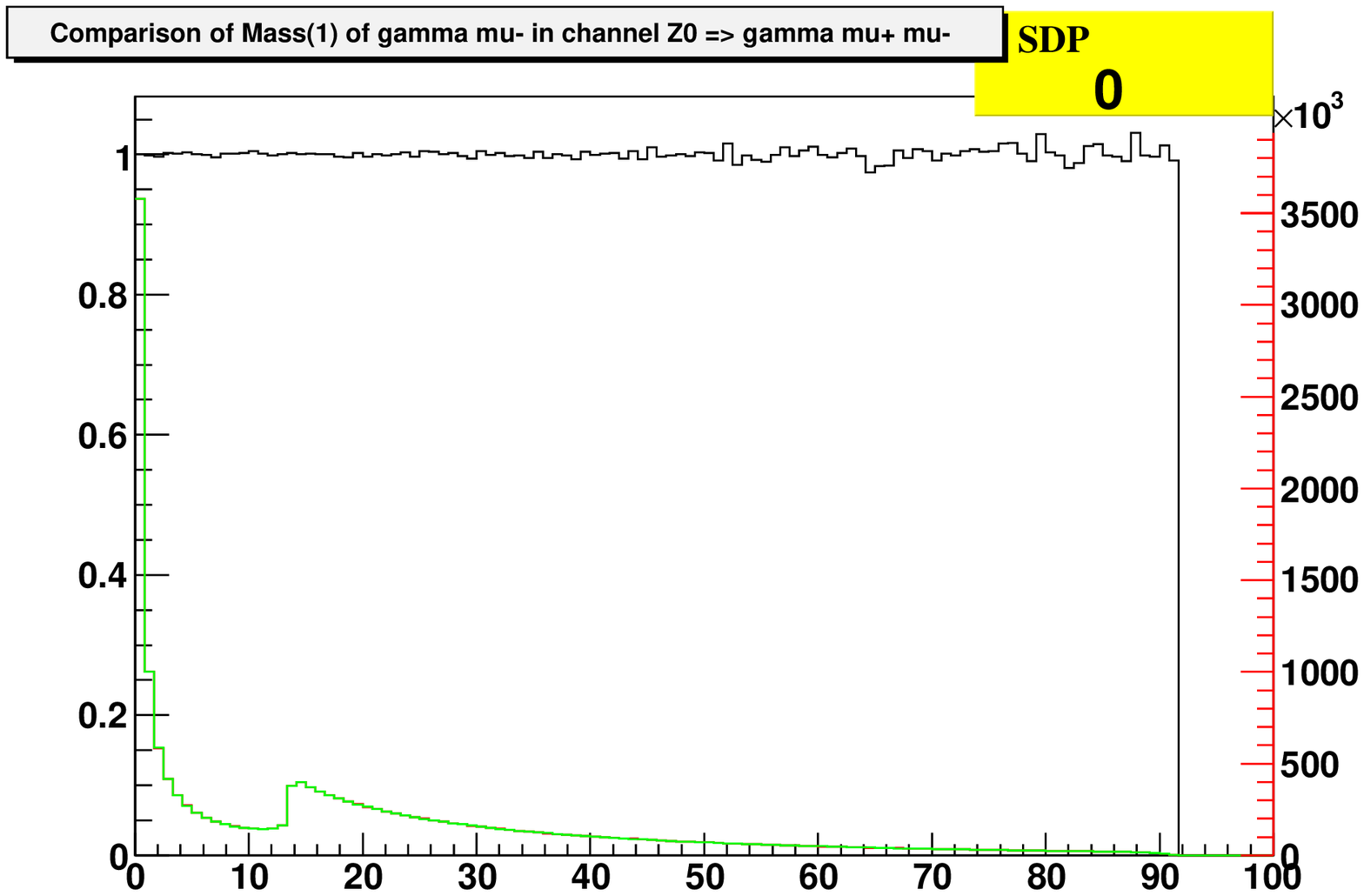}} }
{ \resizebox*{0.49\textwidth}{!}{\includegraphics{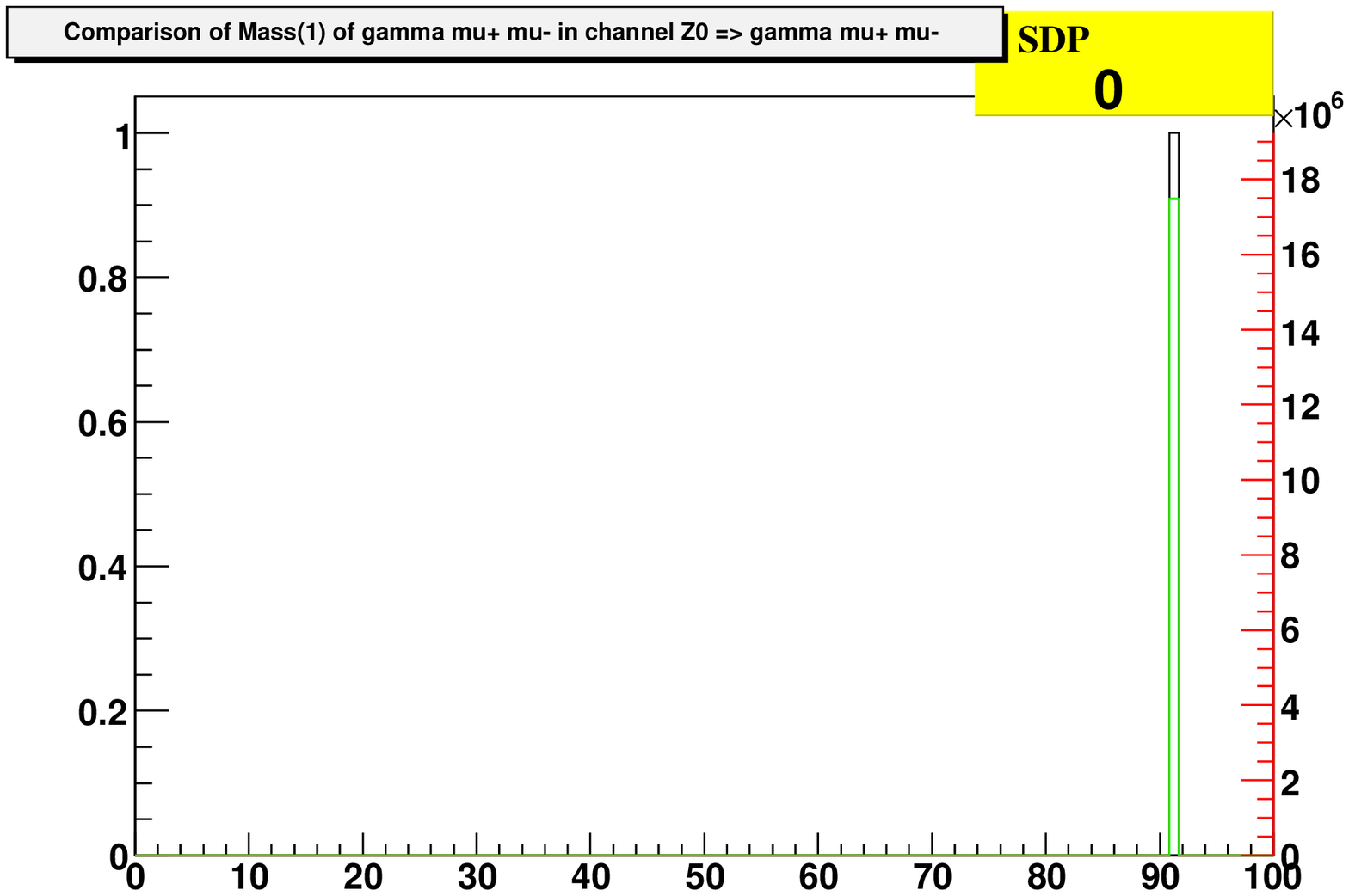}} }

The agreement for the branching fractions (of events with and without photons of energy larger than 1 GeV) is better than 0.01\% now! 
This excellent  agreement indeed confirms that the theoretical
effects missing in the standard version of PHOTOS are negligibly small. It is equally important 
that it provides, a powerful technical test of the generator. The  
kinematical variables used in PHOTOS differ from those of KORALZ;
four-vectors are used instead of angles
to parametrize the intermediate steps of the generation.
The differences could have indicated, say,  consequences of aggregation of rounding errors.
Keeping in mind that 
similar levels of agreement  for muons 
was achieved for the multiphoton  version of 
PHOTOS and KKMC in the case 
of $Z\to e^+ e^-$ decay,  
we can  confidently claim that PHOTOS has numerical stability
under  control. This was not the case for the early 
versions of the program, and reaching that level of technical reliability
required a major effort.

\begin{center}
{\bf 5. Algorithm for multiple-photon generation}
\end{center}
Before presentation of the  results for  multiple-photon generation from PHOTOS, 
 let us comment on those
technical details of the PHOTOS algorithm,
 that are important in the implementation of the NLO contribution to the correcting weight.
The iteration algorithm, as explained first in ref. \cite{Barberio:1994qi}, and recently also
 in refs. \cite{Golonka:2005dn,Golonka:2005pn}, did not require changes for the case of multiple 
photon generation. Nevertheless, the following details have to be clarified for the proper implementation of
the NLO weight, given by formula (\ref{wgt1}). All identical terms present in the
 numerator and denominator, expressions   (\ref{X-mustraal}) and 
(\ref{X-fotos}) respectively,
were cancelled out at the analytical level%
 \footnote{The cancelled-out terms  could have been calculated using slightly different 
kinematical variables; the  differences would have appeared only in the case of more than one hard photon 
 present
in the final state. In such a case the ratio of the terms  would not be equal to 1.
These effects generally go beyond the NLO, and in fact  our choice was motivated by the comparisons 
with the second-order matrix-element calculation, but without necessary details. That is why 
an appropriate   discussion of this choice
 would require detailed presentation of the second-order matrix element.
It would have to be similar, for example,  to the discussion of the extrapolation procedure as 
that described in ref. \cite{Was:2004ig}. }. 
The weight  (\ref{wgt1})
is always calculated for the 
single-photon configuration. If there are other photons 
generated in the previous steps of the iteration,
their momenta are absorbed into the momenta of the final-state fermions. The constraint on
 the direction and the opening angle between the photon under consideration and the direction of the
charged emitter \cite{Barberio:1990ms}, is assured.

\vspace{.5cm}

\begin{center}
{\bf 6. Numerical results of the tests performed with multiple-photon radiation}
\end{center}

Let us now turn to the tests of PHOTOS running in multiple-photon option.
For that purpose we will use \emph{test1} as defined in Section 3, and samples of 
$10^8$ events generated from KKMC with exponentiation and the second-order 
matrix element and the multiple-photon radiation version of PHOTOS, without NLO terms.
The results are included in the table and  plots below.


\vspace{0.3cm} 
{\centering \begin{longtable}{|c|c|c|c|} 
\hline 
Decay channel &\multicolumn{2}{|c|}{ Branching ratio $\pm$ rough errors} & Max. SDP\\ 
      & \textcolor{red}{KKMC} & \textcolor{green}{PHOTOS} & \\ 
\hline 
\hline \( Z^{0} \rightarrow \mu^{-} \mu^{+}  \) & \textcolor{red}{83.9176 $\pm$  0.0092\%} &\textcolor{green}{  83.8372 $\pm$  0.0092\%} & 0.00000 \\ 
\hline 
\hline \( Z^{0} \rightarrow \mu^{-} \mu^{+} \gamma  \) & \textcolor{red}{16.0824 $\pm$  0.0040\%} &\textcolor{green}{  16.1628 $\pm$  0.0040\%} & 0.00409 \\ 
\hline 
\end{longtable}\par} 
\vspace{0.3cm}

\section*{Decay Channel: $Z^{0} \rightarrow \mu^{-} \mu^{+} \gamma $}
{ \resizebox*{0.49\textwidth}{!}{\includegraphics{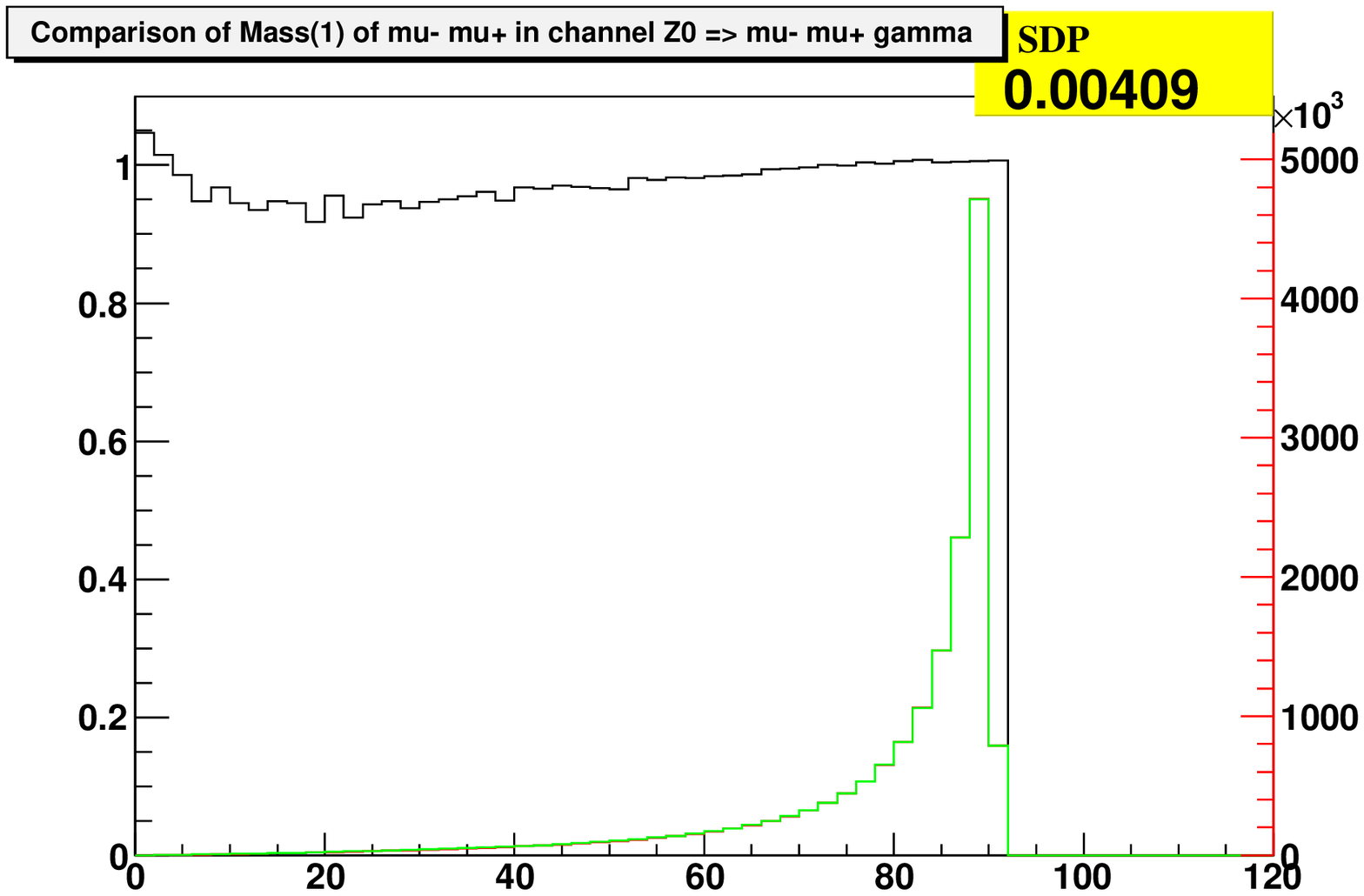}} }
{ \resizebox*{0.49\textwidth}{!}{\includegraphics{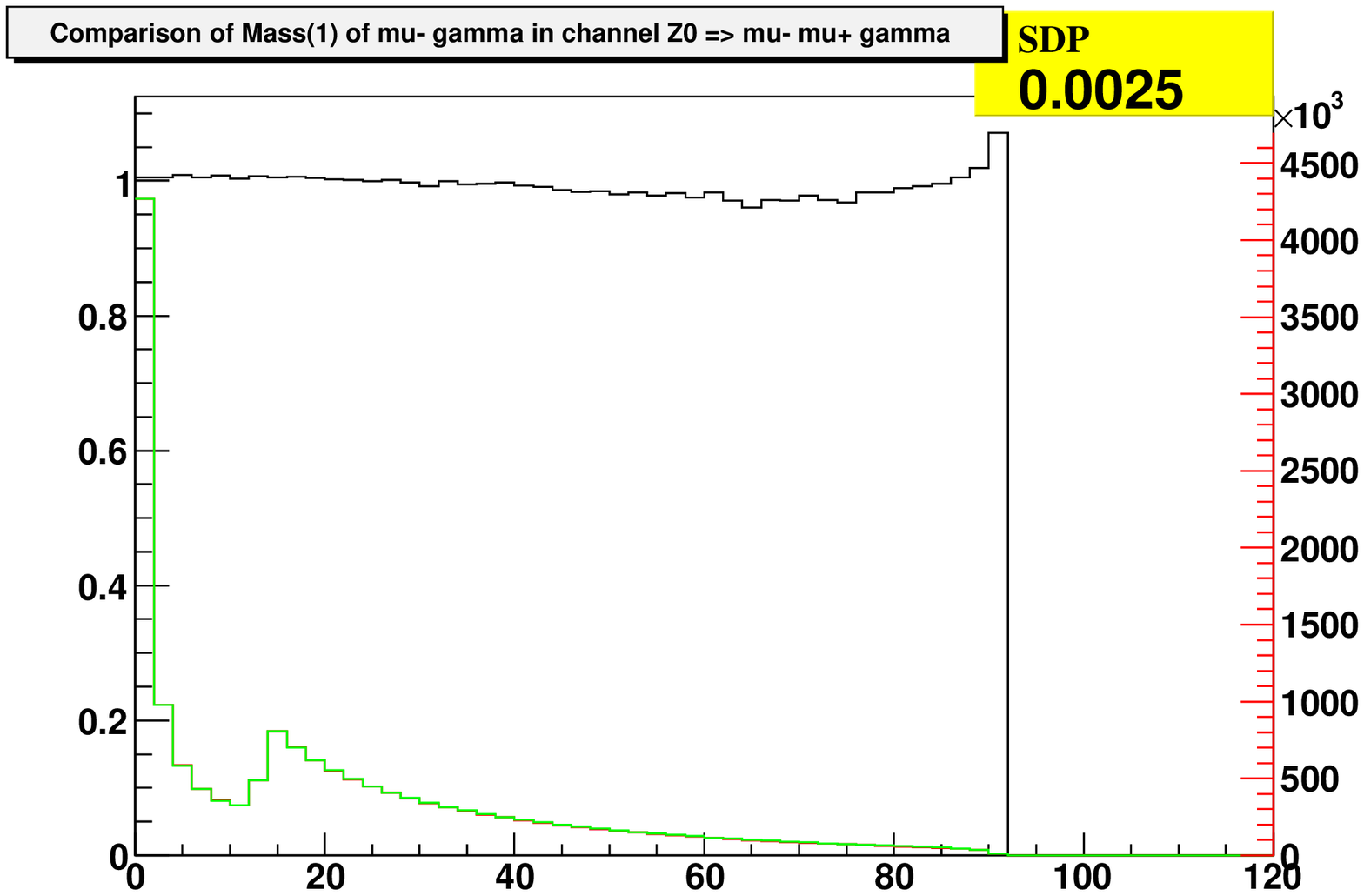}} }
{ \resizebox*{0.49\textwidth}{!}{\includegraphics{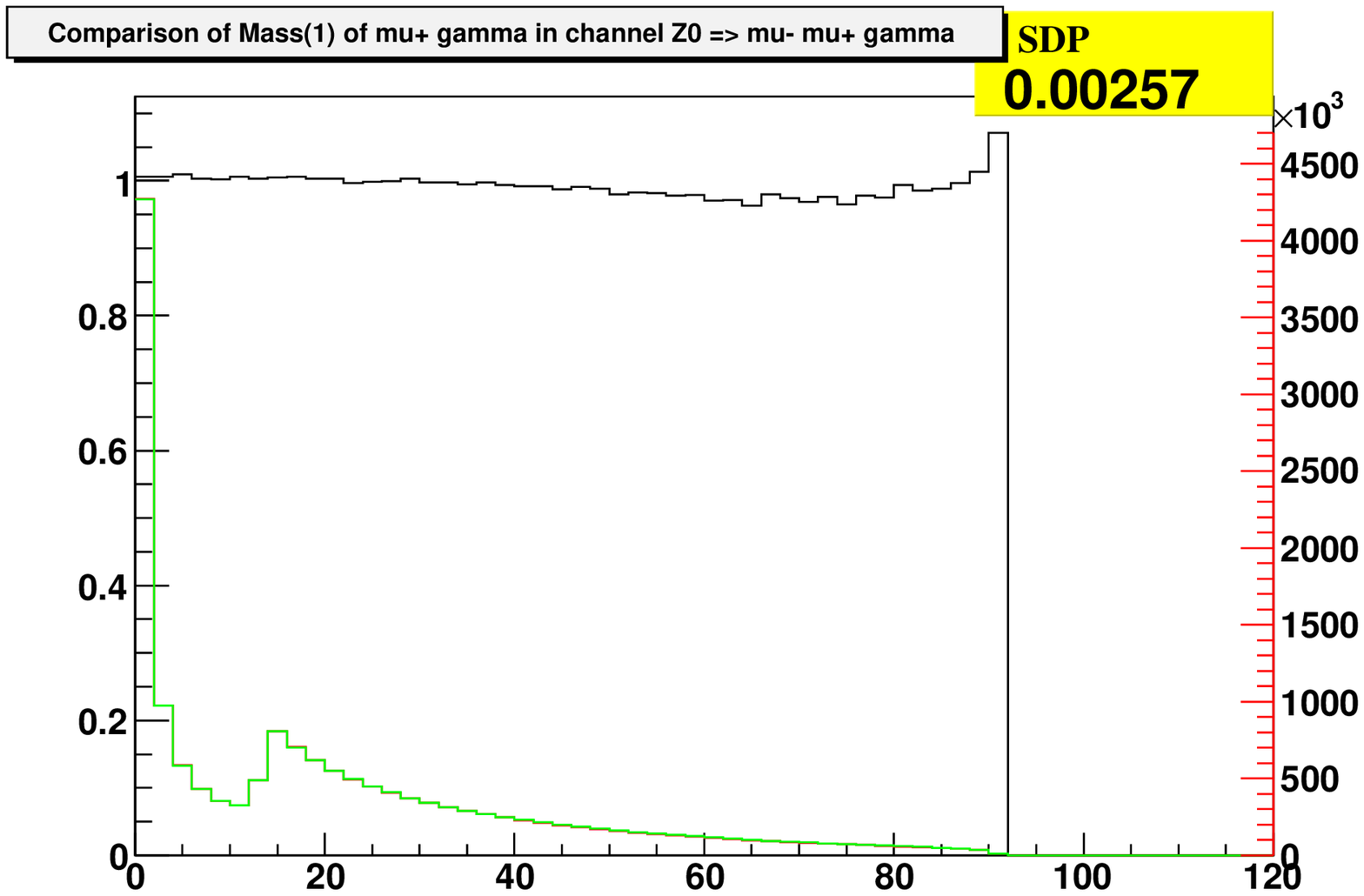}} }
{ \resizebox*{0.49\textwidth}{!}{\includegraphics{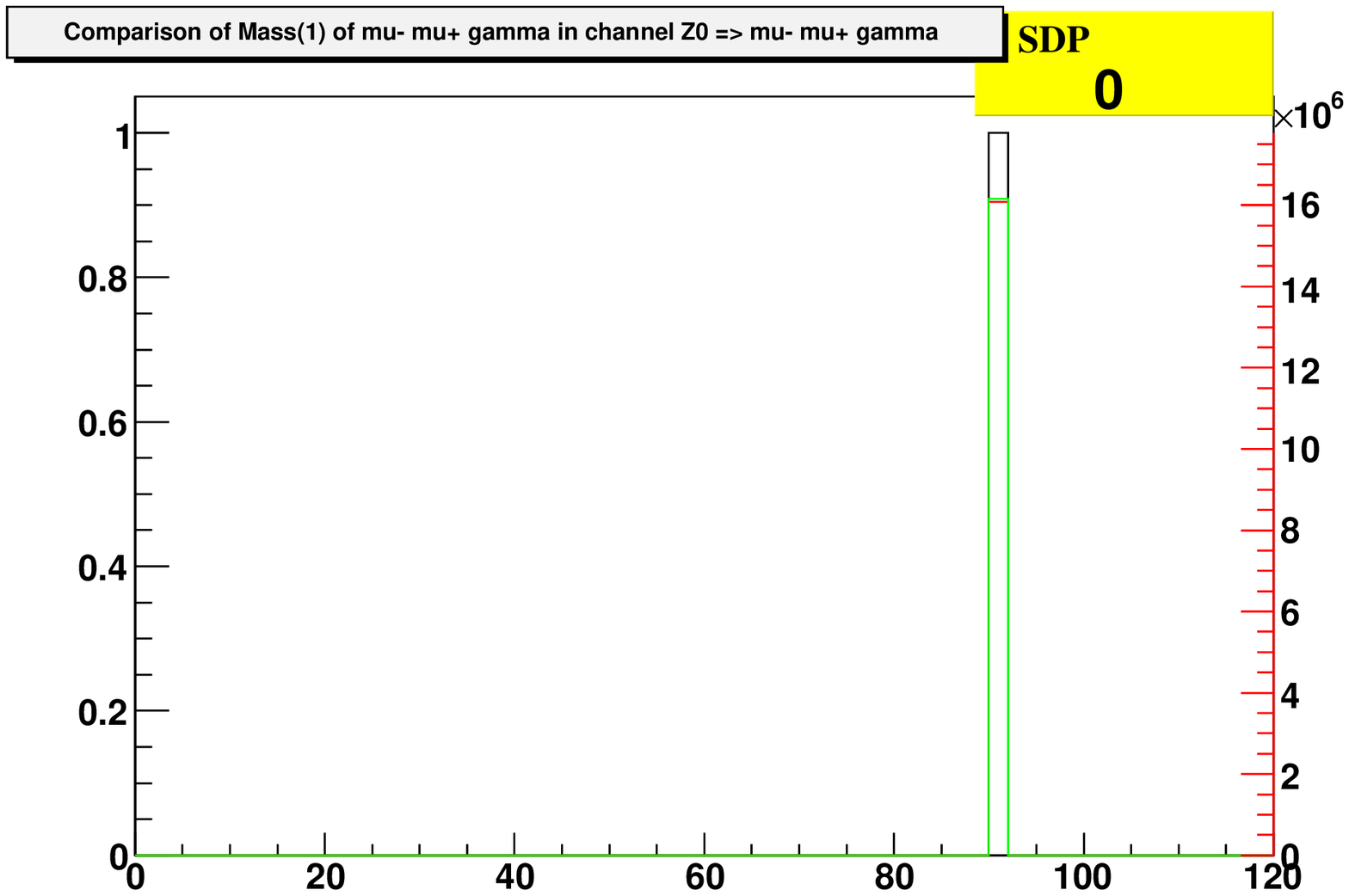}} }

The pattern of differences between the results of PHOTOS and KKMC runs resembles 
the one present in the plots for the comparisons performed at the first order.
Again, the black curves of  the histogram ratios for the KKMC and PHOTOS results are not consistent with 1
for configurations with hard photons and in the regions where histograms are nearly 
at  0.  The differences reach few per cent  in the
 corners of the phase space contributing few per mille to the total rate. The discrepancies  are  again 
smaller than 0.1\% with respect to the total rate.

Once the NLO weight in PHOTOS is activated,  the already small differences become even smaller,
by a factor of about 50,  measured with the SDP. The differences  are practically 0
with $10^8$ samples, which can be seen in the table and the plots below.

\vspace{0.3cm} 
{\centering \begin{longtable}{|c|c|c|c|} 
\hline 
Decay channel &\multicolumn{2}{|c|}{ Branching ratio $\pm$ rough errors} & Max. SDP\\ 
      & \textcolor{red}{KKMC} & \textcolor{green}{PHOTOS} & \\ 
\hline 
\hline \( Z^{0} \rightarrow \mu^{-} \mu^{+}  \) & \textcolor{red}{83.9176 $\pm$  0.0092\%} &\textcolor{green}{  83.9312 $\pm$  0.0092\%} & 0.00000 \\ 
\hline 
\hline \( Z^{0} \rightarrow \mu^{-} \mu^{+} \gamma  \) & \textcolor{red}{16.0824 $\pm$  0.0040\%} &\textcolor{green}{  16.0688 $\pm$  0.0040\%} & 0.00003 \\ 
\hline 
\end{longtable}\par} 
\vspace{0.3cm} 

\section*{Decay Channel: $Z^{0} \rightarrow \mu^{-} \mu^{+} \gamma $}

{ \resizebox*{0.49\textwidth}{!}{\includegraphics{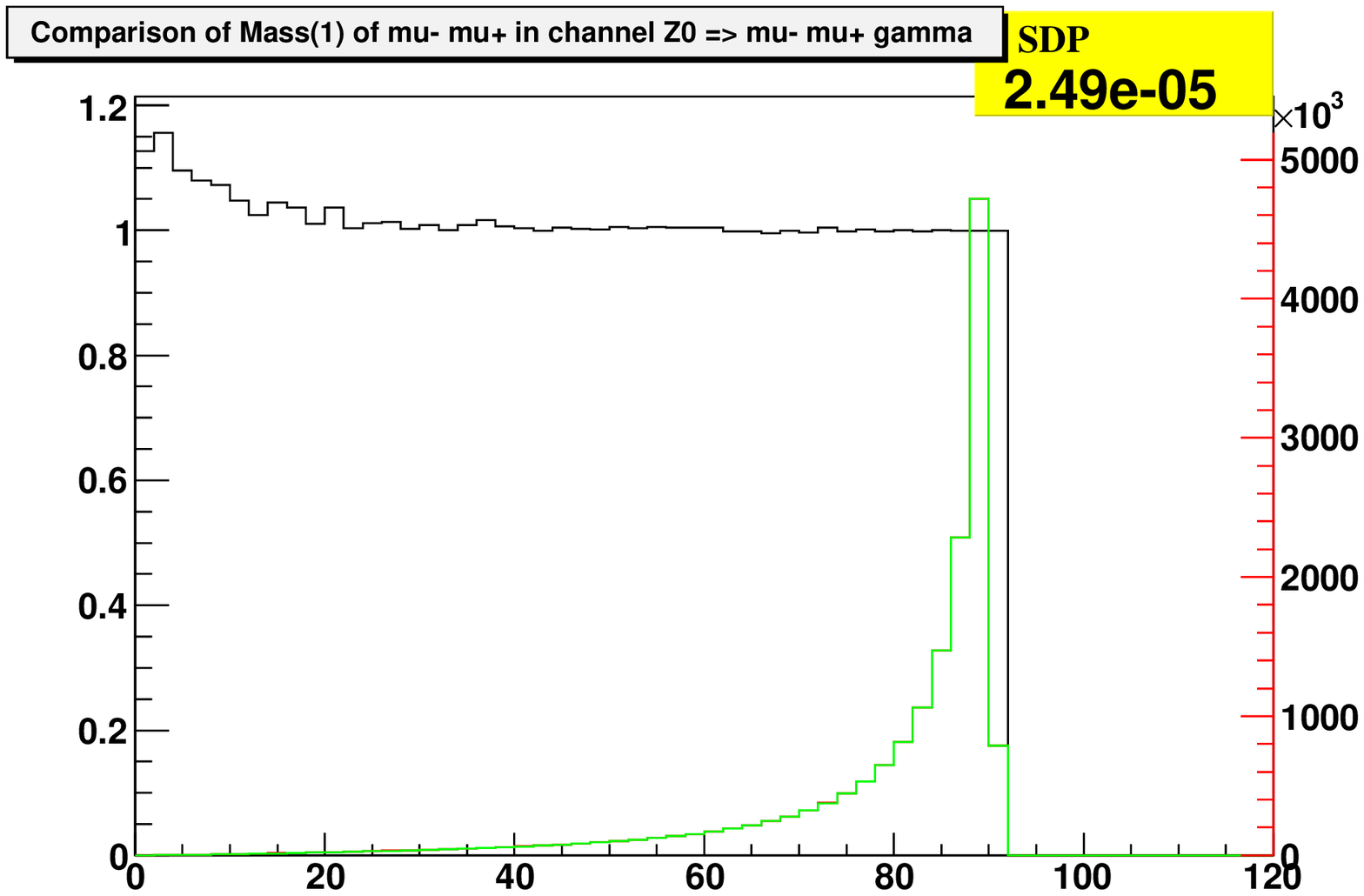}} }
{ \resizebox*{0.49\textwidth}{!}{\includegraphics{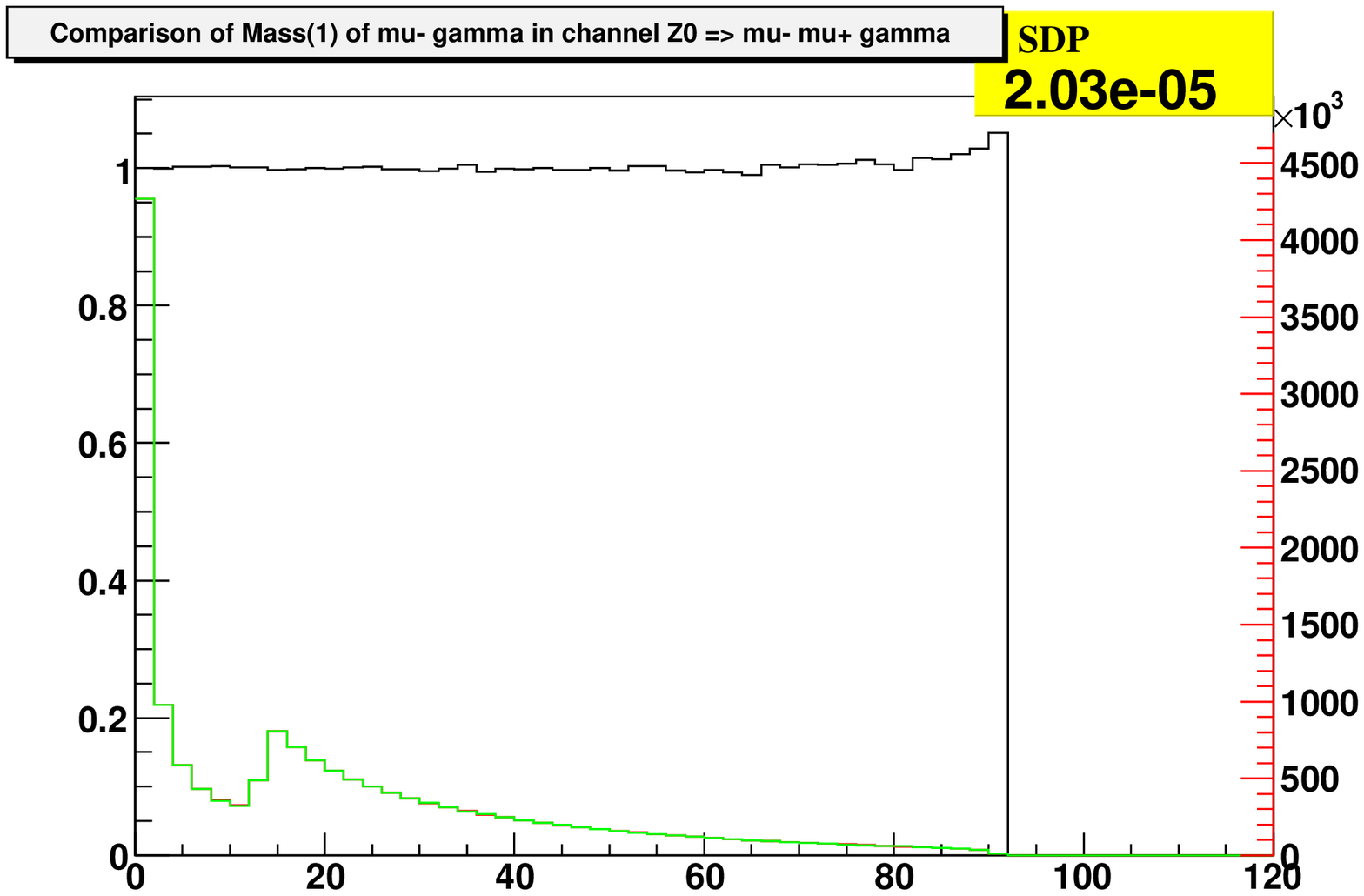}} }
{ \resizebox*{0.49\textwidth}{!}{\includegraphics{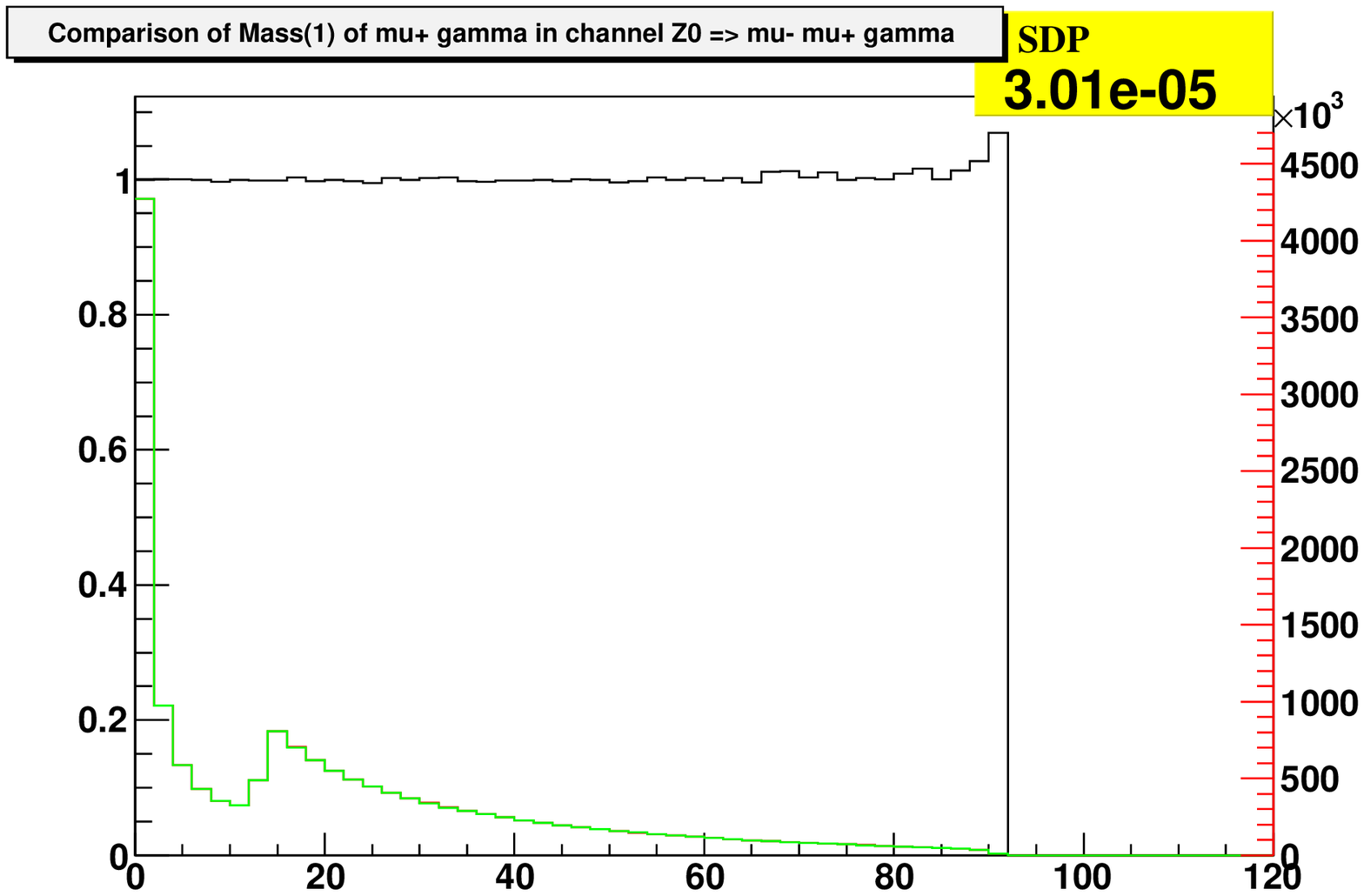}} }
{ \resizebox*{0.49\textwidth}{!}{\includegraphics{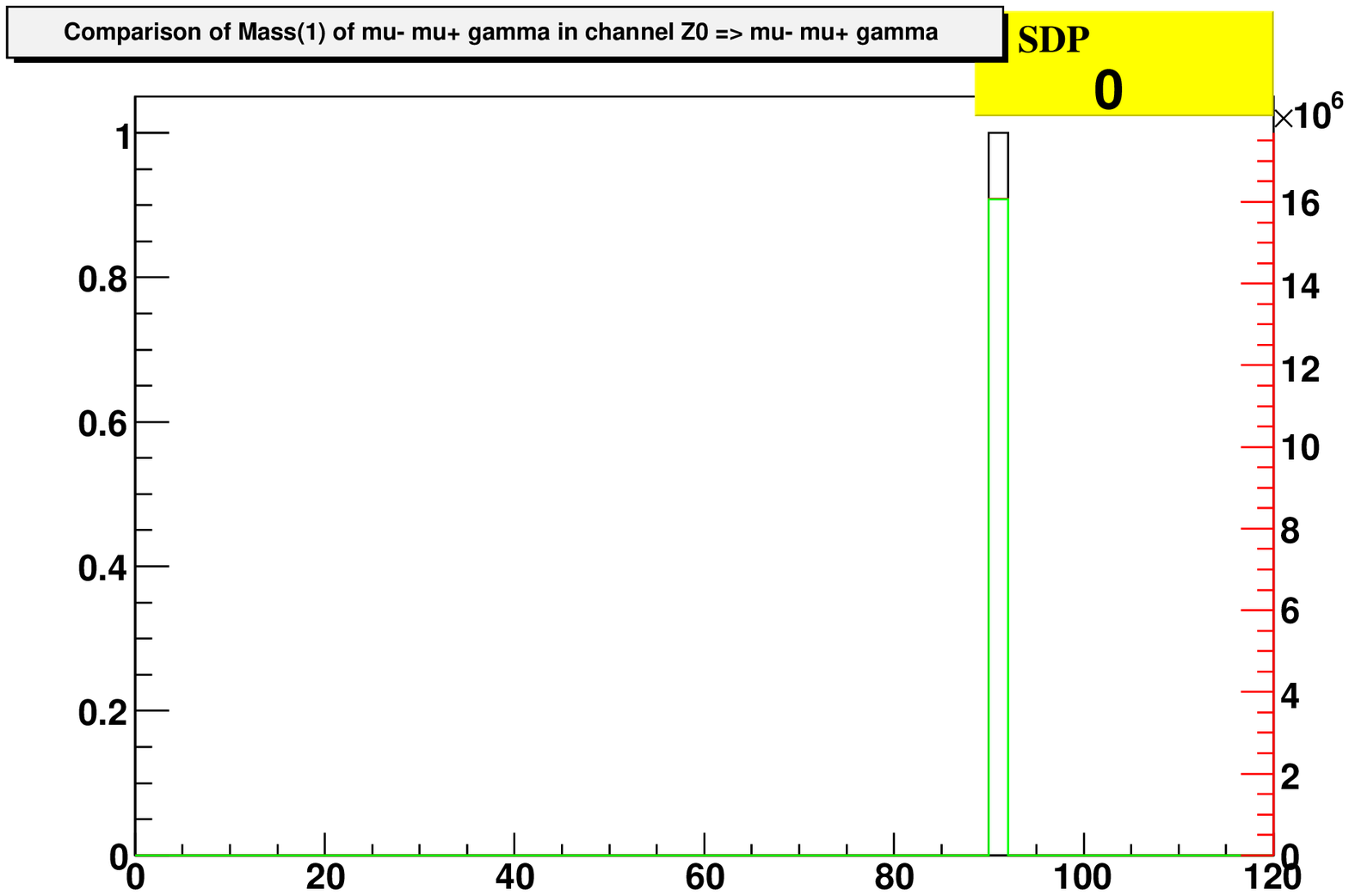}} }

 This  confirms that the main source of residual discrepancy between KKMC and 
PHOTOS, both running in exponentiated versions,  was due to the NLO term missing
in the previous group of results and activated now.

Let us now turn to \emph{test2}, where configurations
of up to 2 hard photons are analysed. To this end,
we will present at first the  results of a comparison of  the standard multiple-radiation version of PHOTOS 
with that of KKMC, followed by the comparison of multiple-radiation version PHOTOS with NLO weight and KKMC
(again with  second-order matrix element and exponentiation).

\vspace{0.3cm} 
{\centering \begin{longtable}{|c|c|c|c|} 
\hline 
Decay channel &\multicolumn{2}{|c|}{ Branching ratio $\pm$ rough errors} & Max. SDP\\ 
      & \textcolor{red}{KKMC} & \textcolor{green}{PHOTOS} & \\ 
\hline
\hline \( Z^{0} \rightarrow \mu^{-} \mu^{+}  \) & \textcolor{red}{83.9177 $\pm$  0.0092\%} &\textcolor{green}{  83.8372 $\pm$  0.0092\%} & 0.00000 \\ 
\hline 
\hline \( Z^{0} \rightarrow \mu^{-} \mu^{+} \gamma  \) & \textcolor{red}{14.8164 $\pm$  0.0038\%} &\textcolor{green}{  14.8676 $\pm$  0.0039\%} & 0.00232 \\ 
\hline  
\hline \( Z^{0} \rightarrow \mu^{-} \mu^{+} \gamma \gamma  \) & \textcolor{red}{ 1.2659 $\pm$  0.0011\%} &\textcolor{green}{   1.2952 $\pm$  0.0011\%} & 0.00918 \\ 
\hline 
\end{longtable}\par} 
\vspace{0.3cm} 

\section*{Decay Channel: $Z^{0} \rightarrow \mu^{-} \mu^{+} \gamma $}
{ \resizebox*{0.49\textwidth}{!}{\includegraphics{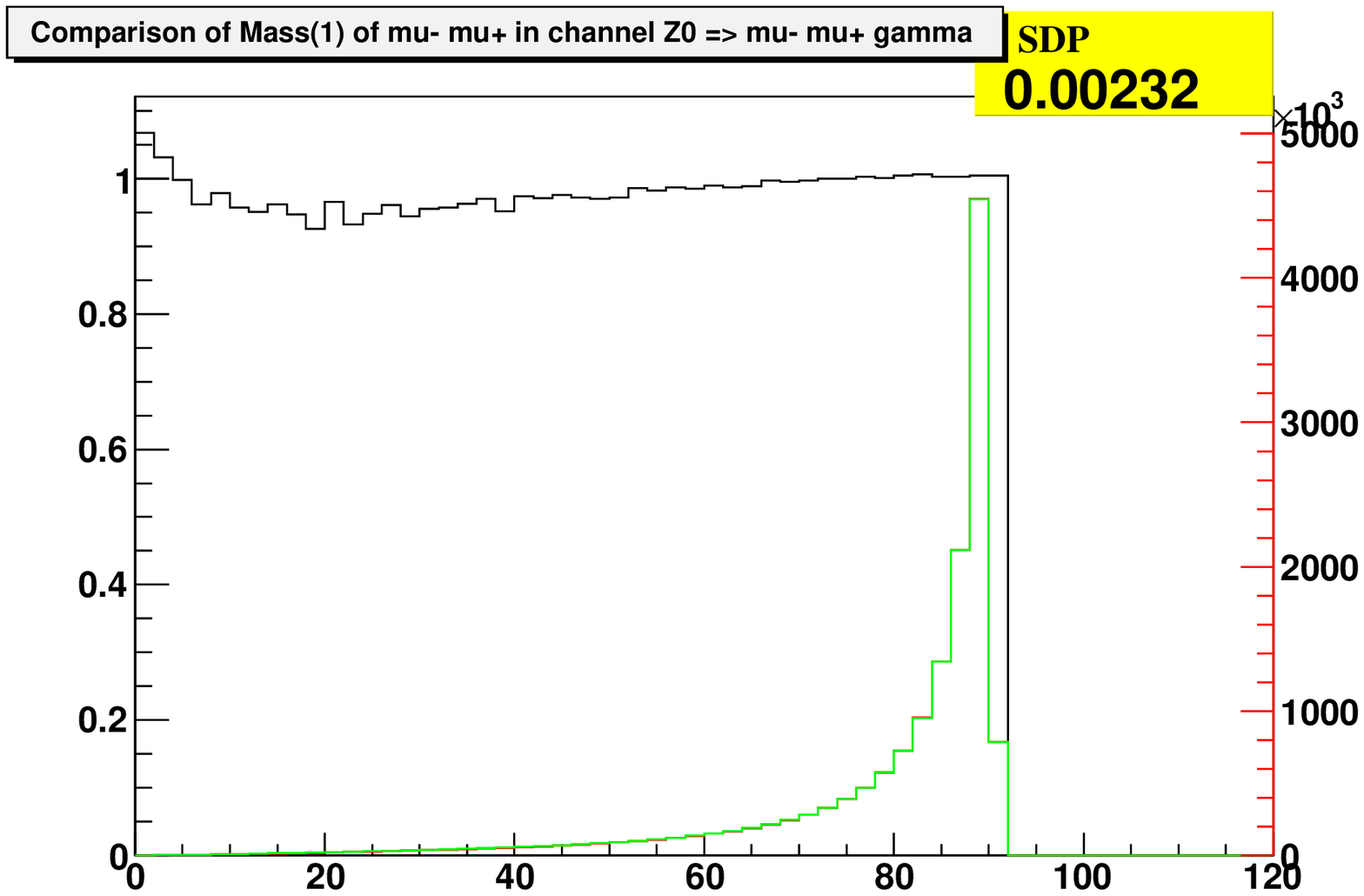}} }
{ \resizebox*{0.49\textwidth}{!}{\includegraphics{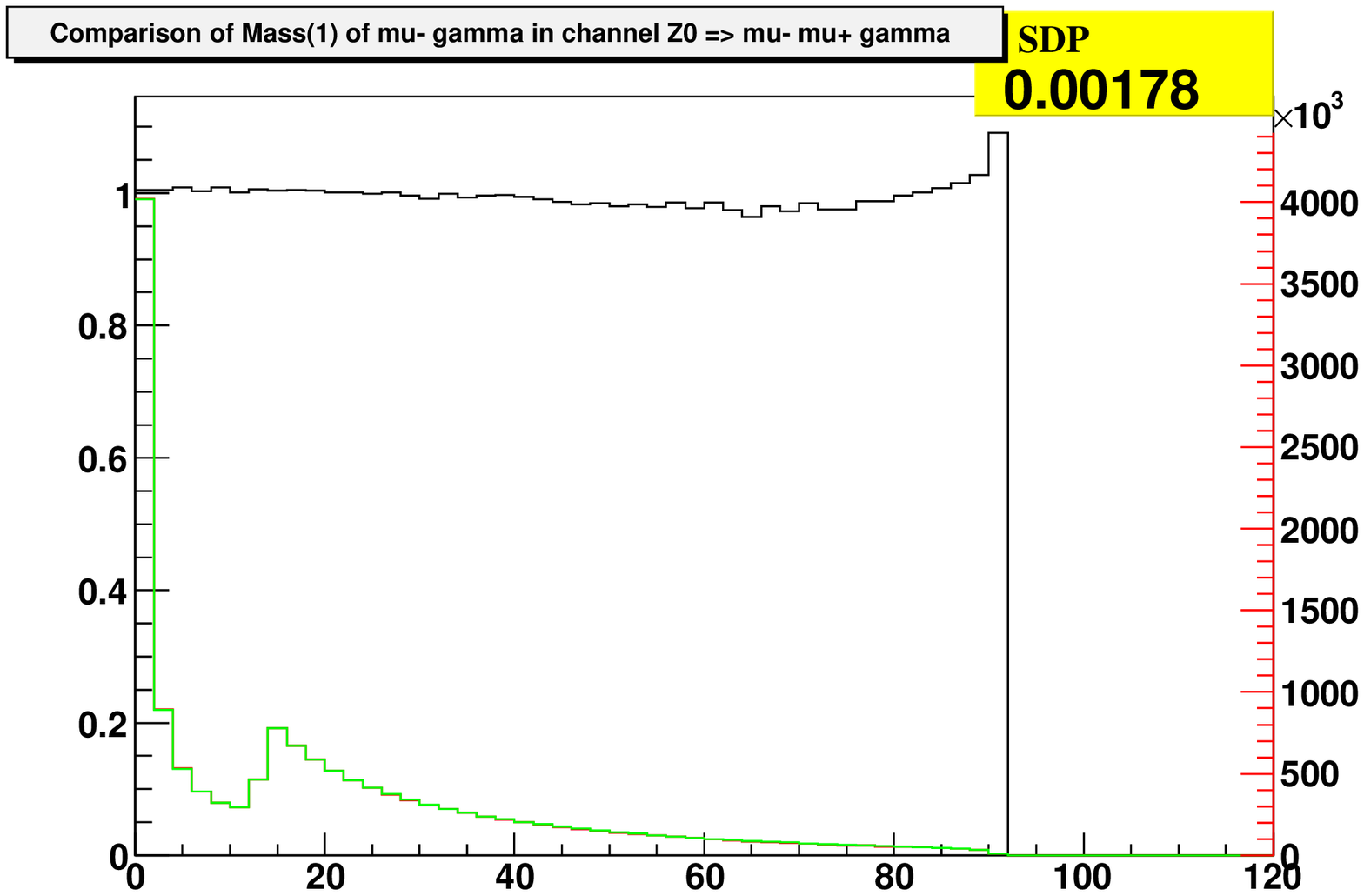}} }
{ \resizebox*{0.49\textwidth}{!}{\includegraphics{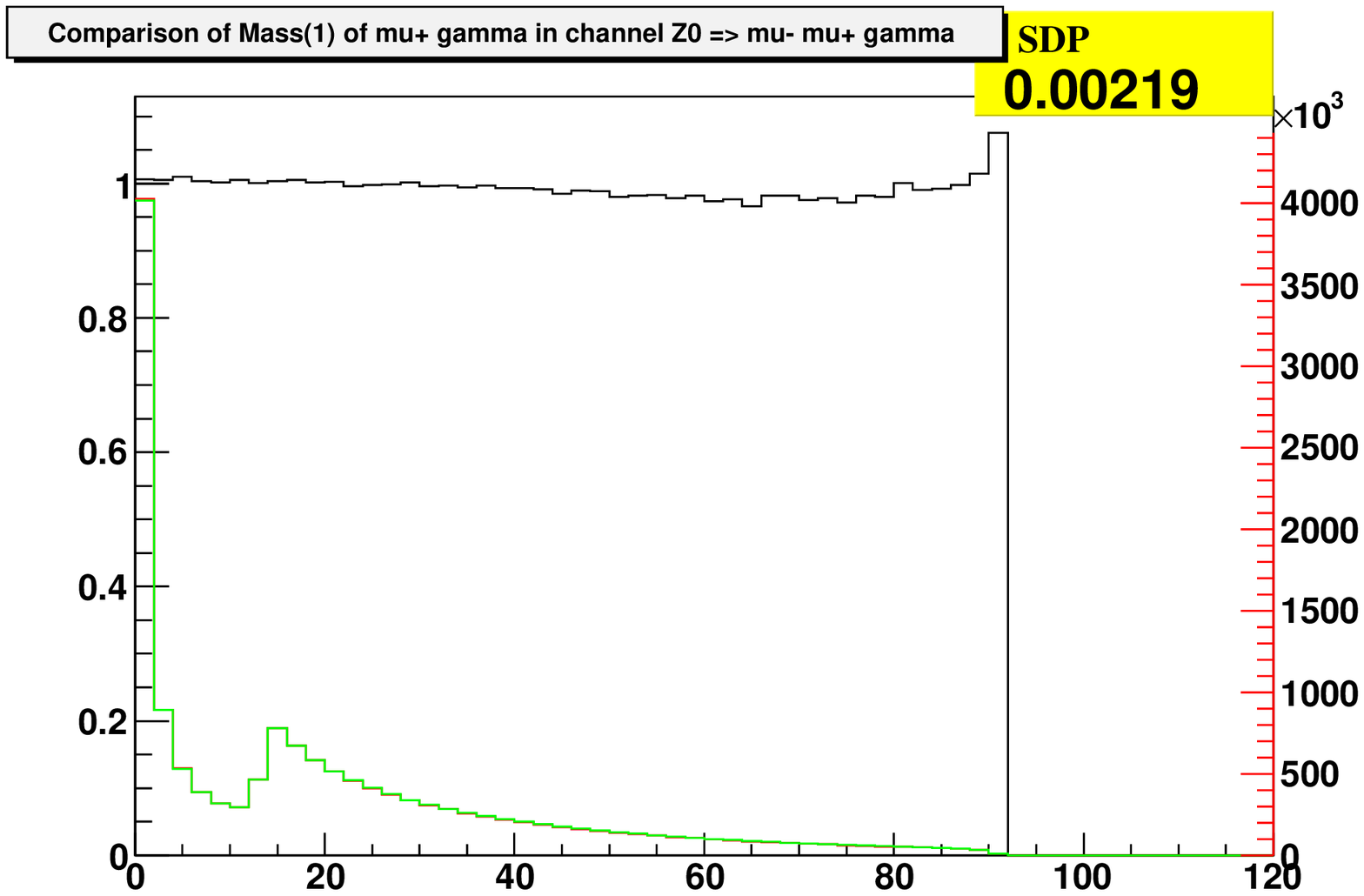}} }
{ \resizebox*{0.49\textwidth}{!}{\includegraphics{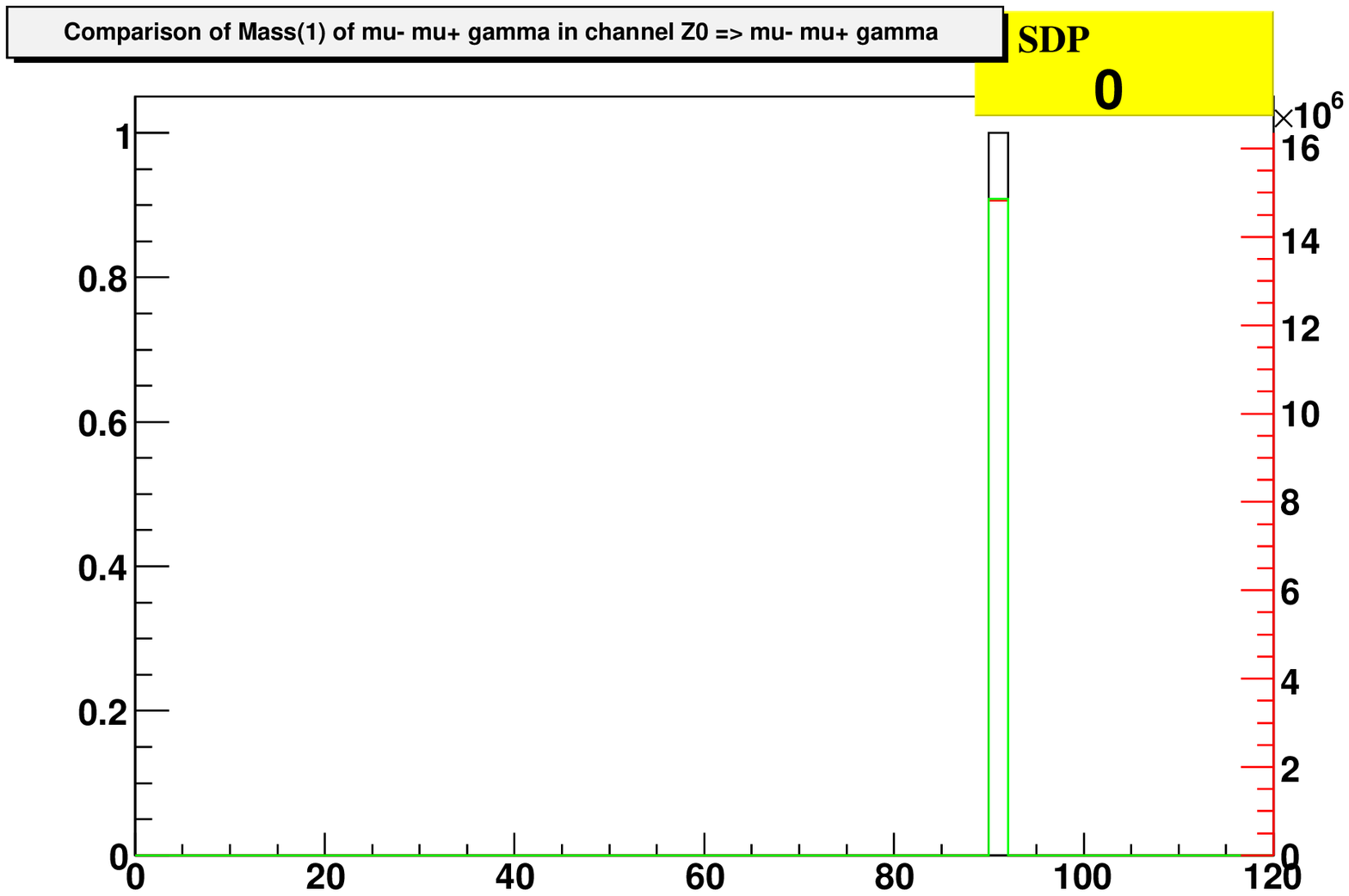}} }

\section*{Decay Channel: $Z^{0} \rightarrow \mu^{-} \mu^{+} \gamma \gamma $}
{ \resizebox*{0.49\textwidth}{!}{\includegraphics{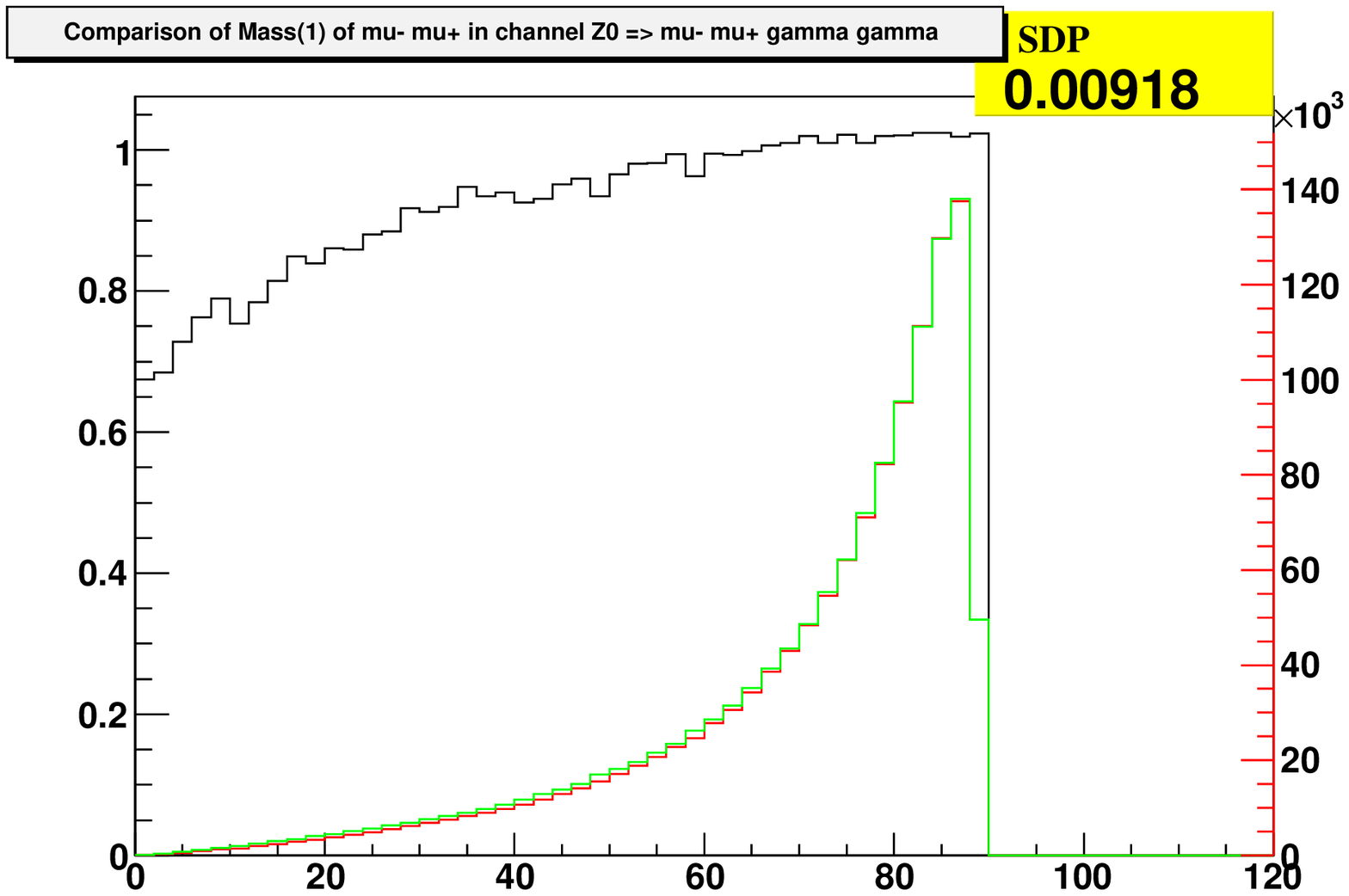}} }
{ \resizebox*{0.49\textwidth}{!}{\includegraphics{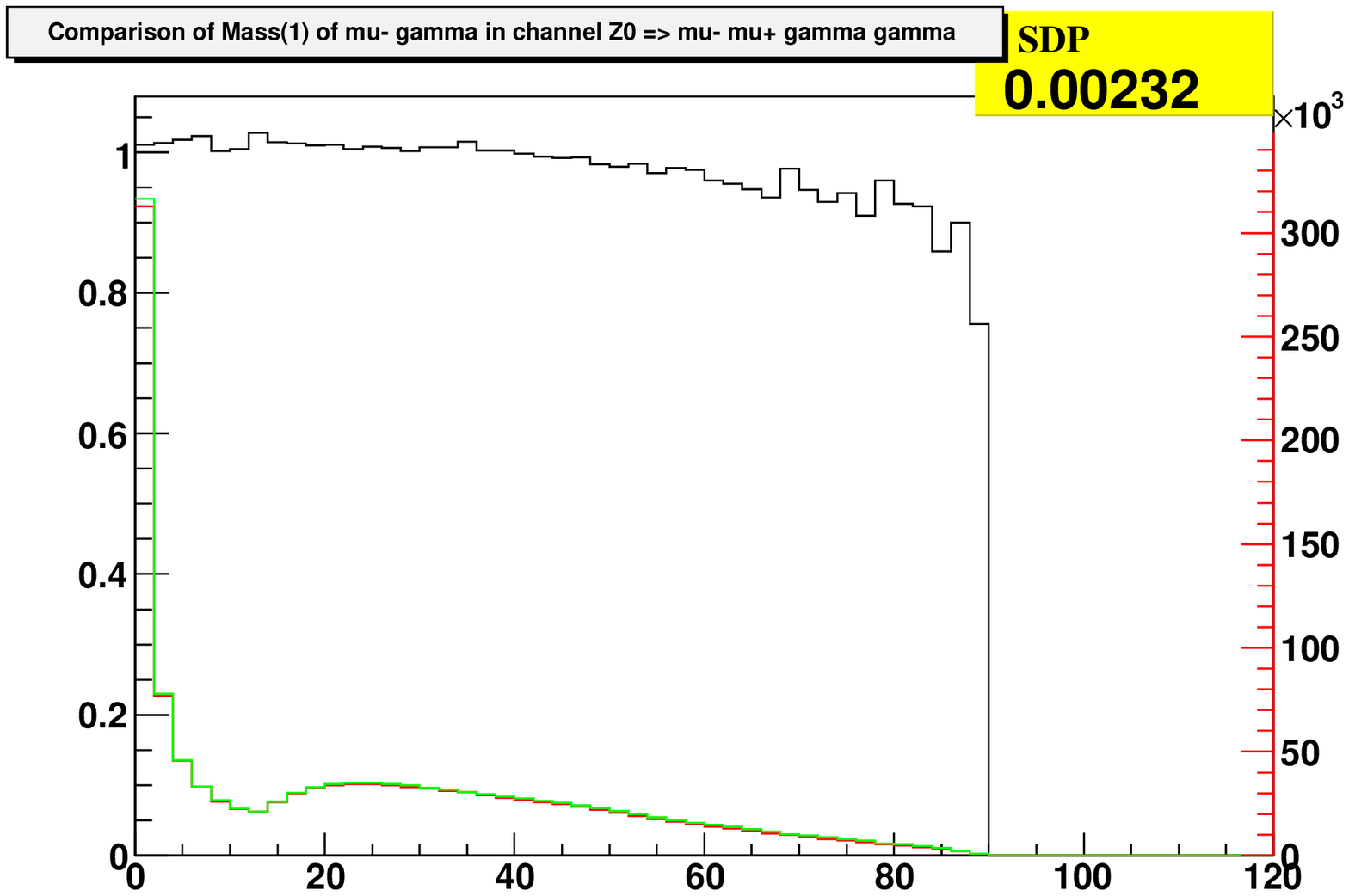}} }
{ \resizebox*{0.49\textwidth}{!}{\includegraphics{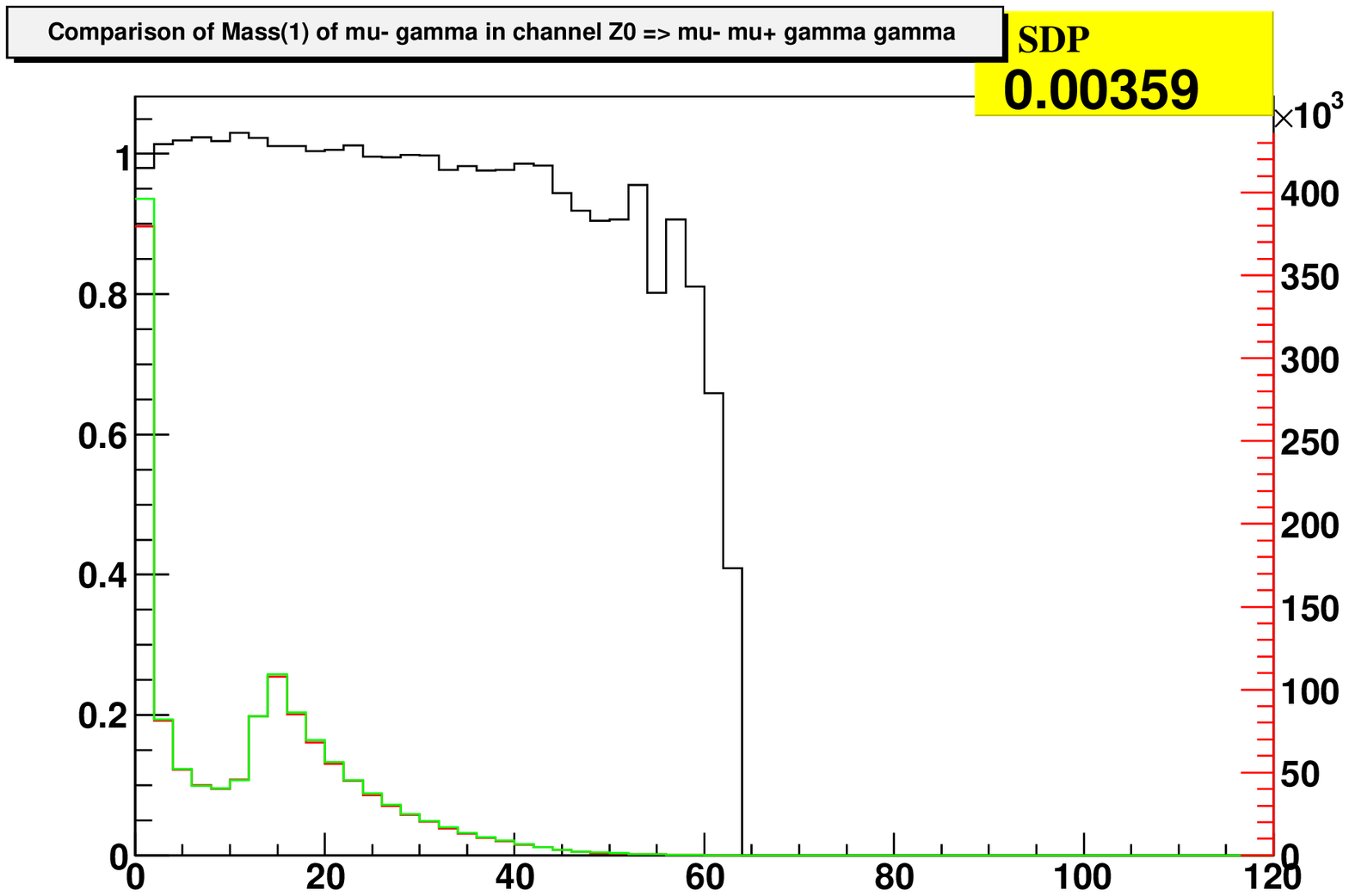}} }
{ \resizebox*{0.49\textwidth}{!}{\includegraphics{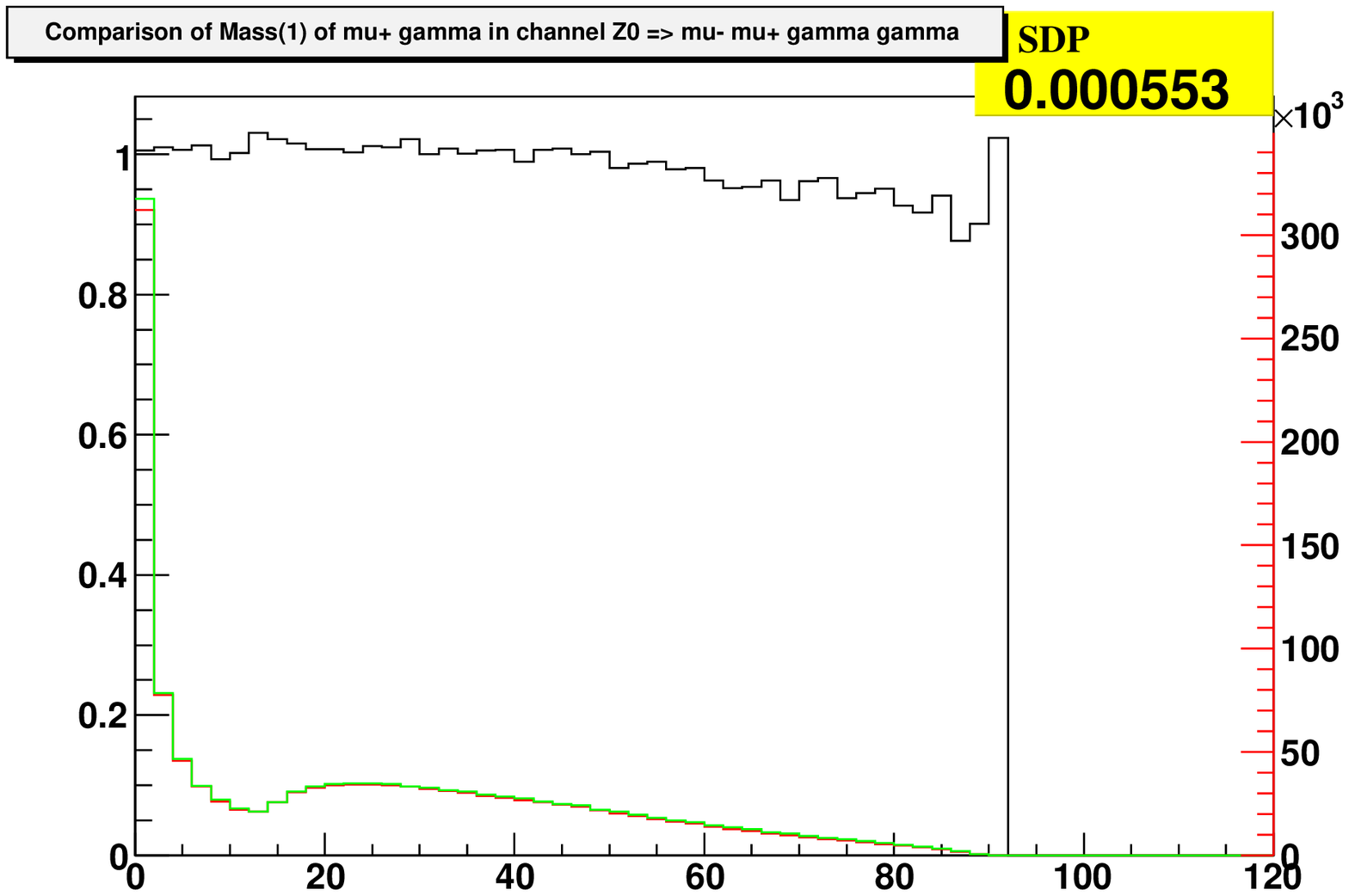}} }
{ \resizebox*{0.49\textwidth}{!}{\includegraphics{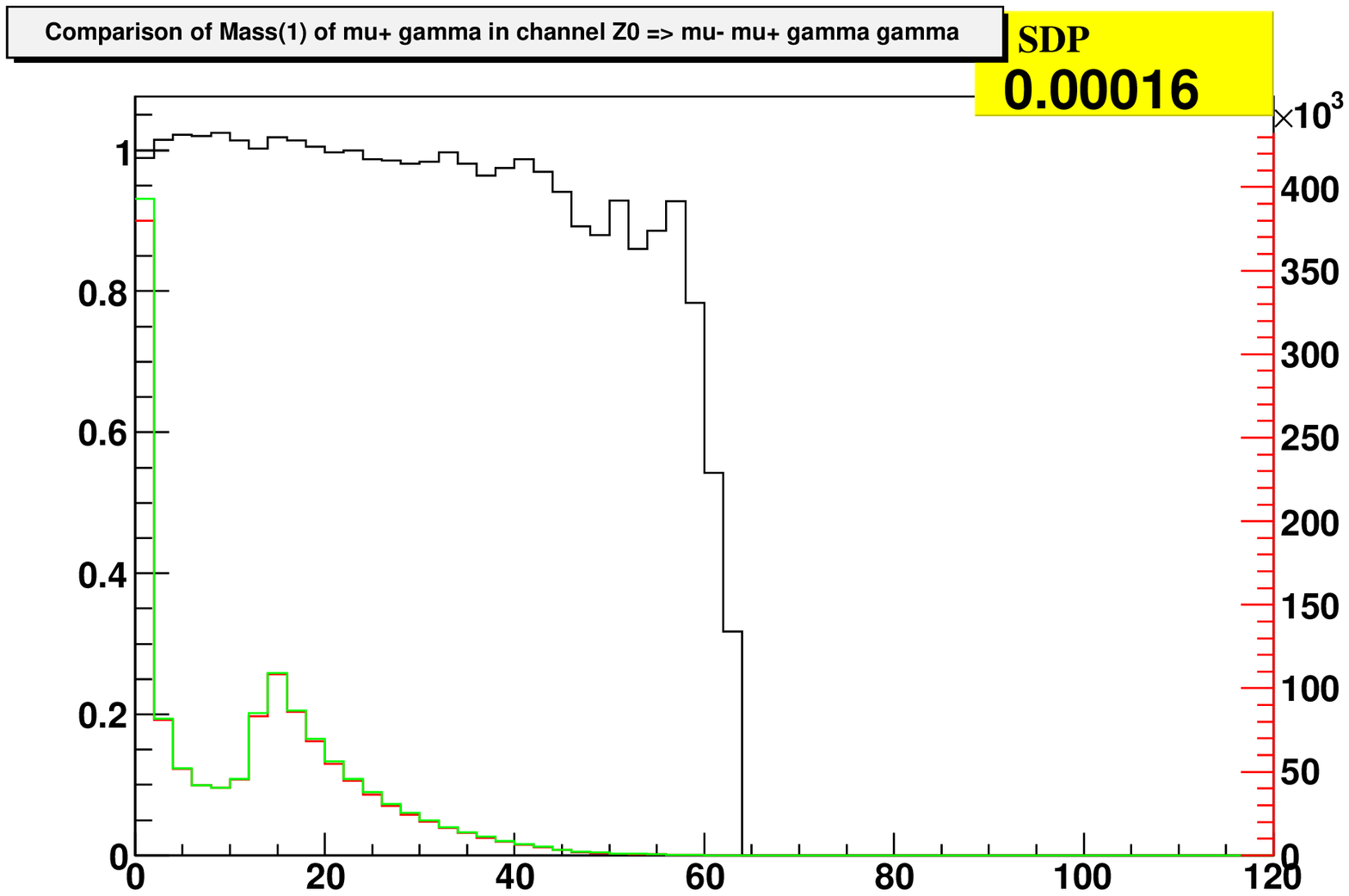}} }
{ \resizebox*{0.49\textwidth}{!}{\includegraphics{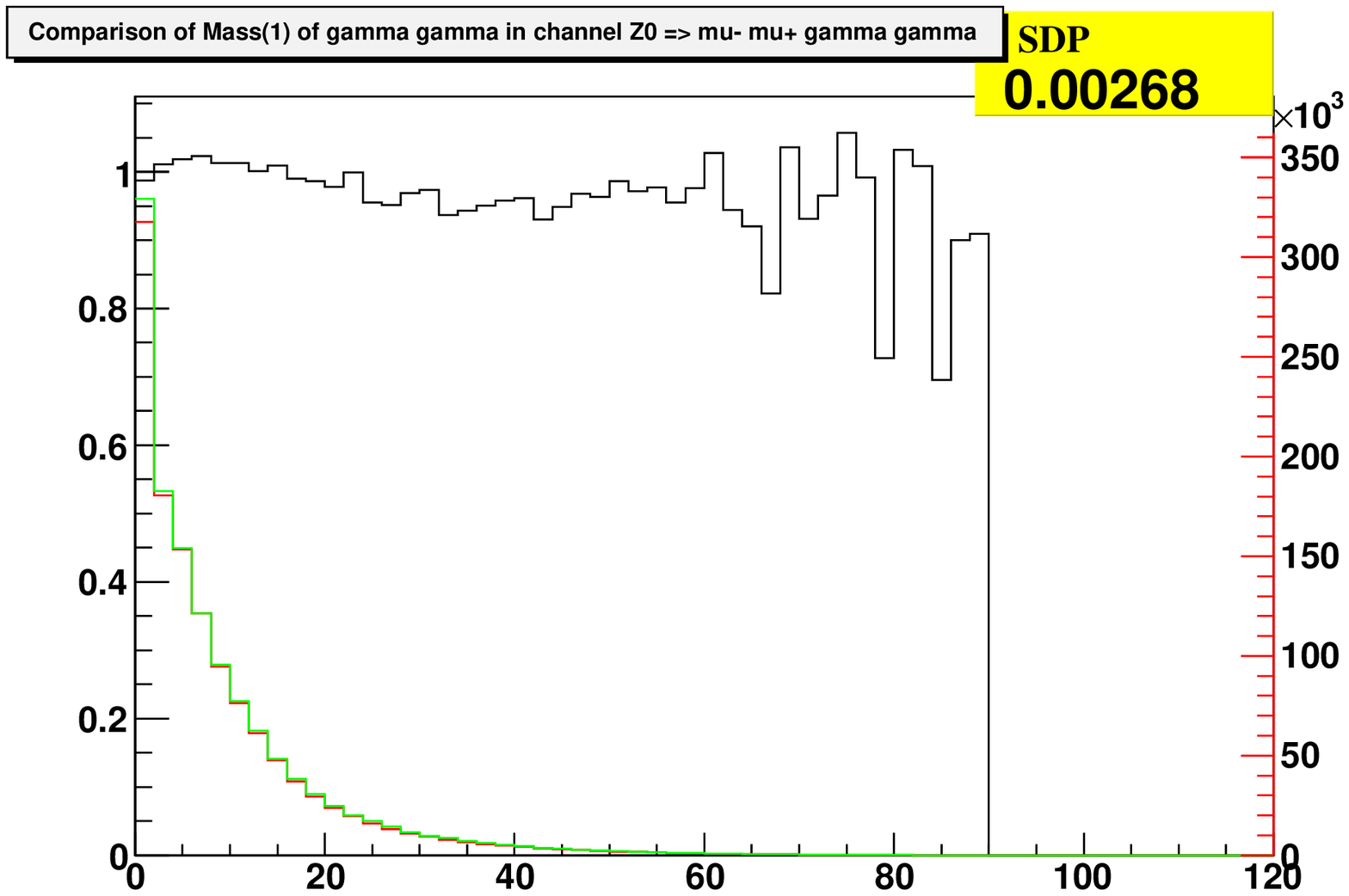}} }
{ \resizebox*{0.49\textwidth}{!}{\includegraphics{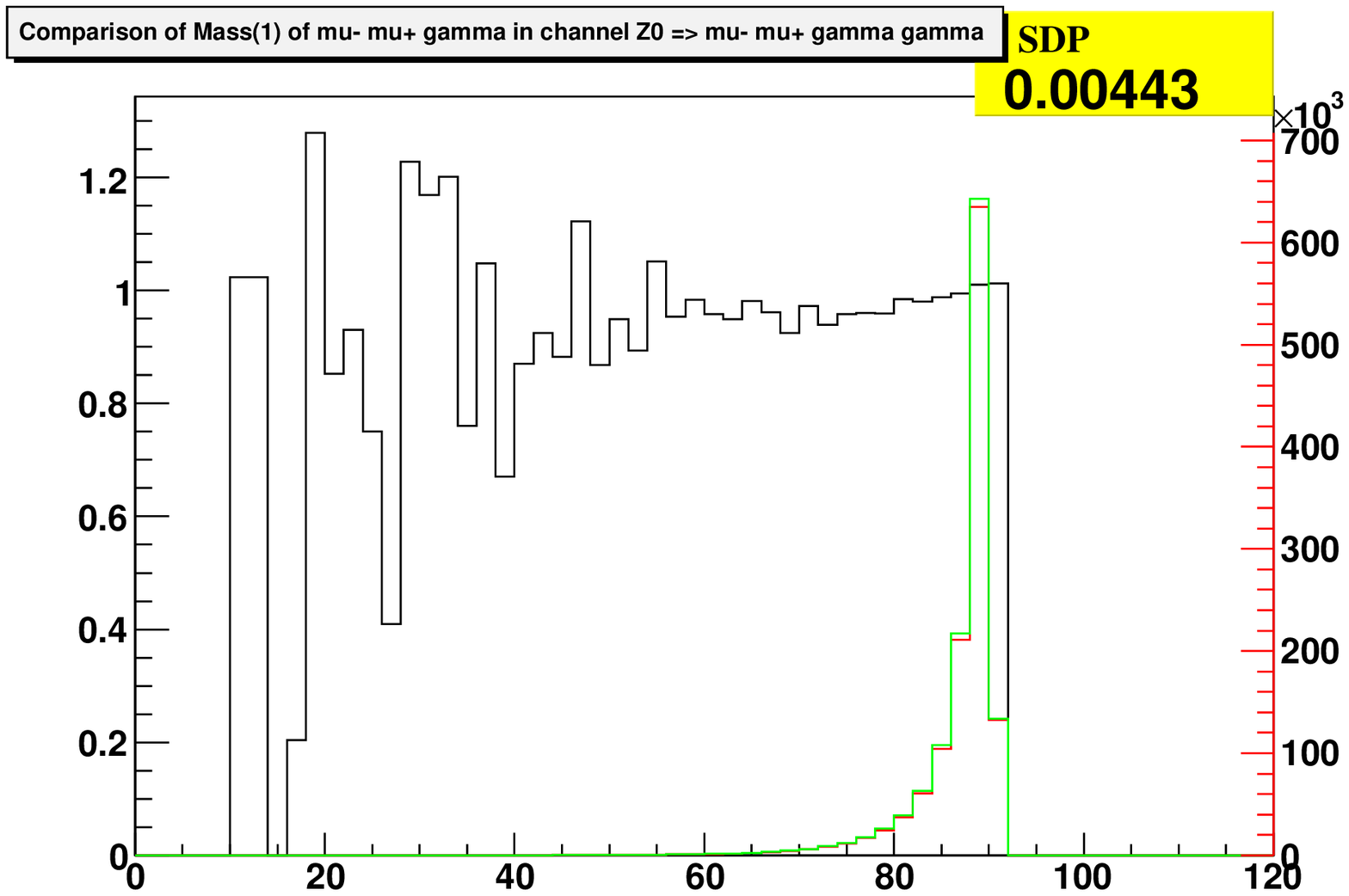}} }
{ \resizebox*{0.49\textwidth}{!}{\includegraphics{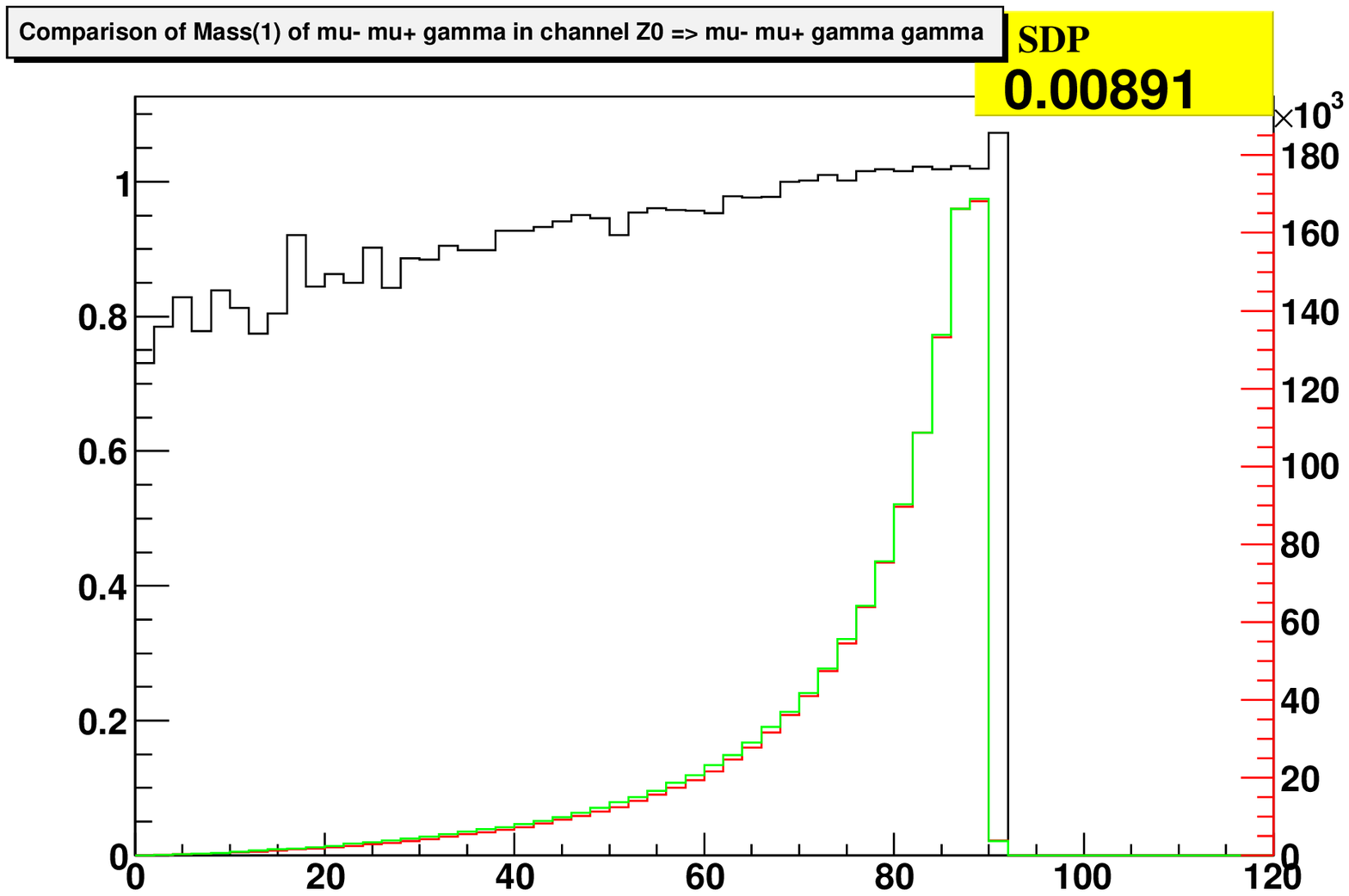}} }
{ \resizebox*{0.49\textwidth}{!}{\includegraphics{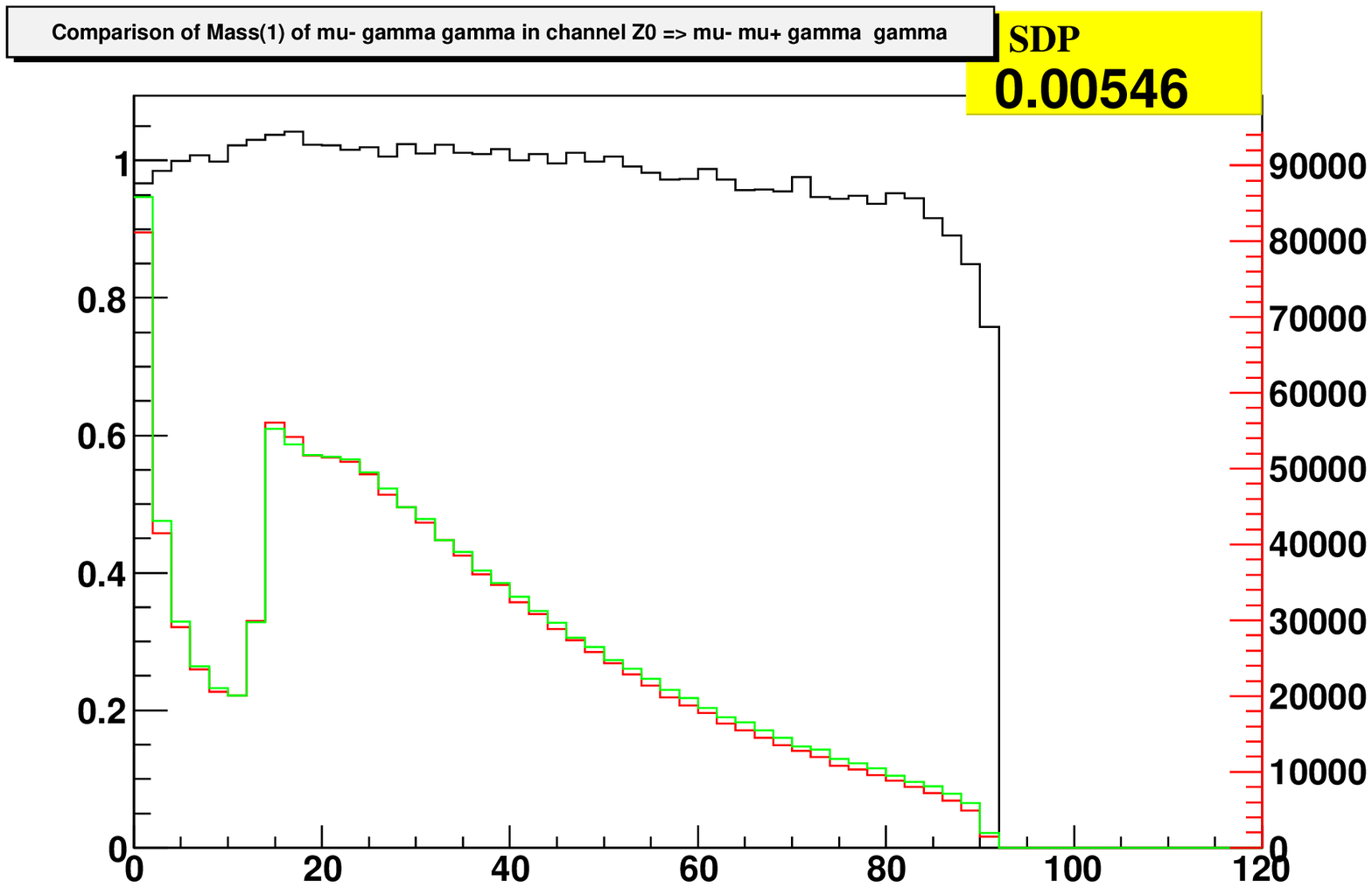}} }
{ \resizebox*{0.49\textwidth}{!}{\includegraphics{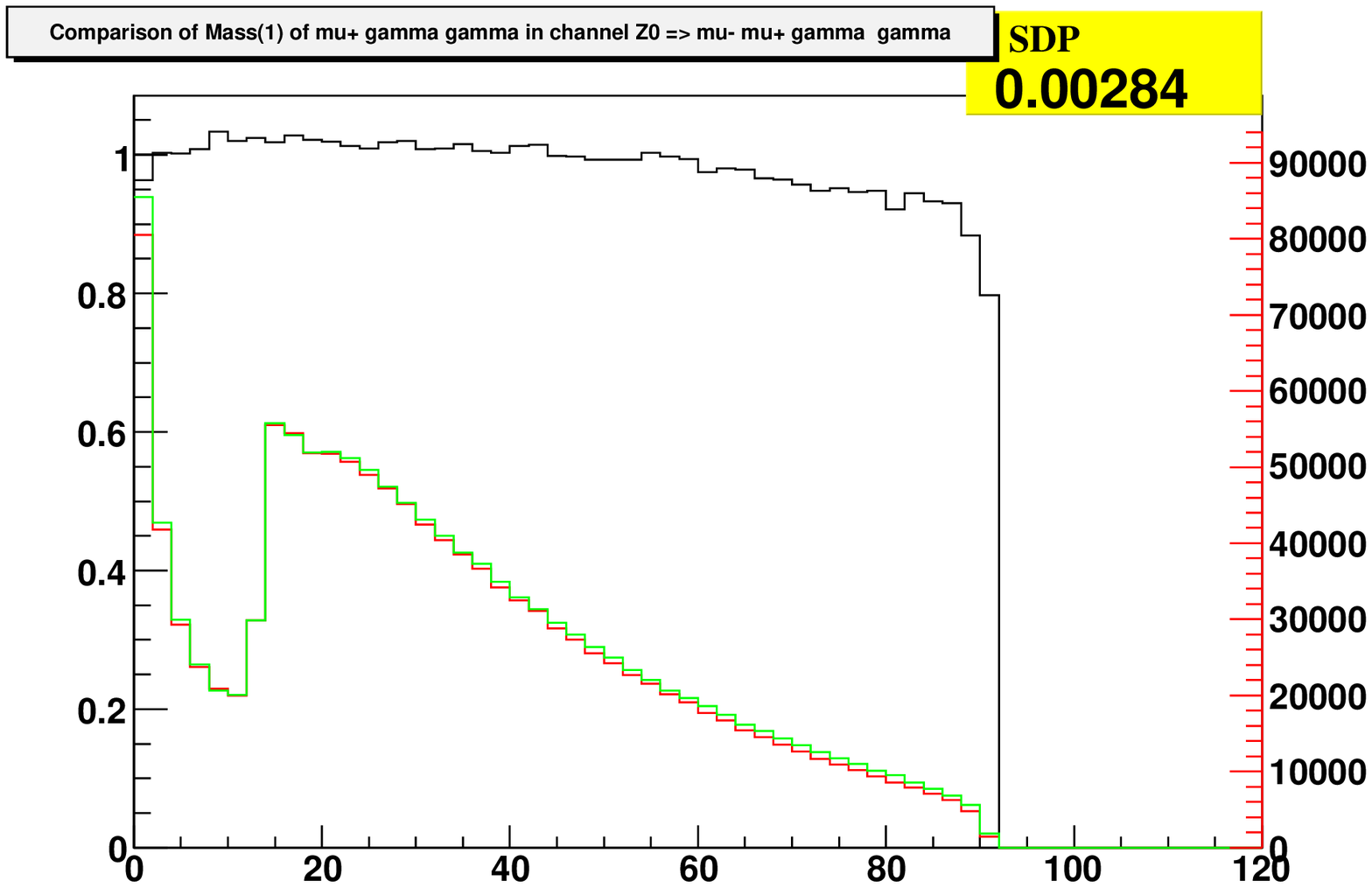}} }

One can see that already for the standard PHOTOS the agreement is good.
Residual  deficiencies are small for both slots of the phase space: 
 single hard photon, 
and  two hard-photon.
Let us present now what kind of changes the inclusion of  NLO terms in PHOTOS brings to the 
results of \emph{test2}.

\vspace{0.3cm} 
{\centering \begin{longtable}{|c|c|c|c|} 
\hline 
Decay channel &\multicolumn{2}{|c|}{ Branching eatio $\pm$ rough errors} & Max. SDP\\ 
      & \textcolor{red}{KKMC} & \textcolor{green}{PHOTOS} & \\ 
\hline 
\hline \( Z^{0} \rightarrow \mu^{-} \mu^{+}  \) & \textcolor{red}{83.9177 $\pm$  0.0092\%} &\textcolor{green}{  83.9303 $\pm$  0.0092\%} & 0.00000 \\ 
\hline 
\hline \( Z^{0} \rightarrow \mu^{-} \mu^{+} \gamma  \) & \textcolor{red}{14.8164 $\pm$  0.0038\%} &\textcolor{green}{  14.7829 $\pm$  0.0038\%} & 0.00005 \\ 
\hline 
\hline \( Z^{0} \rightarrow \mu^{-} \mu^{+} \gamma \gamma  \) & \textcolor{red}{ 1.2659 $\pm$  0.0011\%} &\textcolor{green}{   1.2868 $\pm$  0.0011\%} & 0.00293 \\ 
\hline 
\end{longtable}\par} 
\vspace{0.3cm}

\section*{Decay Channel: $Z^{0} \rightarrow \mu^{-} \mu^{+} \gamma $}
{ \resizebox*{0.49\textwidth}{!}{\includegraphics{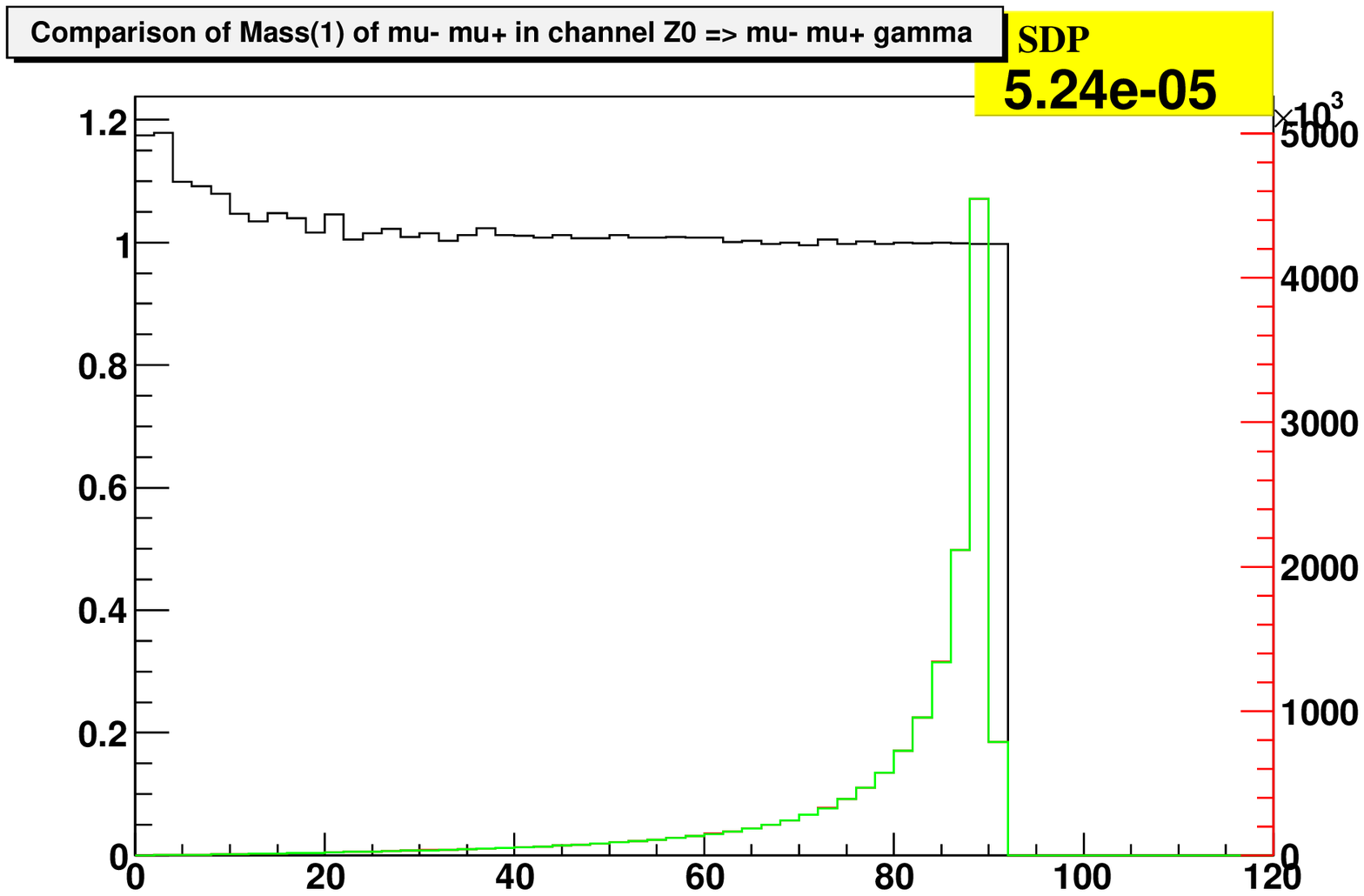}} }
{ \resizebox*{0.49\textwidth}{!}{\includegraphics{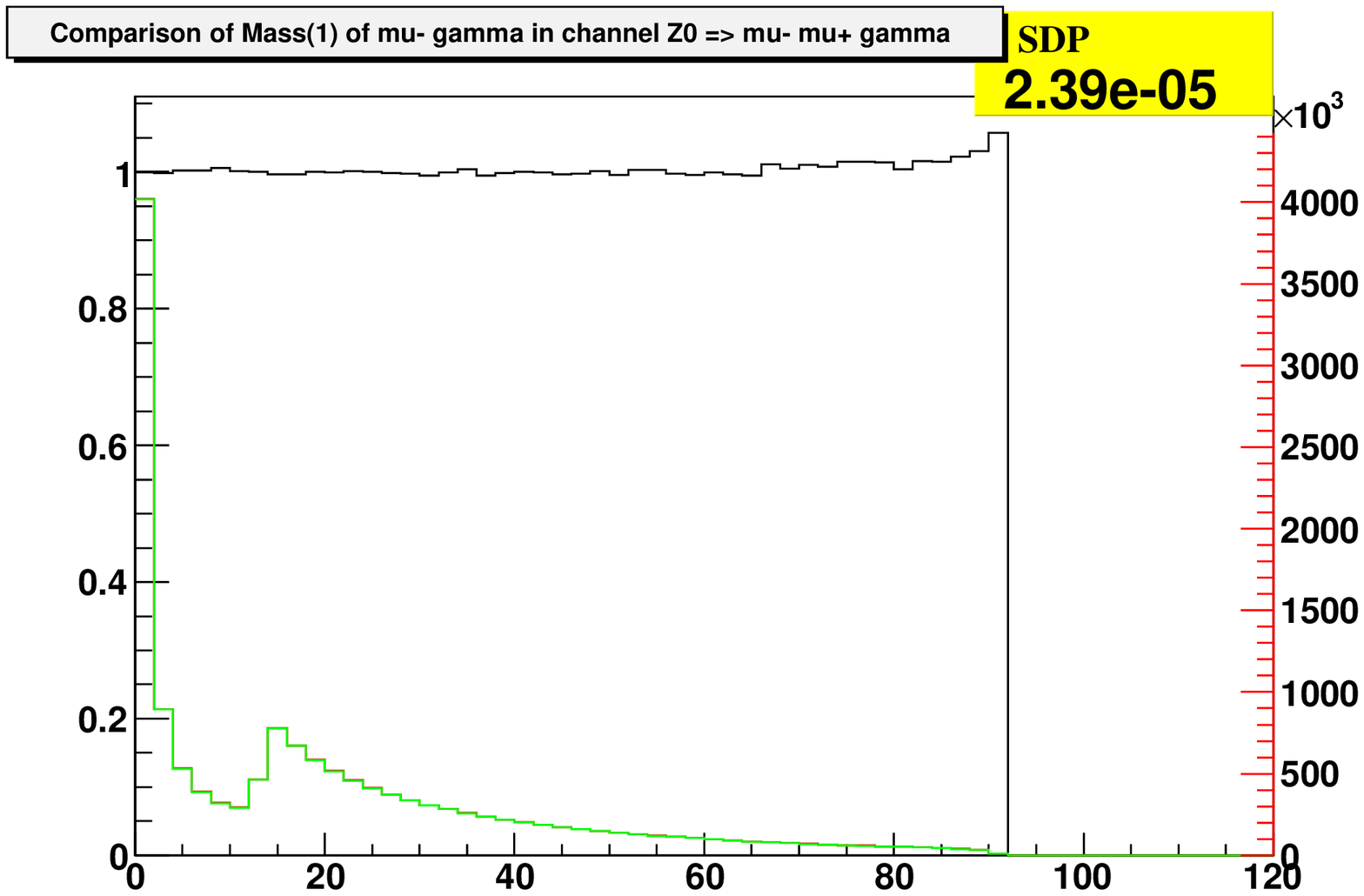}} }
{ \resizebox*{0.49\textwidth}{!}{\includegraphics{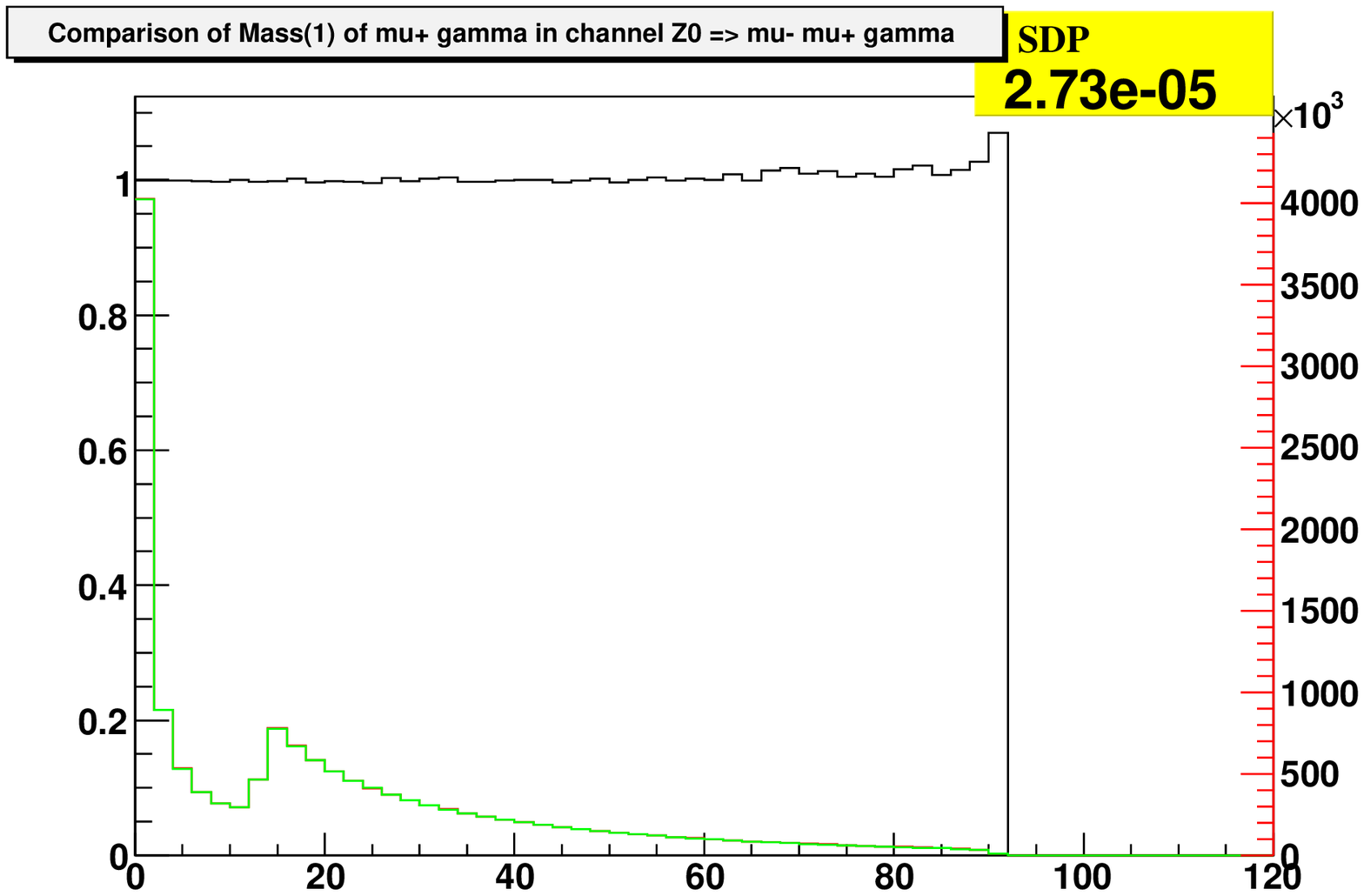}} }
{ \resizebox*{0.49\textwidth}{!}{\includegraphics{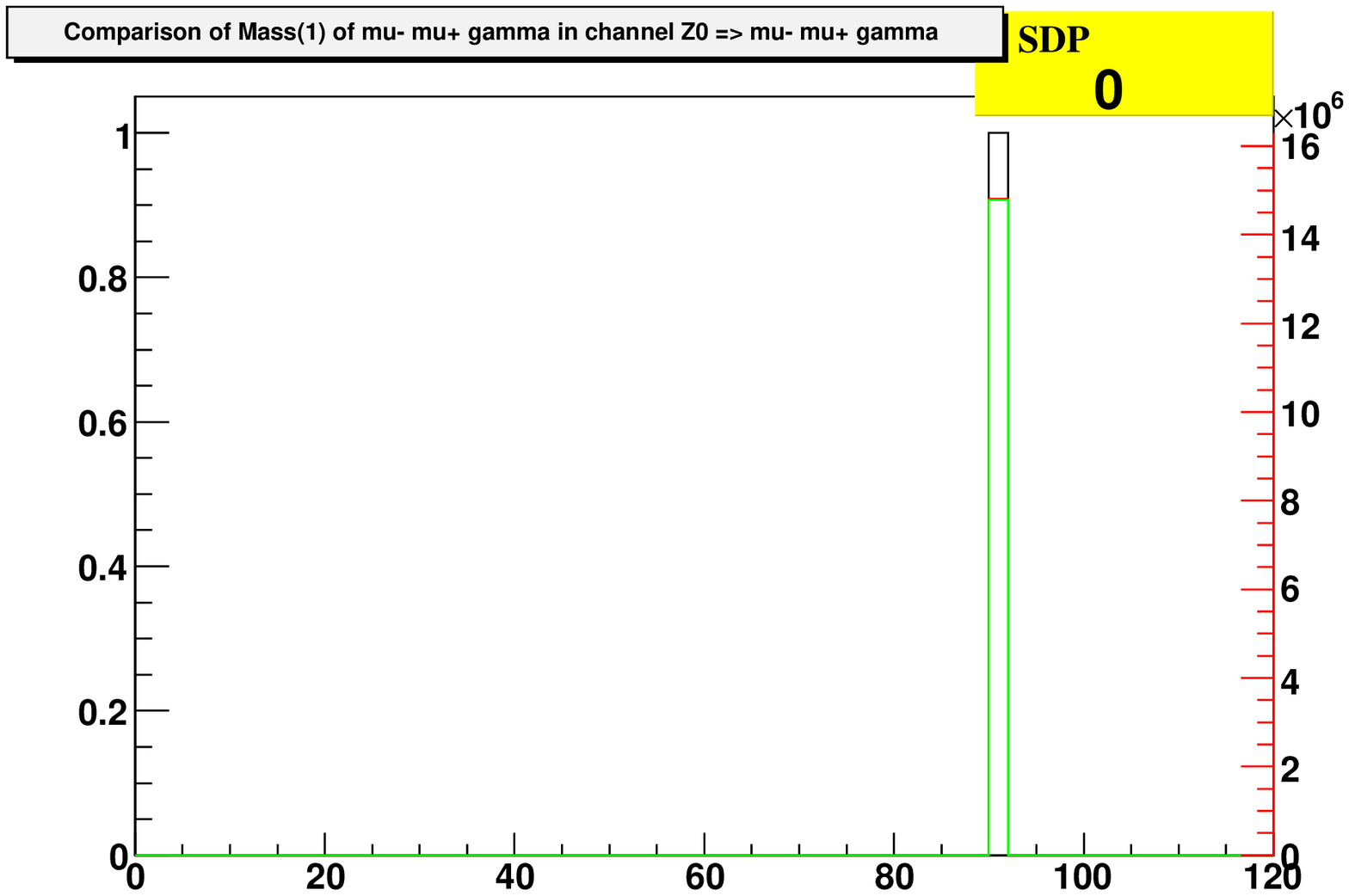}} }
\section*{Decay Channel: $Z^{0} \rightarrow \mu^{-} \mu^{+} \gamma \gamma $}
{ \resizebox*{0.49\textwidth}{!}{\includegraphics{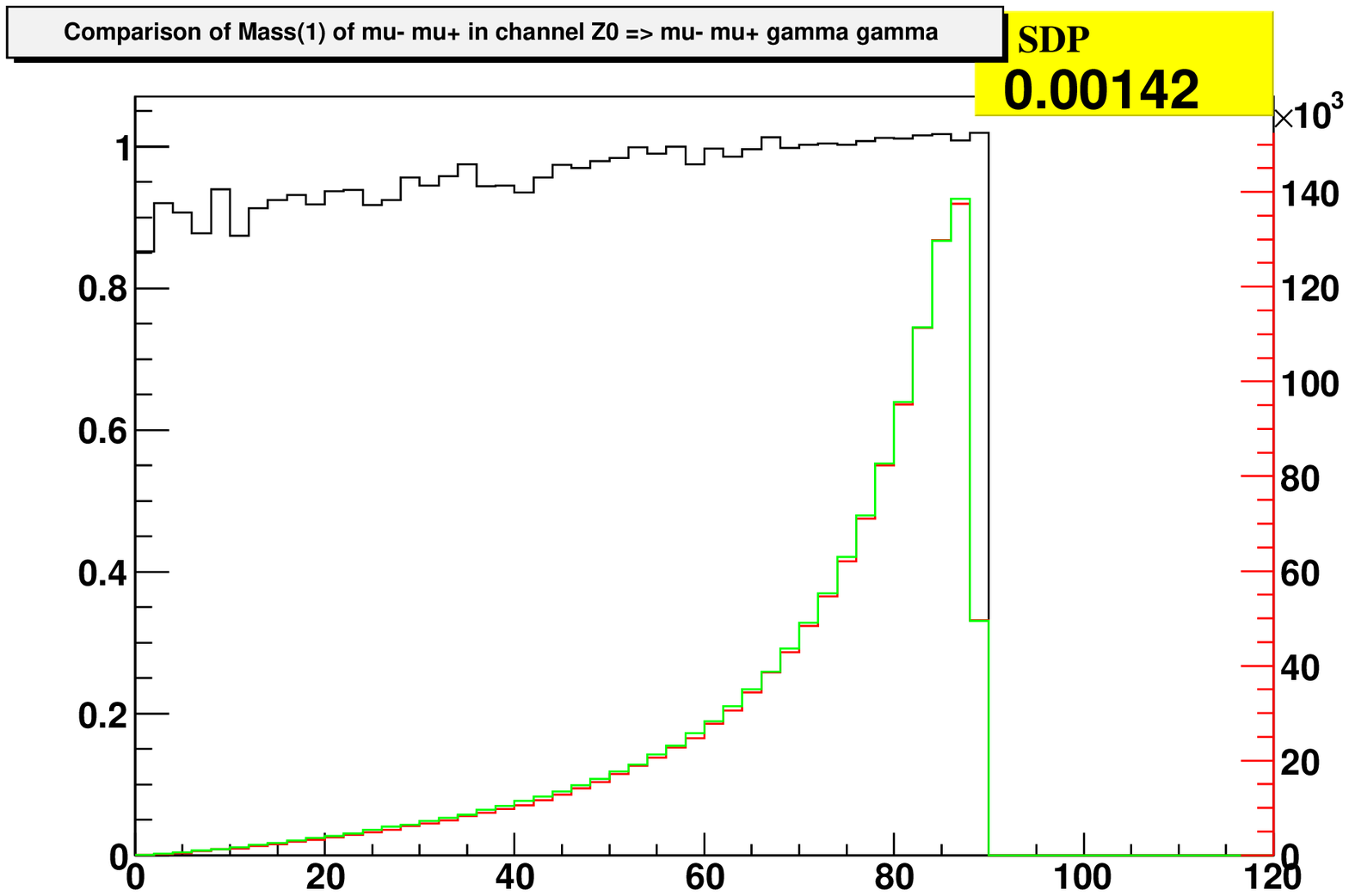}} }
{ \resizebox*{0.49\textwidth}{!}{\includegraphics{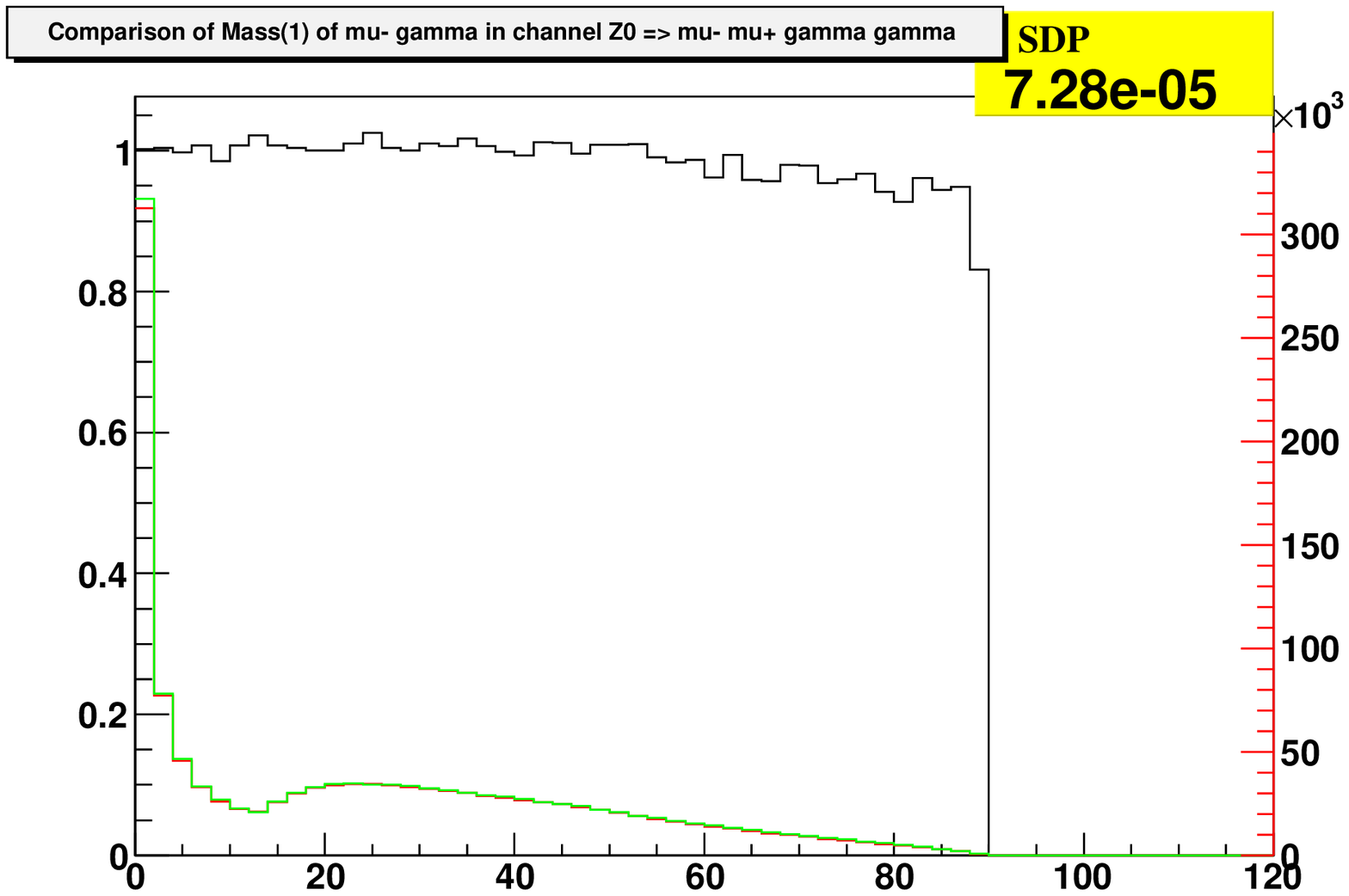}} }
{ \resizebox*{0.49\textwidth}{!}{\includegraphics{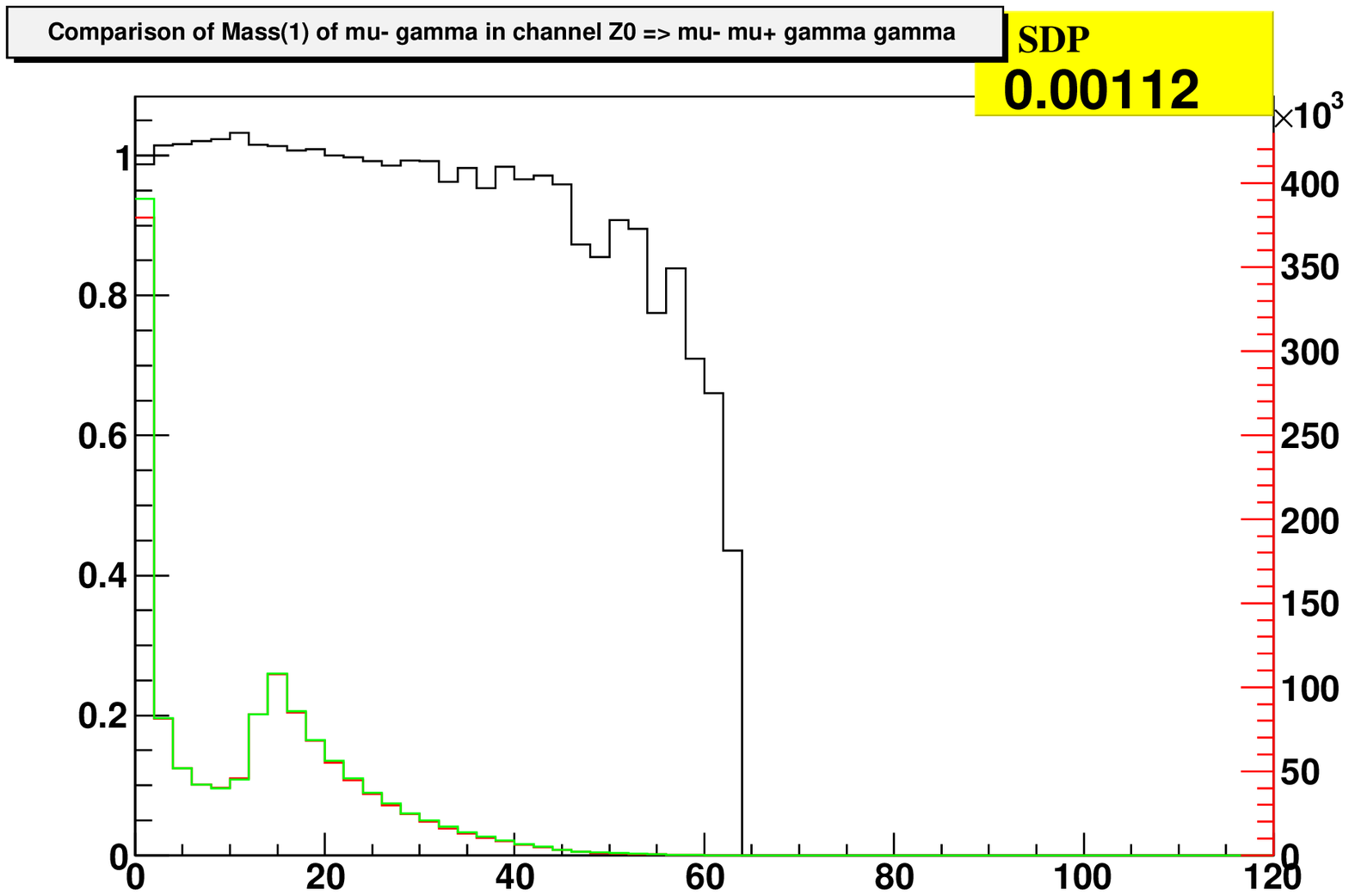}} }
{ \resizebox*{0.49\textwidth}{!}{\includegraphics{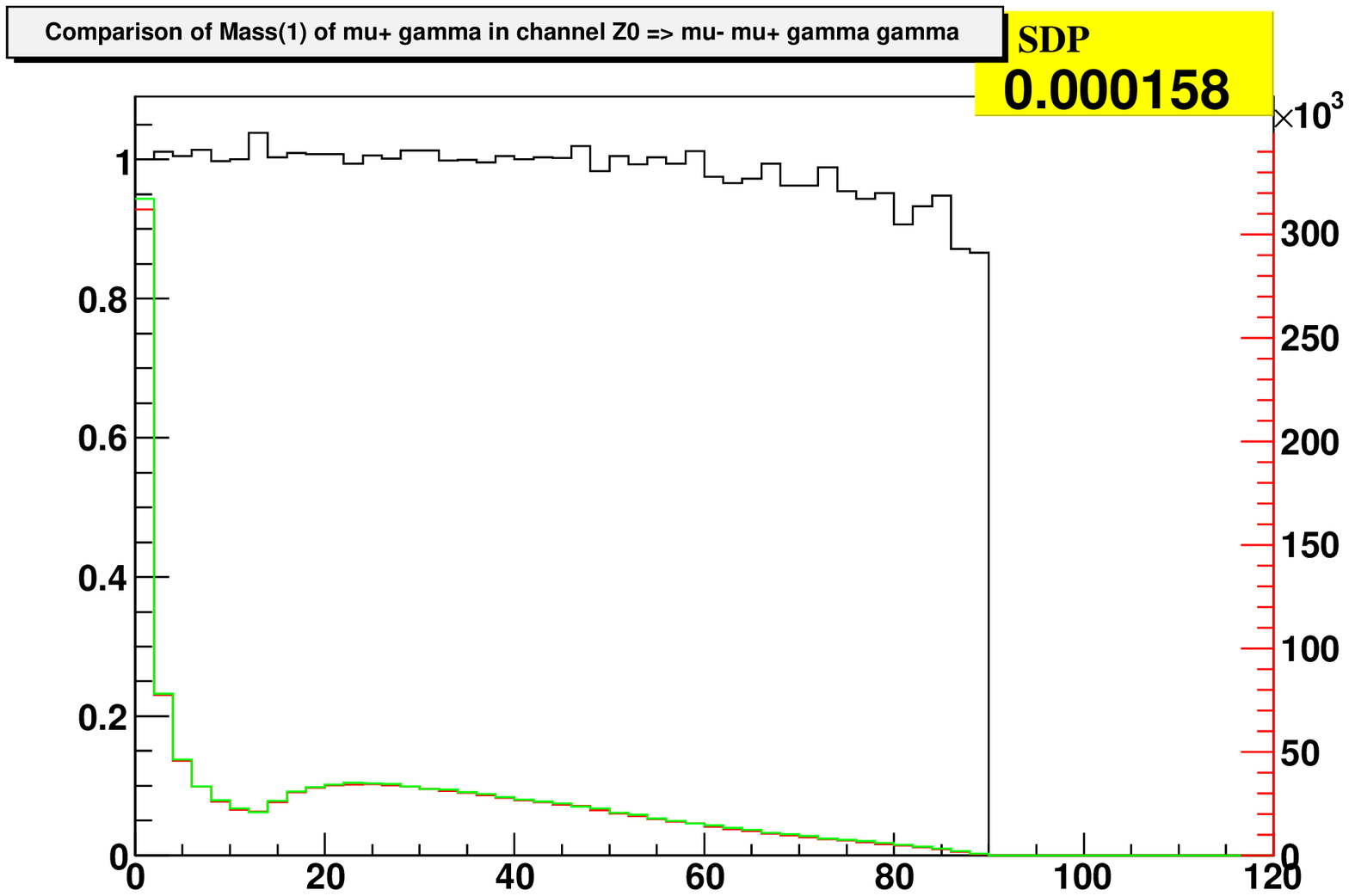}} }
{ \resizebox*{0.49\textwidth}{!}{\includegraphics{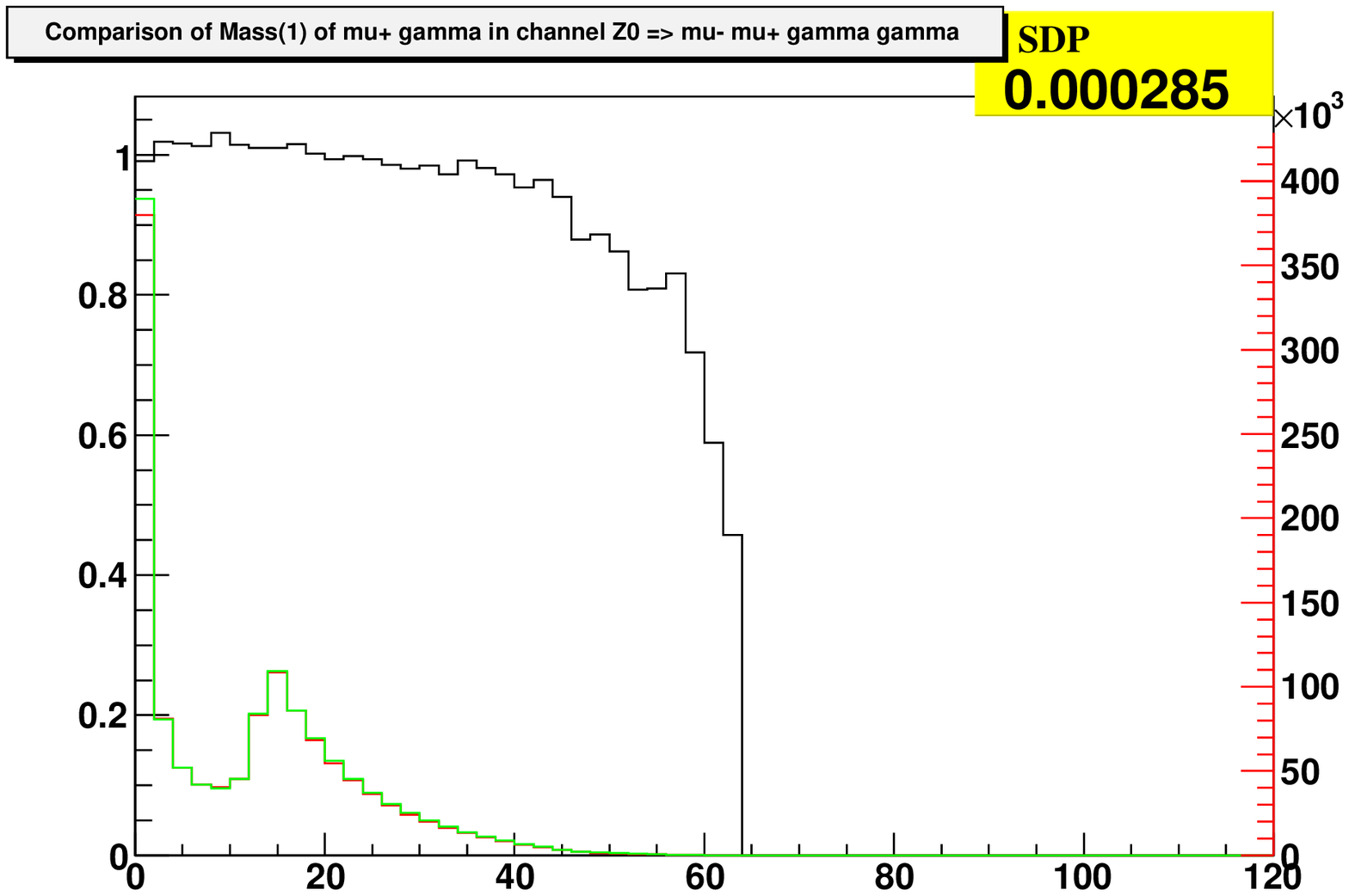}} }
{ \resizebox*{0.49\textwidth}{!}{\includegraphics{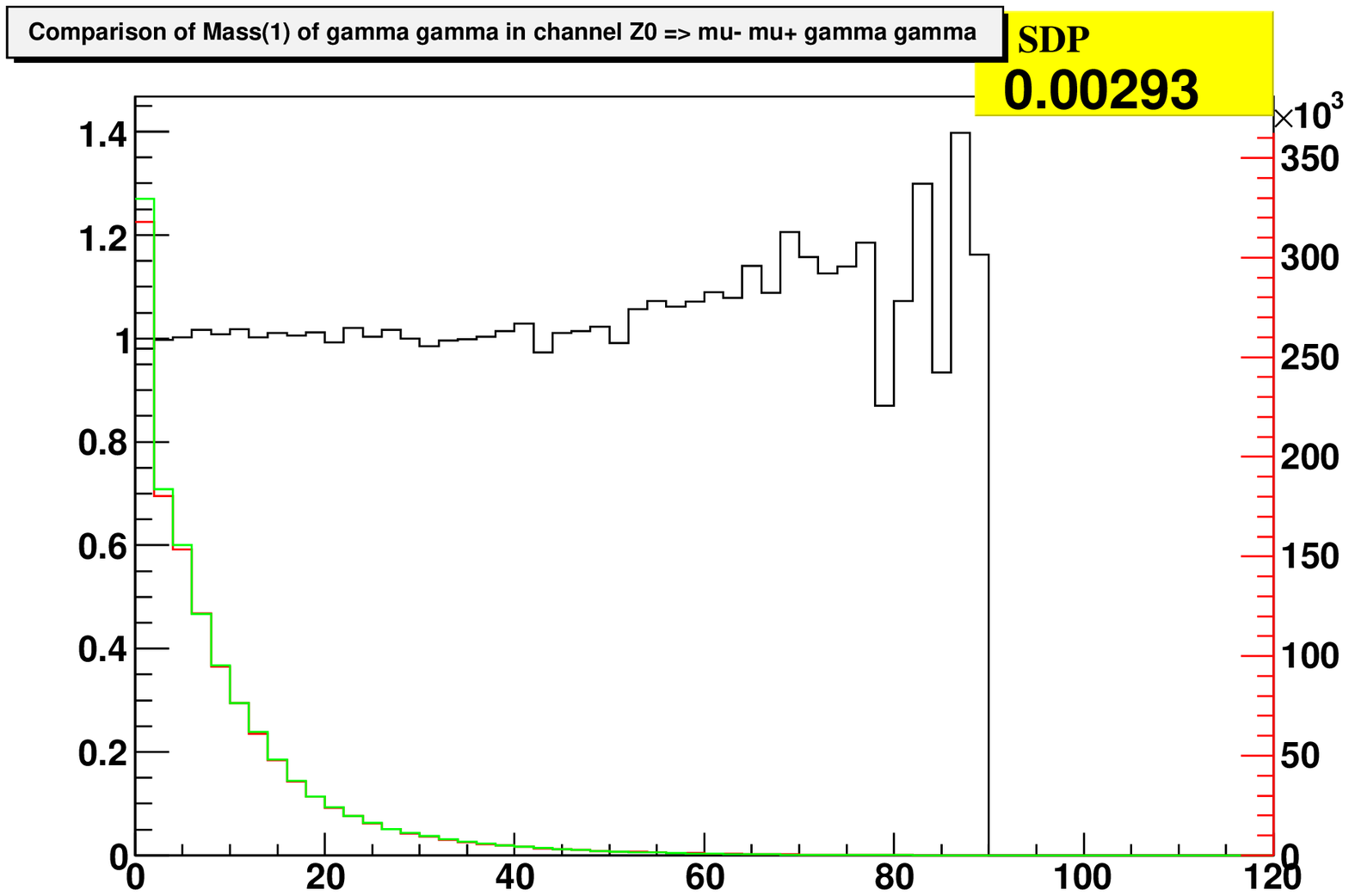}} }
{ \resizebox*{0.49\textwidth}{!}{\includegraphics{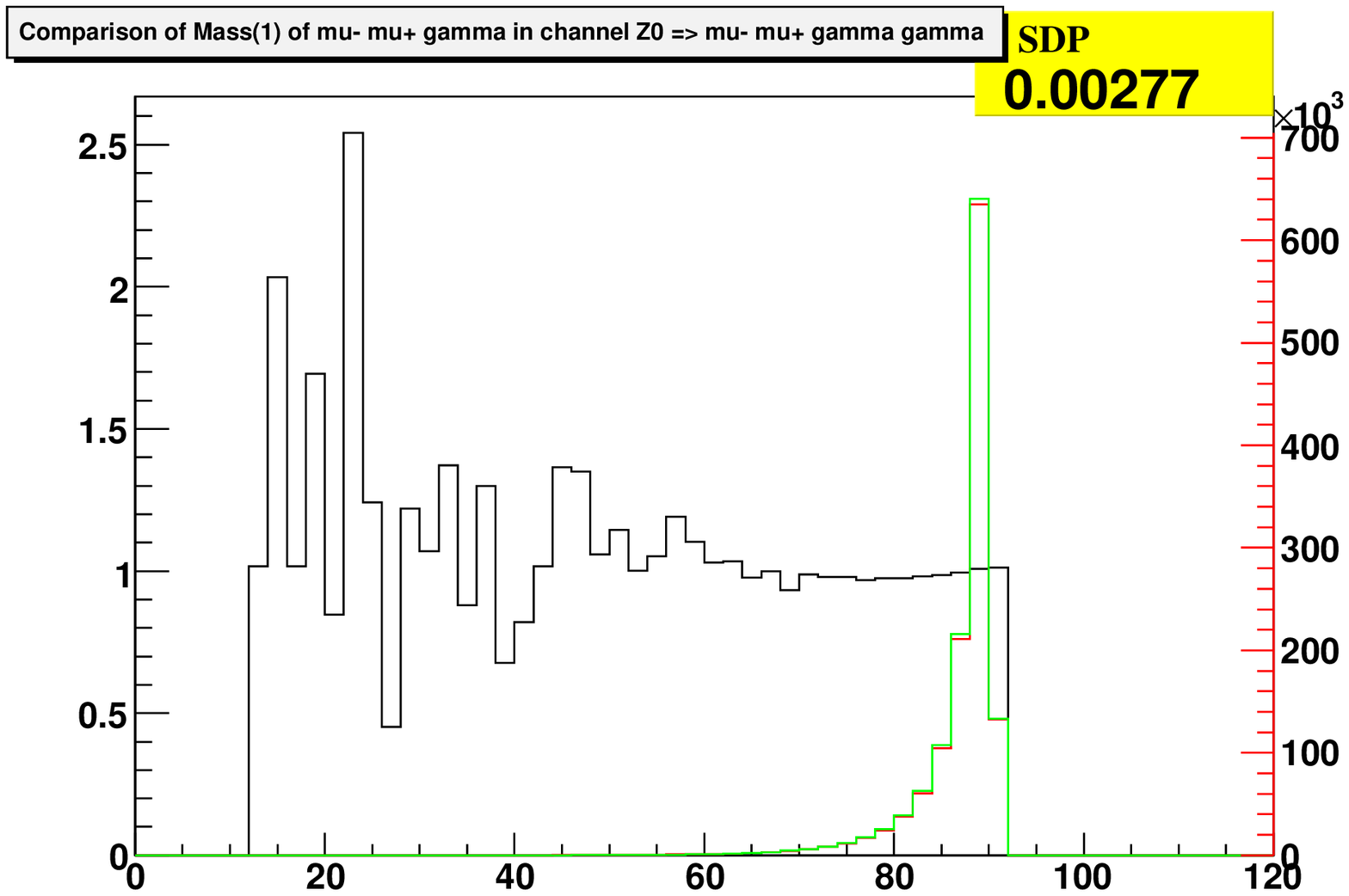}} }
{ \resizebox*{0.49\textwidth}{!}{\includegraphics{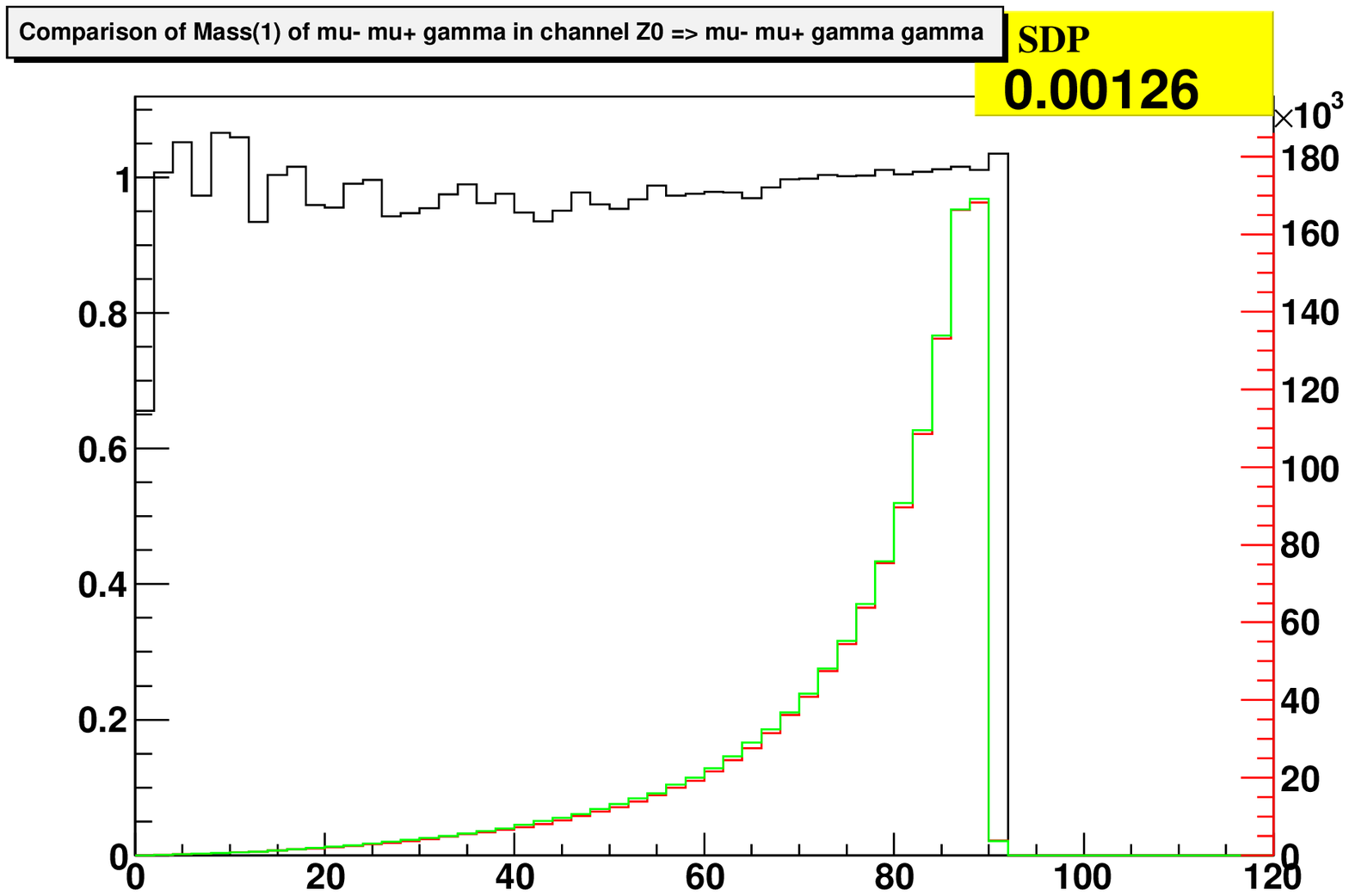}} }
{ \resizebox*{0.49\textwidth}{!}{\includegraphics{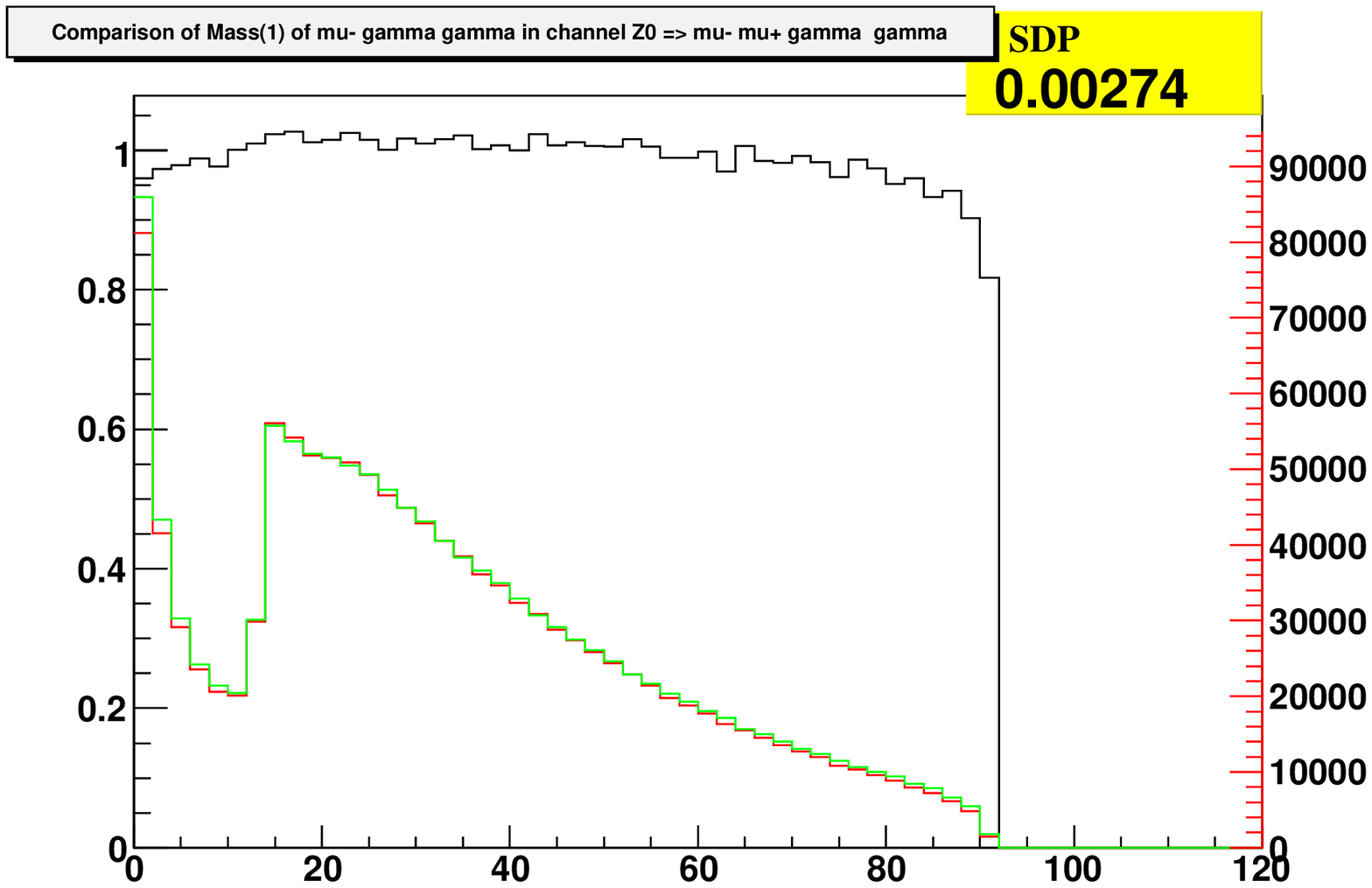}} }
{ \resizebox*{0.49\textwidth}{!}{\includegraphics{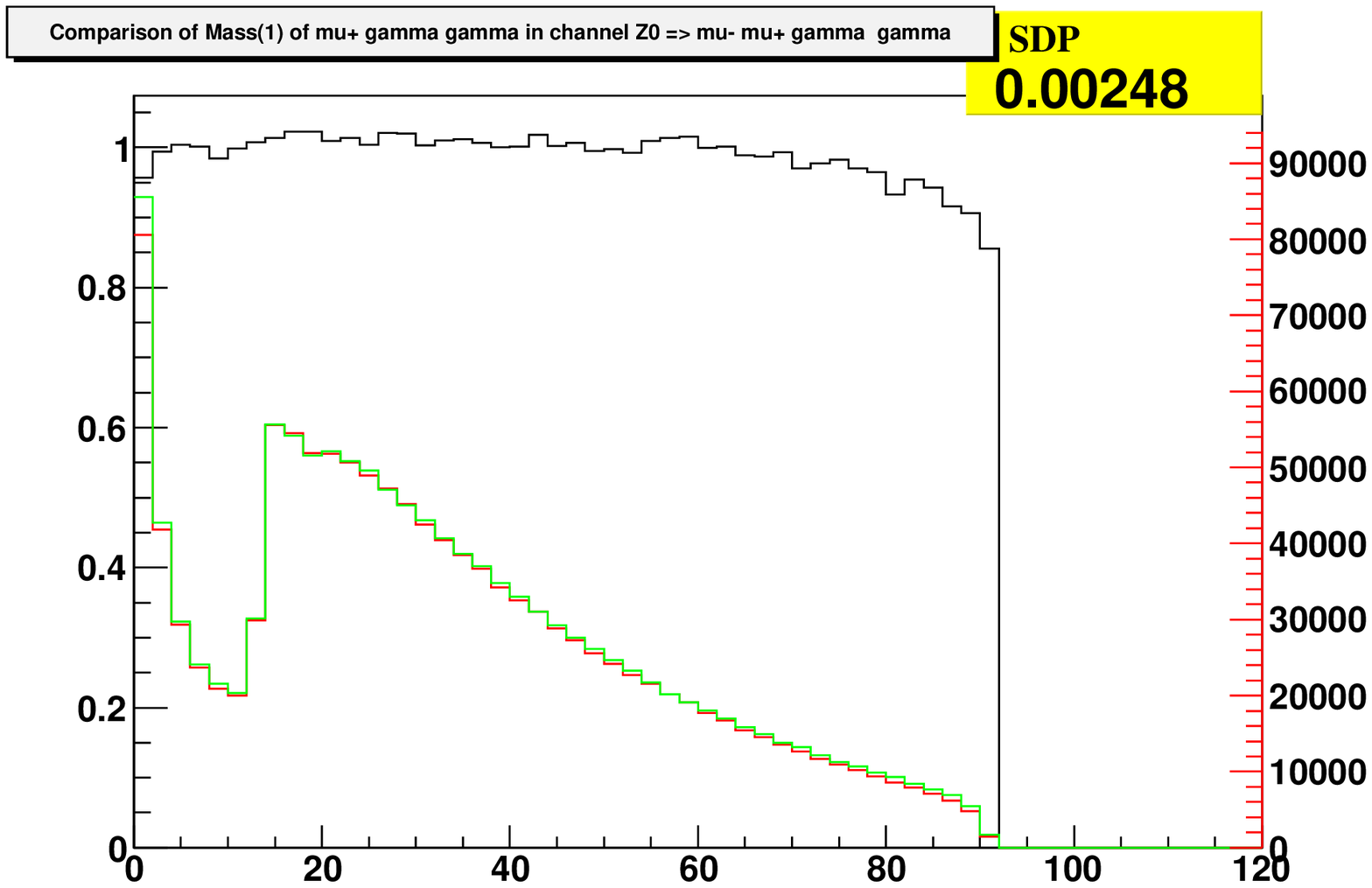}} }

For the single hard-photon distributions, the differences 
diminished significantly, again  a factor of about 50! This was to 
be expected. Even for the 
distributions of the phase-space slot with  two hard photons, the  differences diminished.
The SDP  decreased by a factor of about 3. 
 This is not as striking as for the single hard-photon
configuration, but  it is of no surprise: the complete second-order matrix 
element is missing. The improvement by a factor of 3 provides, however, a strong indication that the
algorithm of iteration used in the generation of consecutive photons work well from the
point of view of NNLL level as well. 
The acoplanarity plots presented in the previous paper \cite{Golonka:2005pn} also demonstrated some of the NNLL aspect of the algorithm. That is why we are not going to discuss
this point here, but we would rather leave it to future 
 discussion of the NNLL content of our algorithm: this aspect goes 
beyond the  purpose of the present paper and the scope of interest of most of PHOTOS users. 

Much as is described in the present paper,
a new  contribution to the PHOTOS correcting weight would  be needed for the NNLL case.   However no changes
in the phase-space algorithm would  {\it a priori} be required. The techniques 
of  gauge-invariant separation of the amplitudes into parts, 
as used for instance  in ref. \cite{Was:2004ig}, will probably be necessary.
They proved to be instrumental 
in the implementation of the second-order
matrix elements for $e^+e^- \to \nu_e \bar \nu_e \gamma \gamma $ into   KKMC. The 
exclusive exponentiation scheme of the KKMC Monte Carlo is prepared for $s$-channel processes.

We have to admit that once the NLO terms are switched on in PHOTOS, 
the difference between its results and those of the second-order matrix-element 
generator KKMC  are at the limit of being recognized, even if samples
of $10^8$ events are used. For the case of the two-photon test,  differences 
due to the missing second-order matrix element in PHOTOS 
can be  observed; yet they are  too 
small, and don't have  enough structure, to understand their possible origin.
A part of the differences may even originate from the third-order LL 
terms (after integration), which are missing in KKMC but generated in PHOTOS in the process of iteration.

In any case,
 even for PHOTOS running with standard options, the differences 
affect only a tiny fraction of the $Z$ decay phase space.  Thus, we do not consider it to be of much 
interest to continue the discussion of the missing terms.
Nevertheless, from a more fundamental side, we are disappointed by the fact that the comparisons did not provide 
numerical insight into the structure of the differences. 
The particularly interesting aspects of the study in the context of the extension of the algorithm for QCD did not bring any constructive indications so far.

\begin{center}
{\bf 7. Summary}
\end{center}
\vspace{.5cm} 
To quantify the size of the NLL effects, which are normally 
missing in PHOTOS, we reinstalled them back into the program,
using the original complete first-order expression for $Z$ decay. After the NLO correcting weight was installed, 
the differences between PHOTOS and  KORALZ
were  below the statistical error of $10^8$ events and for all the distributions 
used in the tests. Both PHOTOS and KORALZ were run at fixed first order without exponentiation.
The agreement provided a technical cross-check test for the two simulations. 
For the case of multiple-photon radiation in PHOTOS, a comparison 
with the  KKMC  generator\cite{kkcpc:1999} (exponentiation and  second-order matrix element used) was performed.
The implementation of the NLO terms in PHOTOS indicated, in the results of our universal test, 
an improvement  by a factor of about 50 for the observables 
sensitive to a single hard photon in the final states and  remained at the level of better 
than 0.1\% on the total rate for all other
cases we examined. Because of the smallness of the residual differences, it was difficult 
to understand their structure and origin in the final states with two hard photons.

The improvement in the agreement due to the introduction of the NLO correcting weight 
came at a price. Even though the weight is analytically simple and 
generation of weight 1 events remained possible, the calculation of the weight 
required information on Born-level coupling constants of the intermediate 
$Z/\gamma^*$. Also, the direction
of the beam was necessary in the calculation of the weight. These requirements
threatened the modular organization of the PHOTOS solution, as used
in the large Monte Carlo generation chains of experimental collaborations. Numerically, the introduced improvements are
 rather small and the deficiencies of standard PHOTOS are 
localized  in  the corners 
of bremsstrahlung phase-space populated by photons of very high energies and angularily,  well separated
from the final-state muons. Those regions of the phase space  
weigh less than 0.005  to the total
rate and the differences in that region approach 20\% of their size. The effects are thus  less  
than 0.1\% of the total rate of the $Z$ decay to muons. That is why we do not think it 
justified
to complicate the PHOTOS algorithm and to enable the use of the NLO correcting weight
in the general case.

The analysis presented here concentrates not only on the numerical  results for the final-state 
bremsstrahlung in $Z$ decay, but also on various aspects of a mathematical organization 
of the program for calculation of radiative corrections in $Z$ production
and decay. Separation of radiative corrections
into parts: (i)  embodied in effective couplings of the hard-scattering process,
(ii) final-state QED bremsstrahlung, and (iii) initial-state 
bremsstrahlung, eventually with initial-state hadronic interactions, were mentioned as well.
The effects of QED initial--final-state bremsstrahlung interference were to a large degree neglected.
Such an approach is reasonable in the leading-pole approximation for the $Z$,
but at a certain precision level the effects may need to be taken care of. For the
time being the results of  ref. \cite{Jadach:1988zp}
can be used instead.   

Thanks to the analytic form of the kernel used in PHOTOS for the single-photon emission, the analysis presented here may easily be 
 extended to other 
decay channels, if high precision is required and a calculation of matrix element 
 is available. The study for the case of $B$ meson decay into a pair  
of $\pi^\pm (K^\pm)$   is near completion 
\cite{Giso2006}.
In this case the questions of reliability of  scalar QED for the calculation of photons
of high $p_T$, with respect to charged scalars,
 need to be addressed. A natural extension of the study of the systematic error 
in PHOTOS simulations, as presented here, would be the discussion
of  bremsstrahlung in $W$  and Higgs boson decays.

The decays of the $W$ and Higgs bosons 
are probably the only ones where formal studies of the NL terms,
similar to the ones presented in this paper, can be performed. As 
 was the case with $Z$ decays, those cases will also be limited
to  the leading-pole approximation.

For other decay channels, the correction weight can be 
applied as well; however, in most cases the part of the weight going beyond soft and/or collinear
regimes may need to be constructed 
 with the help of the fits to the data.  Let us stress that the unique design of PHOTOS, enabling 
the use of the same 
kernel for multiple-photon radiation (exponentiated) 
mode and  at fixed first, second, third, and fourth orders,
establishes a convenient environment for such fits of form factors  to the data.
At the same time the analytical form of these form factors
can be taken from the first-order 
analytic calculations based on effective theory, or from any other model.

On the technical level it is worth mentioning that the  NLO correcting 
weight of PHOTOS can be used as an internal correcting weight.

Finally, let us stress that the approximations introduced in PHOTOS affect the matrix elements and not the 
phase space. The generation of the latter is based on the tangent space constructed from
eikonal approximation but used also for hard photons, even of energies above the available maximum enforced
by energy--momentum conservation. Only in the second step are phase-space constraints enforced. This is similar
to the case of the classical exclusive exponentiation. The energy momentum constraints
are introduced for each individual photon, step by step, and conformal symmetry is not used 
in that procedure. 

In principle, if necessary, complete higher-order matrix elements can
 be incorporated with the help of correcting  weights.
This point goes  beyond the scope of the present paper. This is equally true for the possible extensions
to simulations in QCD. 

\vskip 3mm

{\bf Acknowledgements.}
Useful discussions with Dmitri Bardin, Borut Kersevan, Maarten Bonekamp and Daniel Froidevaux are acknowledged. 
Special thanks go to Torbj\"orn Sj\"ostrand whose critical remarks convinced us
of the necessity and urgency to perform the presented work.

\providecommand{\href}[2]{#2}

\end{document}